\documentclass[12pt]{article}

\usepackage[utf8]{inputenc}
\usepackage{graphicx}
\usepackage{setspace}
\usepackage[margin=1in]{geometry}
\usepackage{titlesec}
\usepackage{titling} 
\usepackage{authblk} 
\usepackage[
    backend=biber,
    style=nature,
    sorting=none,
    natbib=true,
    defernumbers=true
]{biblatex}
\usepackage{subcaption}
\usepackage{caption}
\usepackage{svg}
\usepackage{pbox}
\usepackage{amsmath}
\usepackage{tcolorbox}
\usepackage{enumitem}
\tcbuselibrary{listings}
\usepackage{tcolorbox}
\usepackage{afterpage}
\usepackage{cellspace}
\usepackage{hyperref}

\newtcblisting{promptbox}[1]{%
  title=#1,
  listing only,
  colback=white!95!gray,
  colframe=gray,
  width=\columnwidth,
  arc=0mm,
  boxrule=0.5mm,
  colbacktitle=white!95!gray,
  coltitle=black,
  fonttitle=\bfseries,
  left=6pt,
  right=6pt,
  top=6pt,
  bottom=6pt,
  listing options={basicstyle=\scriptsize\ttfamily, breaklines=true}
}

\DeclareFieldFormat{labelnumber}{\textsuperscript{#1}}
\DeclareFieldFormat{citationlabel}{\textsuperscript{#1}}
\let\cite\supercite

\titlespacing\section{0pt}{12pt plus 4pt minus 2pt}{6pt plus 2pt minus 2pt}

\title{Billions of Sketches Reveal Hidden Cultural Variation in Human Concepts}

\author[1]{Arianna Pera$^*$}
\author[2]{Mauro Martino$^*$}
\author[2]{Nima Dehmamy}
\author[3]{Douglas Guilbeault}
\author[1,$\dagger$]{Luca Maria Aiello}
\author[4,$\dagger$]{Andrea Baronchelli}

\affil[1]{IT University of Copenhagen, Copenhagen, Denmark}
\affil[2]{MIT-IBM Watson AI Lab, Cambridge, MA, USA}
\affil[3]{Stanford University, Stanford, CA, USA}
\affil[4]{City St George's University of London, London, UK}

\makeatletter
\let\old@maketitle\maketitle
\renewcommand{\maketitle}{
  \old@maketitle
  \vspace{-1em}
  \begin{center}
    \footnotesize{$^*$These authors contributed equally to this work.}\\
    \footnotesize{$\dagger$ Email: luai@itu.dk, andrea.baronchelli.1@citystgeorges.ac.uk}
  \end{center}
}
\makeatother

\begin{document}

\maketitle

\begin{refsection}

\begin{abstract}
\noindent Claims about the universality of human concepts have been predominantly assessed through linguistic similarity across languages and cultures. However, words are effective as communication devices because they compress rich experiential variation into shared conventions, potentially obscuring hidden individual and cultural differences in how concepts are mentally represented. Here, we analyse 2.6 billion human-made sketches of common concepts from 236 countries and territories to examine conceptual structure through people's visual imagination. Consistent with recent work on image-based cognition, we find that single concepts unfold into multiple distinct visual exemplars, revealing latent information about similarities and differences in conceptual structure across cultures. This variation is strongest for concepts involving haptic interaction, suggesting that visual imagery reflects variation in embodied experience as much as conventional definitions. Comparing embedding models of sketches with word embedding models across languages, we find that their geometries diverge, with visual representations preserving rich semantic and cultural structure that language models compress. Cross-cultural similarities derived from sketches align 32\% more closely with established cultural distances than do text-based measures. Together, these results suggest that patterns of human conceptual universality may depend critically on the modality through which concepts are measured, with large-scale sketching providing a direct, high-resolution probe of conceptual diversity across embodied and cultural dimensions of thought.
\end{abstract}

\vspace{1em} 

\section{Introduction}
How universal are human concepts? Recent advances in natural language processing have made it possible to explore this longstanding question empirically by identifying similarities between linguistic concepts across languages and cultures, as captured by the statistical associations of words in large online corpora. Yet the results of these efforts have been markedly mixed. On the one hand, several large-scale textual analyses provide evidence that different languages and cultures conceptualise common aspects of experience in highly similar ways, ranging from colours and actions to natural objects and kinship, supporting the existence of largely universal concepts~\cite{berlin1991basic, kemp2018semantic, lewis2023local, xu2020conceptual, liang2024shared, zaslavsky2018efficient, san2018universal, kemp2012kinship}. On the other hand, studies using similar techniques report striking cross-cultural variation in the conceptual structure of highly familiar and seemingly basic domains of everyday life, including emotions, body parts, and food~\cite{thompson2020cultural, jackson2019emotion, tjuka2024universal, thompson2018quantifying}. However, it remains unclear whether these inconsistencies reflect genuine variation in the level of universality in conceptual structure across cultures, or rather variation at the level of word-based representations, which can co-exist with other forms of variation in conceptual structure, for example in image-based or affect-based associations. A defining feature of word-based communication is that it is a lossy medium of expression ~\cite{guilbeault2021experimental}: by design, words compress enormous amounts of detail and variation from sensory experience to communicate abstractions that generalize across contexts (e.g., humans perceive many more colors than are captured by the dozen or so color words in common parlance). This raises the question of whether, to complement word-based analyses, insights can be gained into cultural variation by examining people's conceptual associations within a richer, less lossy medium like images, which can preserve perceptual variation. Indeed, research on embodied cognition shows that human concepts consist both of word-based and image-based sensory information ~\cite{barsalou2010grounded, lakoff2012explaining, bergen2012louder, lewis2021characterizing}, suggesting a fuller representation of cultural variation in conceptual structure can be developed by comparing and integrating measures of conceptual similarity across words and images. 

At present, automated analyses of textual data are especially well-represented in the cognitive and social sciences due, in large part, to the abundance of textual data online and the availability of accessible algorithmic methods for analyzing this data to make robust psychological and cultural predictions. Yet, it has long been recognized in cognitive science that words capture an important but only partial view of conceptual structure. Numerous studies indicate that language and thought are not equivalent~\cite{fedorenko2024language, malt2024representing}; for example, a substantial body of work on embodied cognition finds that sensorimotor representations and implicit image-based associations are also a foundational aspect of conceptual structure ~\cite{barsalou2010grounded, lakoff2012explaining, bergen2012louder, guilbeault2020color, nadler2025statistical, mcneill1992hand, xu2009symbolic, willems2007language, ozyurek2007online}. Indeed, numerous studies observe that the brain activates detailed sensorimotor images while reading and interpreting textual data~\cite{bergen2012louder, fernandino2022decoding, bechtold2023brain}.
Together, this evidence suggests that non-linguistic representations may capture patterns of cultural variation and universality in conceptual structure that can complement word-based analyses. In recent years, a number of studies have turned to analyzing the images that humans produced when representing everyday concepts, yielding valuable insights into conceptual structure \cite{mukherjee2025drawings, zhu2025crosscontextual, long2023developmental, long2024parallel, yu2016sketch, xu2022deep, lewis2021characterizing}. However, to date, these studies have relied on datasets that are comparably much smaller (e.g., a recently released dataset of human drawings described as ``large'' contains less than 30k images~\cite{mukherjee2025drawings}) to the analyses enabled by word-based techniques, limiting direct cross-medium comparisons \cite{zhu2025crosscontextual, long2023developmental, long2024parallel, yu2016sketch, xu2022deep}. No comparably large-scale, cross-cultural dataset exists for analysing the visual representations people produce for everyday concepts under shared communicative constraints. In this study, we address this gap by analyzing over 2.6 billion sketches generated by people from 236 countries and territories for the purpose of visually representing everyday concepts.

Our analyses are based on the data from \href{https://quickdraw.withgoogle.com/}{QuickDraw}, an online game where players are asked to sketch various \emph{concepts} such as objects, actions, or animals, within a 20-seconds time limit. In every round of the game, a concept is randomly selected from a predefined list. As users draw, a neural network attempts to recognize their sketches in real time, improving its own classification performance as more drawings are contributed. QuickDraw encourages participation with the message: ``Can a neural network learn to recognize doodling? Help teach it by adding your drawings to the world's largest doodling dataset, shared publicly to help with machine learning research.'' Our dataset contains 2.6 billion sketches across 344 distinct concepts, collected from November 2016 to December 2019. Each sketch includes pixel-level stroke data, the drawing duration (median 9.6 seconds, full distribution in Supplementary Information (SI)), and the user's country inferred from their IP address. Although 41.3\% of the sketches were produced by users in the United States, the dataset spans 236 countries and territories (see SI for the full geographic distribution of the 100 most represented ones). A much smaller sample of this data has been widely used in AI research, from statistical analysis~\cite{fernandez2019quick} to training and testing computer vision models~\cite{ha2017neural, xu2018sketchmate, xu2021multigraph, lamb2020sketchtransfer, xu2022deep}. In recent cognitive science research~\cite{lewis2021characterizing} one paper in particular examines a subsample of this dataset involving a few million sketches, revealing cultural variation in the depiction of concepts and its connection to population size; however, the smaller scale of this dataset limits the ability to robustly generate image-based embeddings for sketches within and across cultures of the kind amenable to comparison with corresponding word-based embedding models. Consequently, no prior uses of the QuickDraw data have systematically compared image-based embeddings of sketches to corresponding word-based embeddings of the same concepts. 
Here, we present a full-scale examination of the entire global QuickDraw dataset, enabling direct comparison and integration with corresponding embedding models of conceptual similarity via word-based measures. 

\section*{Results}

\subsection*{Sketches unfold into multiple recurrent visual forms}
We begin by testing two alternative hypotheses about how concepts are collectively represented~\cite{murphy2016there,goldstein2015cognitive}. Under a one to one mapping, a given concept would converge on a single dominant visual form shared across individuals, consistent with the idea of a single prototype~\cite{rosch1975family,rogers2004structure}. Under a one to many mapping, the same concept would instead give rise to multiple distinct but recurrent visual forms, corresponding to different exemplars~\cite{medin1978context,smith1981categories}. To distinguish between these possibilities, we group sketches according to their visual similarity using a clustering approach. Sketches are compared in a latent visual space that captures structural properties of the drawings, such as shape and orientation (see Methods). The resulting clusters correspond to homogeneous visual attractors, each characterised by a distinct and recognisable appearance, and are clearly separable from sketches that are too heterogeneous or noisy to form stable patterns.

\begin{figure}[t!]
    \centering

    \begin{subfigure}{0.31\textwidth}
        \includegraphics[width=\linewidth]{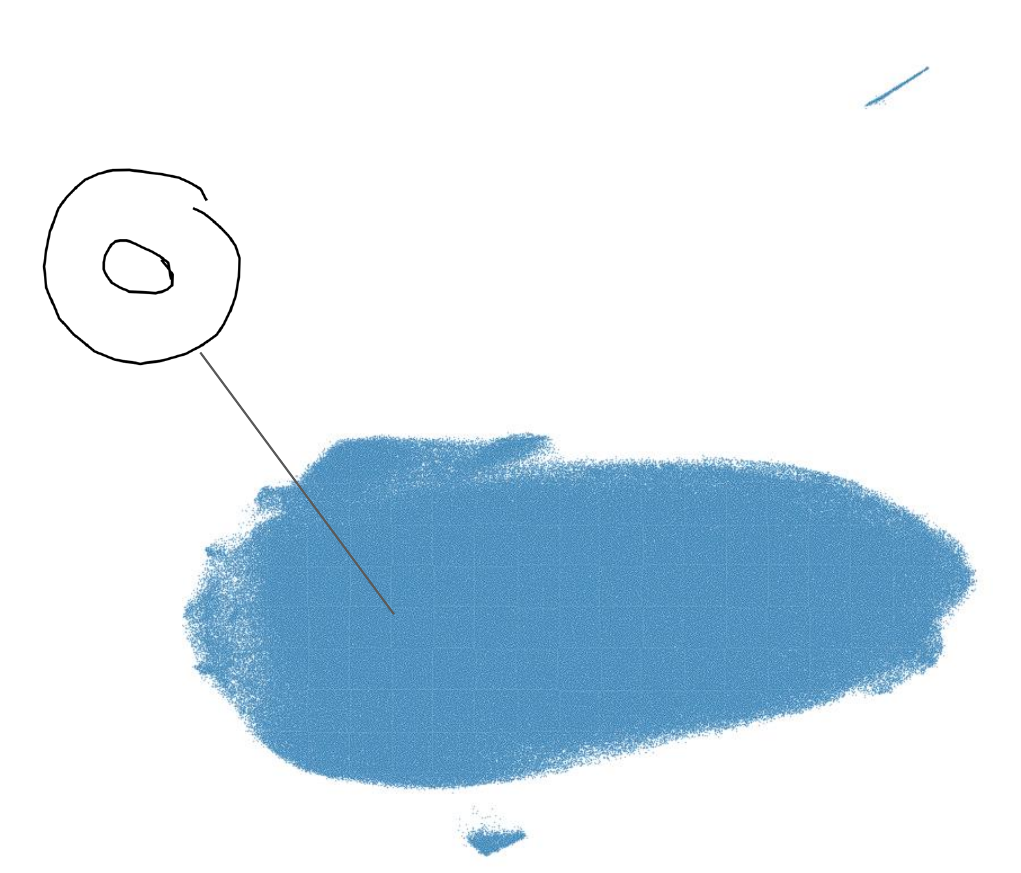} 
        \caption{Donut}
    \end{subfigure}
    \hspace{0.1cm}
    \begin{subfigure}{0.31\textwidth}
        \includegraphics[width=\linewidth]{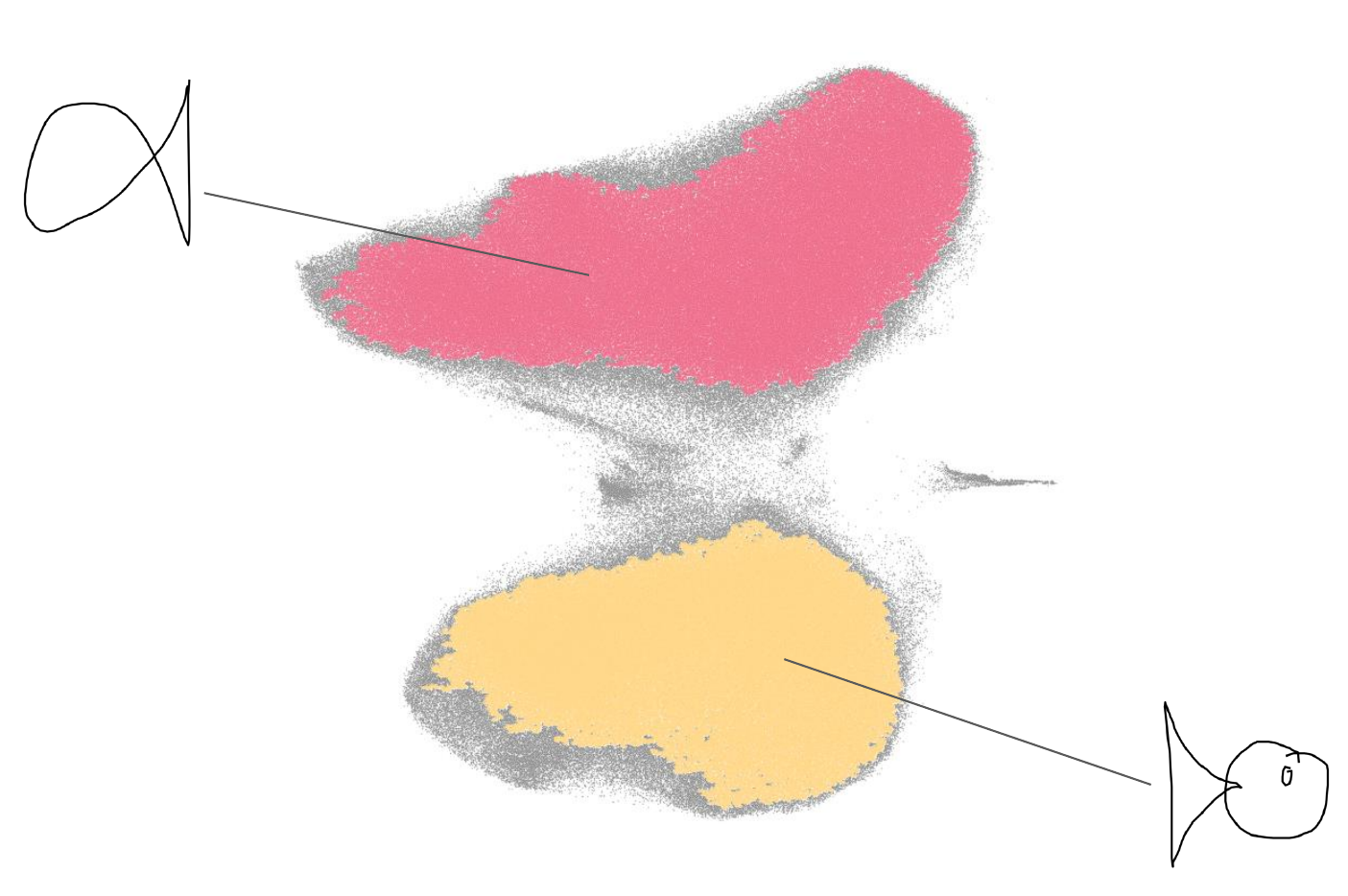}
        \caption{Fish}
    \end{subfigure}
    \hspace{0.1cm}
    \begin{subfigure}{0.31\textwidth}
        \includegraphics[width=\linewidth]{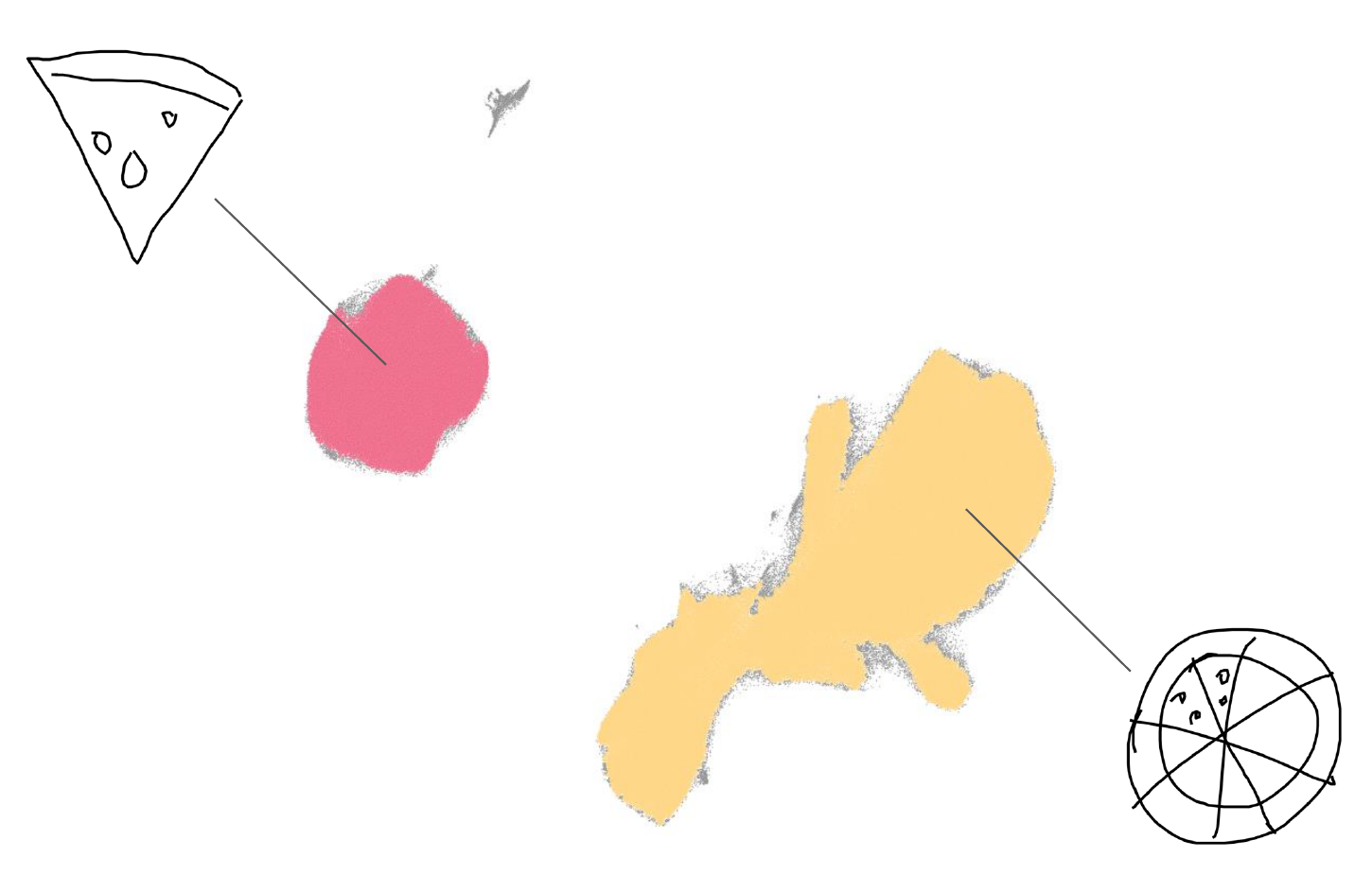}
        \caption{Pizza}
    \end{subfigure}

    \vspace{0.3cm}
    \begin{subfigure}{0.31\textwidth}
        \includegraphics[width=\linewidth]{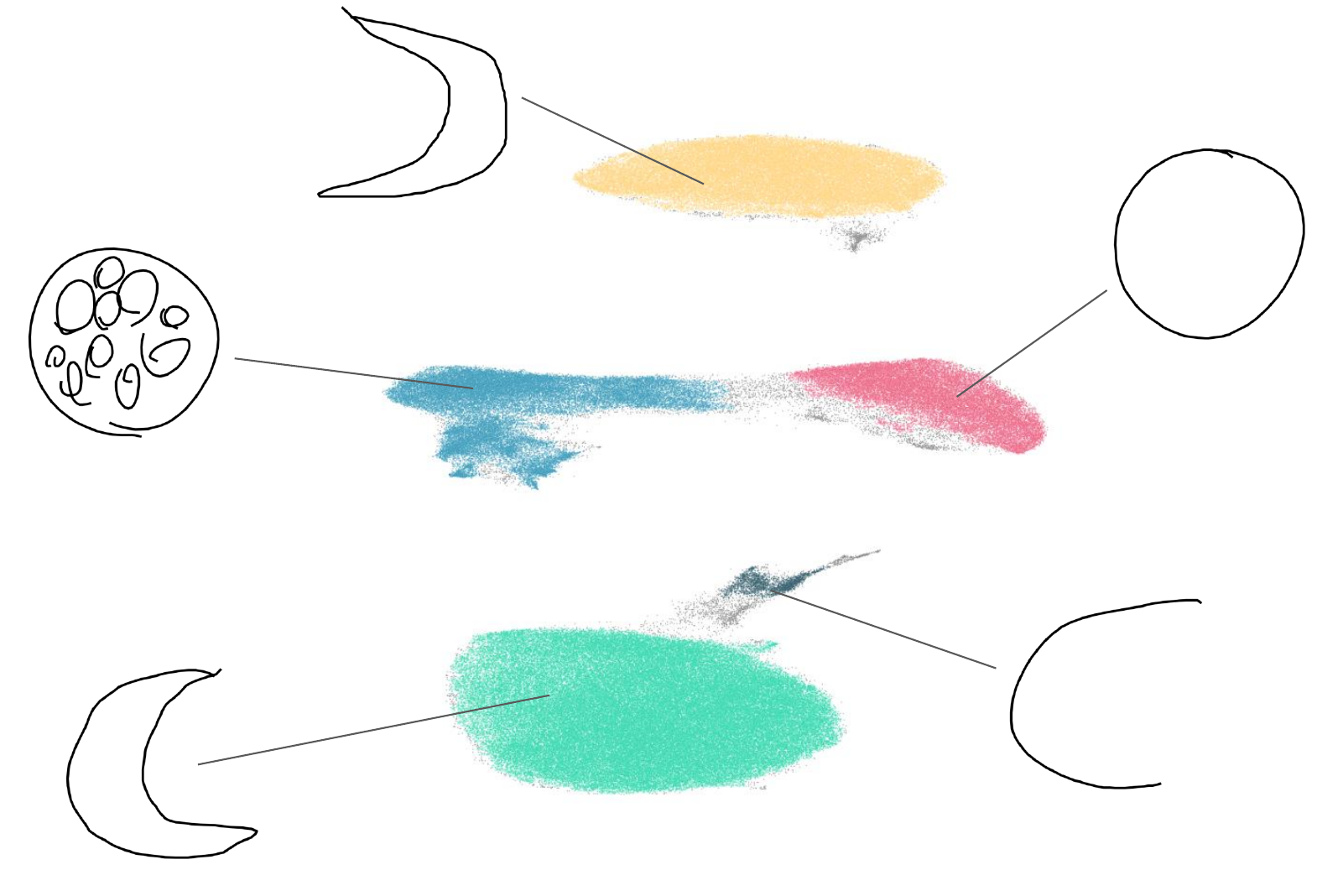}
        \caption{Moon}
    \end{subfigure}
    \hspace{0.1cm}
    \begin{subfigure}{0.31\textwidth}
        \includegraphics[width=\linewidth]{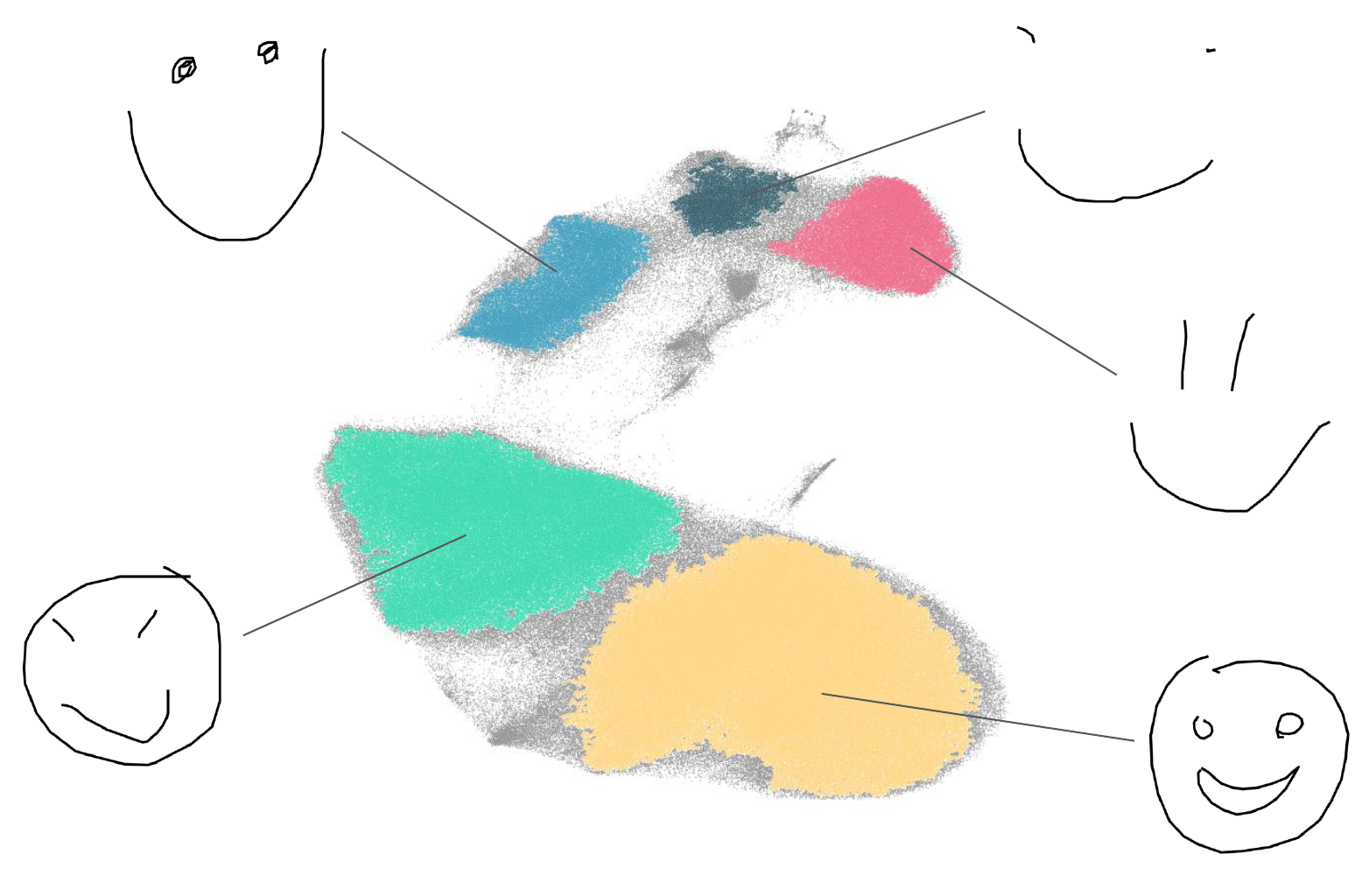}
        \caption{Smiley face}
    \end{subfigure}
    \hspace{0.1cm}
    \begin{subfigure}{0.31\textwidth}
        \includegraphics[width=\linewidth]{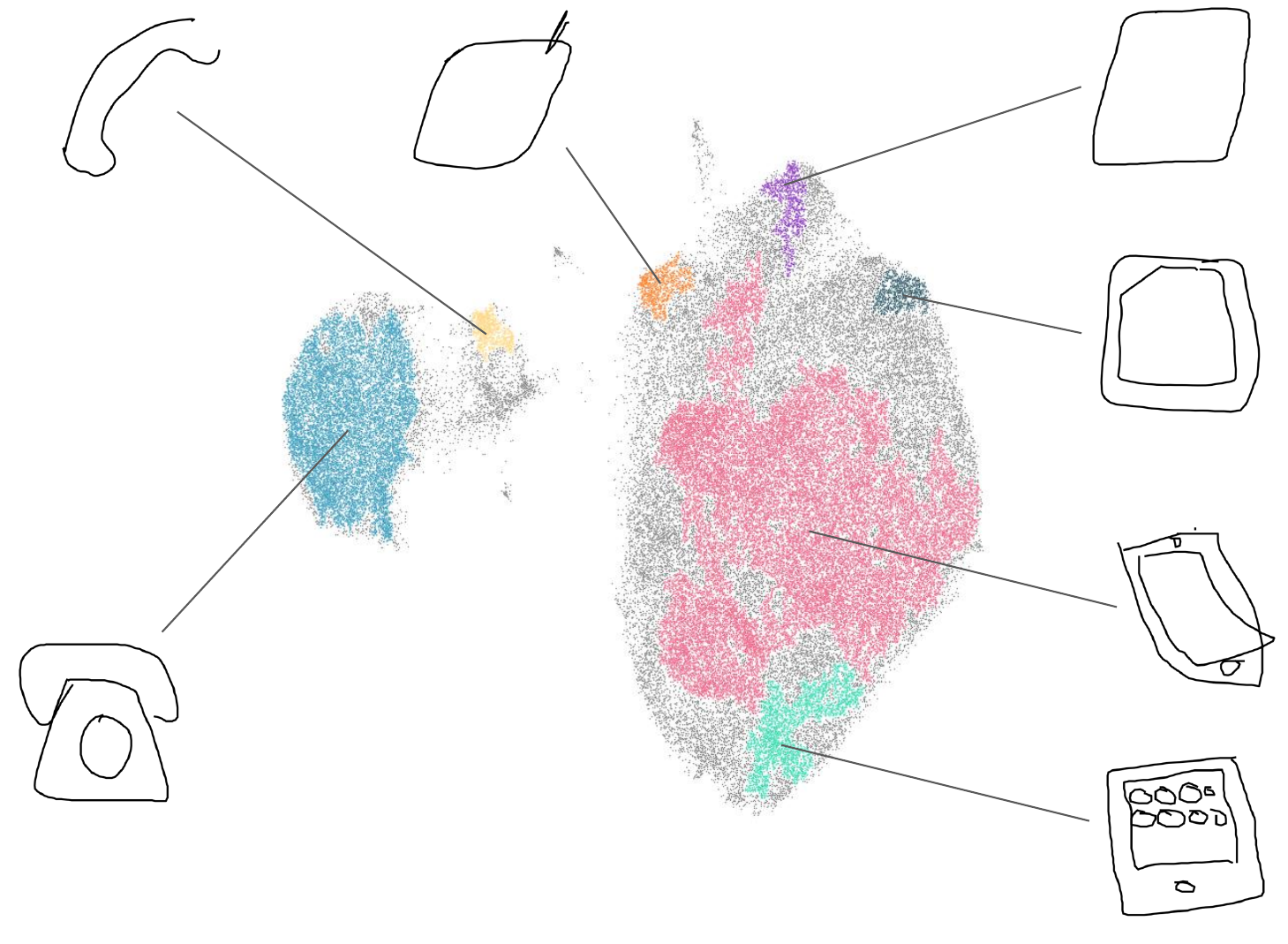}
        \caption{Phone}
    \end{subfigure}

    \caption{\textbf{Sketches of concepts provided by people around the world organize into distinct visual clusters.} Visual clusters for six representative concepts (a–f) selected from the 344 available. Each point represents a drawing projected into a two-dimensional latent visual space. Points are colored according to their algorithmically assigned cluster, with gray points denoting drawings classified as random noise. 
    Each cluster is annotated with an image sampled from sketches within the cluster, revealing the visual forms that clusters represent. Superposition of randomly sampled drawings within the clusters are reported in SI.
    }
    \label{fig:clustering}
\end{figure}

For almost all concepts, sketches are neither completely random nor collapsed into a single prototype, consistent with smaller scale analyses of this dataset \cite{lewis2021characterizing}. Instead, they cluster into a small number of stable exemplar forms. Only a few concepts, such as \emph{donut}, converge on a single dominant form. Others, such as \emph{fish} (left versus right orientation) or \emph{pizza} (slice versus whole), split into two main clusters, while some, such as \emph{phone}, exhibit a richer diversity of forms, including landline designs and multiple smartphone variants. The median number of visual clusters per concept is 2, ranging from 1 (for example, \emph{donut}) to 9 (for example, \emph{watermelon}), with a single outlier at 21 clusters (\emph{crow}) (see the full distribution in SI). Figure~\ref{fig:clustering} illustrates this spectrum, from highly uniform to highly varied representations.

\subsection*{Sketch clusterability is selectively linked to sensorimotor properties}
\begin{figure}[t!]
    \centering
    \includegraphics[width=0.9\textwidth]{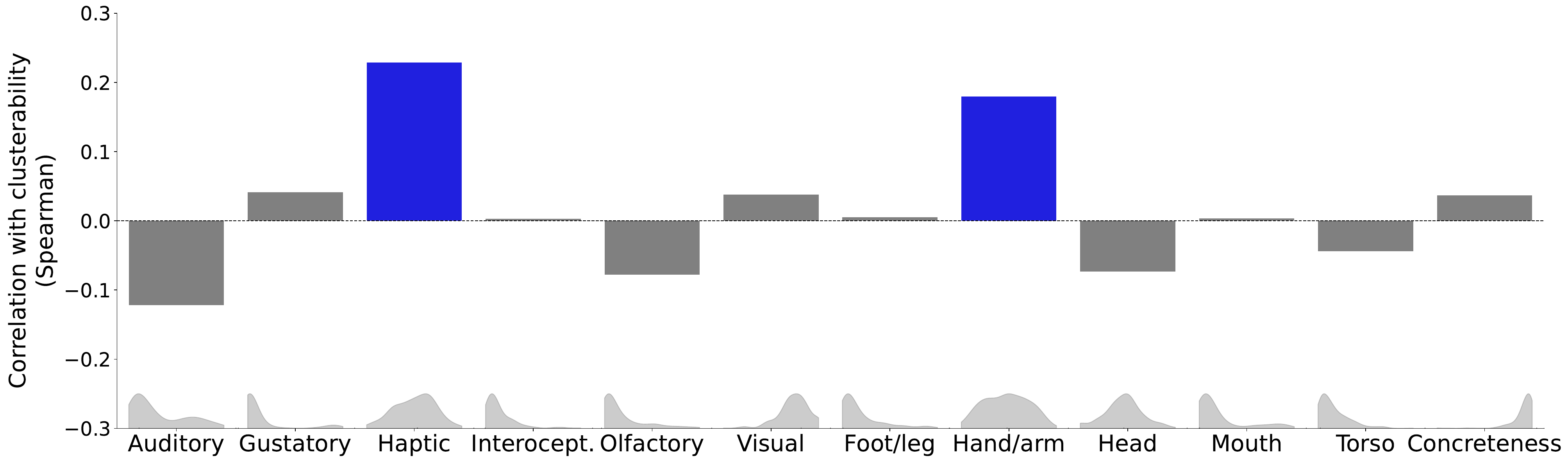} 
    \caption{\textbf{Clusterability of sketches is selectively associated with haptic and sensorimotor conceptual properties.} Correlation between clusterability and conceptual properties of objects. Blue bars correspond to significant values ($\alpha=0.05$, after Bonferroni correction), grey to non-significant ones. The density distribution of property scores is reported on the \emph{x}-axis. To ensure that the correlations were not driven by systematic differences in the distribution of conceptual property scores, we tested the association between the correlation coefficients and the mean values of the conceptual properties scores, finding no evidence of bias (Spearman's $\rho = 0.28$, $p = 0.38$).}
    \label{fig:conceptual_prop}
\end{figure}

Some concepts tend to be drawn in a limited number of recognisable and coherent ways, whereas others are represented more variably or ambiguously. This variation can be quantified through a measure of \emph{clusterability}, defined as the proportion of sketches belonging to well-defined visual clusters relative to those classified as noise. Clusterability is robust to the choice of the latent visual space (see SI).
We investigate whether differences in clusterability can be explained by intrinsic properties of the concepts themselves. Drawing on extensive evidence from psycholinguistics and cognitive neuroscience showing that conceptual experience is deeply embodied~\cite{wilson2002six}, we characterise each concept by its \emph{concreteness} and \emph{sensorimotor experience}. These properties capture, respectively, the extent to which a concept refers to a perceptible entity and the strength of its association with sensory and motor processes. We map each concept to 12 validated numerical scores for concreteness and sensorimotor experience available for thousands of English lemmas~\cite{brysbaert2014concreteness, lynott2020lancaster} (see Methods). Figure~\ref{fig:conceptual_prop} shows that most correlations between clusterability and these conceptual properties are low and non-significant. However, concepts associated with haptic experience, or involving sensory interactions mediated by the hands or arms, tend to form significantly more well-defined clusters. In other words, objects that can be physically manipulated correspond to more coherent visual representations.

\subsection*{Visual representations diverge structurally from linguistic representations}
The observation that sketches of the same concept consistently cluster around multiple visual attractors suggests that collective conceptual representations extend beyond the semantics of the individual words that label them. Importantly, the multiple visual representations we observe do not merely add fine grained detail to the meaning of words. Instead, they reveal a representational space whose structure diverges systematically from that of language. This divergence manifests in two distinct ways.

First, visual clusters corresponding to the same concept are rarely close to one another in the latent visual similarity space. To quantify this, we focus on concepts that split into multiple visual attractors, such as \emph{pizza}, represented as a slice versus a whole pie. Starting from a target visual attractor, we rank all clusters across all concepts by decreasing visual similarity. The most similar visual cluster belongs to the same concept as the target in only 6.3\% of cases and, on average, the first same concept cluster appears only after the top 9\% of the ranked list. This indicates that while alternative representations of the same concept may appear relatively early in the similarity ranking, they are rarely the closest match. For example, the visual cluster closest to the pizza slice exemplar corresponds to the \emph{triangle} concept, whereas the cluster representing a whole pizza pie appears at the 284th position among 344 categories. Importantly, expanding the concept set beyond the categories that are available could introduce additional visually similar alternatives into the similarity ranking, making the observed exemplar separation a conservative estimate.

Second, concepts that are visually similar are rarely semantically close. To quantify this divergence, for each target concept we compute rank correlations between two lists containing all remaining concepts, ranked according to visual similarity or semantic similarity derived from image embeddings and word embeddings, respectively (see Methods). Across all 344 concepts, similarity rankings obtained from the two modalities show consistently low agreement, with a macro average rank correlation of $0.098$ (on a scale from 0 to 1) across multiple correlation metrics and embedding models (see SI for details). This low correspondence indicates that visual representations capture systematic relationships between concepts that are not readily accessible through linguistic representations alone.

\subsection*{Images capture cross-national cultural similarity better than words}

\begin{figure}[t!]
    \centering
    \begin{subfigure}{0.43\textwidth}
        \includegraphics[width=\linewidth]{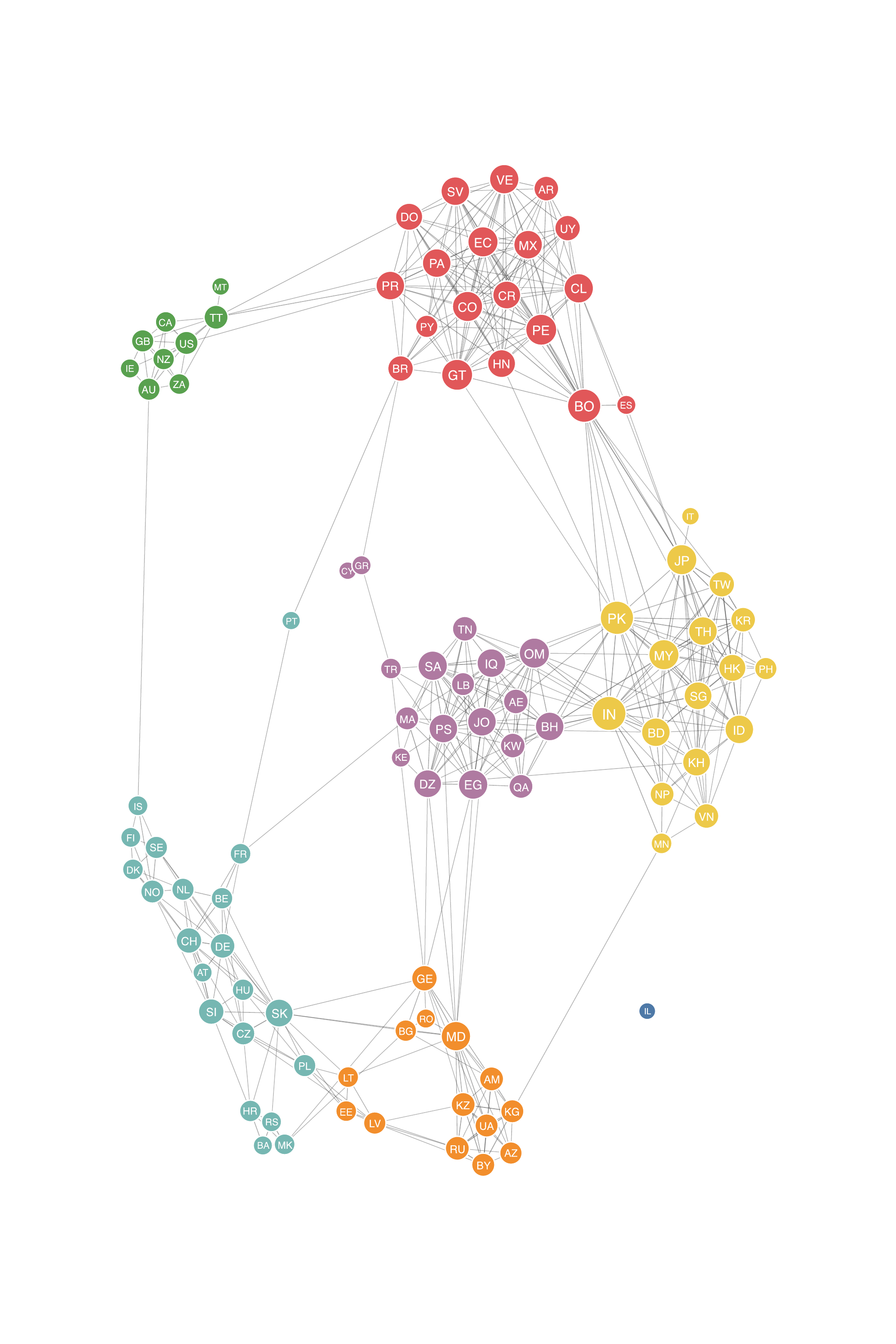} 
    \end{subfigure}
    \hspace{0.5cm}
    \begin{subfigure}{0.43\textwidth}
        \includegraphics[width=\linewidth]{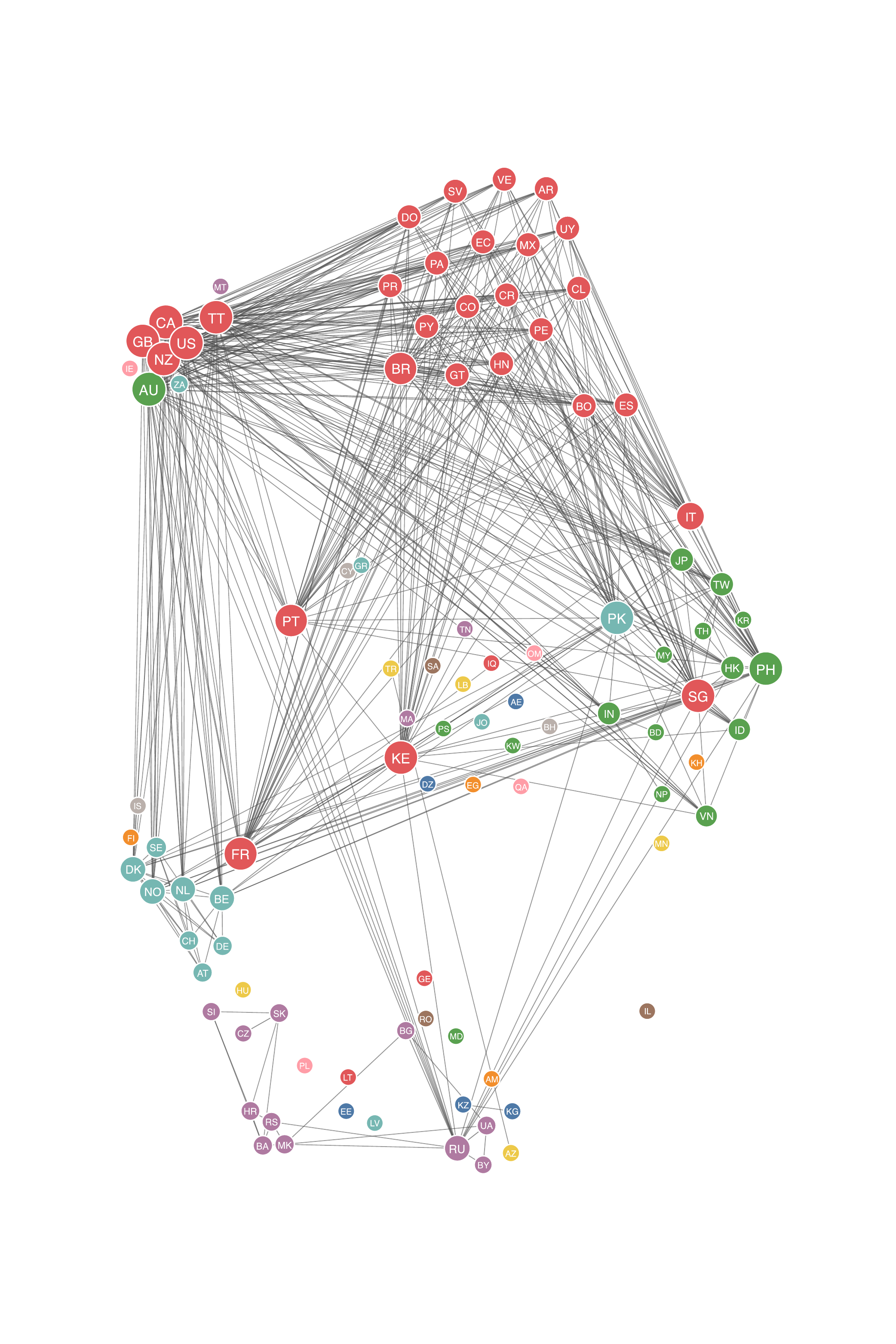}
    \end{subfigure}

    \caption{\textbf{Image- and word-based concept networks exhibit divergent large-scale patterns of inter-cultural distances and clustering.} Networks of countries based on sketch similarity (left) and word similarity (right). The top 100 nodes by number of sketches are shown. Colors denote structural communities of countries that share a high level of similarity with one another within each network, as identified by the Louvain algorithm. To ease the visual comparison, the coordinates of nodes in the word similarity network are the same as in the image similarity network.}
    \label{fig:wordimage_network}
\end{figure}

Overall, sketches do not merely replicate the meaning of words with added noise, but reveal an alternative conceptual geometry of exemplars that language compresses. This divergence reflects not only structural differences between visual and linguistic representations, but also how these modalities differentially encode cultural information. In particular, we find that the collective visual representation of common concepts aligns more closely with established cultural patterns than do their corresponding representations in the semantic space of words. In this way, our findings support the claim that the fullest picture of conceptual structure benefits from integrating and comparing word-based and image-based embedding measures of conceptual similarity.

To quantify the relationship between sketches, words, and culture, we construct two networks of countries based on participants’ country information: one capturing similarity in visual representations and the other capturing similarity in linguistic representations (see Methods). In the sketch similarity network, edge weights reflect how similarly two countries depict concepts across visual exemplar clusters. In the word similarity network, edges are weighted by the average semantic distance between translated concept words in the countries’ primary languages, computed using multilingual word embeddings. Both networks are complete by construction, with all country pairs connected. To focus on the most informative structure, we restrict the analysis to the 100 most represented countries, accounting for 99.5\% of all drawings, and retain only the top 10\% of edges by weight. We then apply community detection using the Louvain algorithm~\cite{blondel2008fast} to identify groups of densely connected countries that share high internal similarity. For the sake of robustness, we also test community detection using the Leiden algorithm~\cite{traag2019louvain}. The two algorithms produce the same results.

Figure~\ref{fig:wordimage_network} illustrates how the structure of the two networks diverges substantially. The image-based network reveals clear regional communities, including English speaking countries, South America, Europe, post-Soviet Eurasia, Africa and the Middle East, and Asia. By contrast, the word-based network exhibits weaker and less coherent clustering. Some regions, such as Eastern Europe and Asia, are visible in both modalities, but appear more faintly in the linguistic network. Together, these results indicate that while visual and linguistic representations of concepts share a limited cross-cultural component, their large-scale structures are largely distinct.

\begin{figure}[t!]
    \centering
    \includegraphics[width=0.6\textwidth]{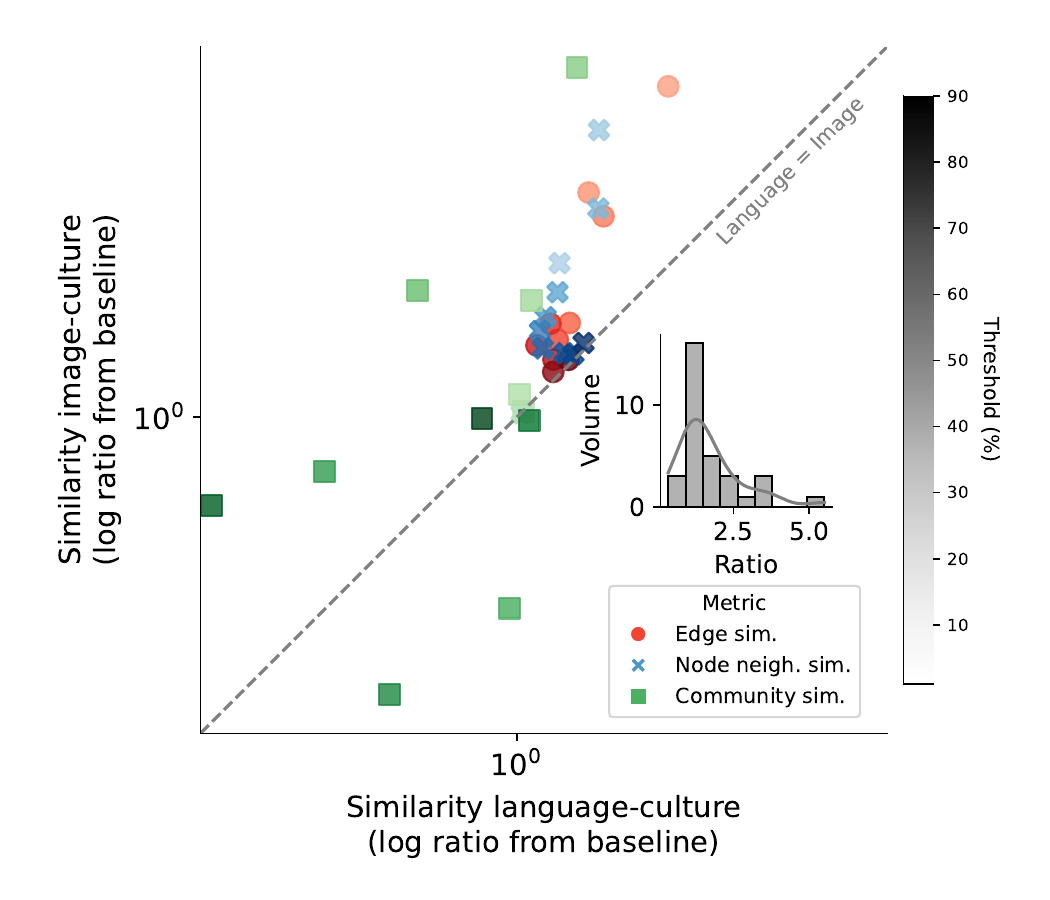} 
    \caption{\textbf{Cultural similarity aligns more closely with image-based than word-based concept networks.} Comparison between image- and language-based network similarity with the cultural network across different metrics, expressed as ratios relative to a baseline defined by a null model. The networks are compared in terms of edge similarity, node similarity, and community overlap. Measurements are repeated on multiple network versions obtained with different edge filtering thresholds (1\%, 5\% and from 10\% to 90\%, in 10\% increments), with lighter shades indicating lower thresholds and darker shades higher ones, as shown in the accompanying grey-scale colormap. Markers above the diagonal indicate a higher similarity between image and culture compared to words and culture. The embedded histogram shows the distribution of the ratio between image-based and language-based similarity with the cultural network across all metrics and thresholds.}
    \label{fig:metrics_similarity_culture}
\end{figure}

Finally, we compare both networks to a third cultural similarity network derived from the World Values Survey~\cite{muthukrishna2020beyond}, a large-scale and validated longitudinal dataset of global values and beliefs. Figure~\ref{fig:metrics_similarity_culture} shows that the cultural network aligns substantially more closely with the sketch based network than with the word-based network across three alternative network similarity metrics. Considering the median across experimental conditions, image culture similarity exceeds language culture similarity by 32\%. 
This difference is statistically significant and robust to the World Values Survey wave and survey dimensions used, as well as to the number of nodes, edge-retention threshold, and network filtering method considered (see SI).

\section*{Discussion}
This study demonstrates that large-scale visual sketching offers a unique window onto human conceptual structure that reveals rich semantic and cultural variation not captured by standard language-based analyses alone. We find that concepts are organised in a richly structured space with individual and cultural heterogeneity largely invisible to word-based embedding models, consistent with the view that words, as abstractions by design, compress away rich perceptual variation. This visual space of perceptual variation nevertheless exhibits (i) stable exemplar organisation, with a median of two visual clusters of distinct exemplars per concept, (ii) systematic divergence from word-based semantic geometry, as reflected in a 0.098 correlation between visual and linguistic similarity rankings, and (iii) substantially closer alignment with cultural variation than text-based measures, with a 32\% improvement in correspondence with World Values Survey distances between cultures. Together, these observations indicate that longstanding debates about conceptual universality and diversity are unlikely to be resolved within a single modality, and that the extent to which concepts appear as universal can depend critically on how concepts are measured and which representational channels are used to access them. These findings, therefore, attest to the multimodal nature of human concepts and the value of developing multimodal-based computational paradigms (e.g., via embedding methods) for large-scale data driven efforts in cognitive science. 

Several limitations of our QuickDraw dataset should be acknowledged. The sample is characterised by a predominance of anglophone users and by the under-representation of certain countries and regions, reflecting patterns of participation and data availability rather than properties of the concepts or cultures themselves. Demographic information such as age, education, and economic background is unavailable, and user location is inferred from IP addresses, introducing additional uncertainty. Participation is also shaped by access to digital devices, internet connectivity, and discretionary time, likely biasing the sample toward more socioeconomically privileged cohorts (though, it is worth noting that a similar problem pervades the training data and responses of popular LLMs~\cite{atari2023humans}). Additional sources of variability include unobserved differences in drawing devices and interfaces, the very short time limit imposed on sketch production, and occasional attempts to circumvent the task, such as writing concept names rather than drawing them.
Each of these factors introduces variability at the level of individual drawings or participants.

At the same time, the unprecedented scale of the dataset fundamentally changes the inferential regime~\cite{michel2011quantitative}. With billions of sketches contributed across a wide range of contexts, it becomes possible to identify stable and recurrent patterns in visual representation that persist despite substantial variability at the level of individual drawings and participants. Scale does not eliminate the limitations associated with sampling bias or incomplete demographic coverage, but it enables robust inference about aggregate structure by revealing regularities that are reproducible across heterogeneous conditions. This level of scale allows systematic properties of conceptual representation, embodiment, and cultural variation to be detected in ways that would not be feasible in smaller or more tightly controlled samples. Future work could complement these findings by designing data collection efforts that intentionally broaden participation across under-represented regions and socioeconomic contexts.

These findings have direct implications for ongoing work in artificial intelligence and its relation to theories of human cognition and culture. Recent advances in large language models have led some to argue that learning statistical associations among words may be sufficient to enable LLMs to acquire human concepts and reasoning, and to robustly model and predict human cognitive processes, including the approximation of perceptions and meanings associated with embodied experience~\cite{stockl2021watching, loyola2023perceptual, piantadosi2024concepts, frank2025cognitive,  marjieh2024large}. Yet, our results show that word-level textual representations mask considerable latent variation in how people visually construe concepts within and across cultures, suggesting that purely language-based analyses alone are limited as models of human conceptual structure. This suggests that multimodal training that incorporates more directly embodied features of human conceptual structure -- such as the visual or other sensory-related imagery that people generate to depict concepts -- may be necessary for artificial systems to approximate the structure and cultural organization of human concepts and their diversity.

\section*{Methods}

\subsection*{Data Processing}
Our dataset contains only sketches that have been recognized by the neural network in the QuickDraw game. Note that the classifier may recognize a sketch before it is finished; while this early-stop is an intrinsic property of the data, it might truncate elements of the drawing that could be linked to cultural aspects.

To analyze the structure of sketch data, we first reconstruct images from their raw pixel coordinates. We then center each image in the coordinate space to allow better comparison.
Each sketch is finally embedded using the DINOv2-small image embedding model, a self-supervised vision transformer trained on large-scale datasets~\cite{oquab2023dinov2}. 
The DINOv2 model generates a 384-dimensional feature representation for each image, capturing high-level semantic and structural information. 
CLIP-ViT-B-32, resulting in 512-dimensional embeddings, is also considered for robustness checks~\cite{radford2021learning}.

To prevent imbalance from overly represented countries (e.g., the US), we down-sample the data by focusing on the 100 most prominent countries and capping the number of drawings per category–country pair at 10,000. In total, we are left with 223,349,940 drawings.
Given the high dimensionality of the embeddings, we perform dimensionality reduction to facilitate clustering. 
First, we apply Principal Component Analysis (PCA) to reduce the embeddings to 40 dimensions while retaining the maximum variance possible. Next, we further reduce the data to 2 dimensions using Uniform Manifold Approximation and Projection (UMAP)~\cite{mcinnes2018umap}, which is particularly effective for preserving the global structure of the data in low-dimensional space.

\subsection*{Clustering}
Once the 2D representations are obtained, we apply Density-Based Spatial Clustering of Applications with Noise (DBSCAN)~\cite{ester1996density} to identify clusters of sketches within each object class. 
DBSCAN is well-suited for this task as it does not require specifying the number of clusters a priori and can identify clusters of varying density while marking outliers as noise. 
The key parameter in DBSCAN is $\epsilon$, which defines the radius of a neighborhood around each core point. 
To determine the optimal value of $\epsilon$, we employ Density-Based Cluster Validity (DBCV)~\cite{moulavi2014density} optimization, which evaluates the quality of different clustering solutions based on density connectivity and intra-cluster compactness. 
To avoid the presence of a large number of very small clusters, we decided to consider sketches part of clusters containing less than 1\% of the total number of sketches as noise. 
This approach ensures that the clustering results reflect the natural groupings within the sketches, allowing us to identify meaningful variations in drawing styles and representations within each object category.

For some concepts, identifying meaningful clusters with DBSCAN proves challenging, as the algorithm often returns a high proportion of noise points without extracting clear groups.
By examining the distribution of noise percentages across all categories, we observe a bimodal pattern (see Supplementary Information (SI)). We use the local minimum between the two modes as a threshold to distinguish \textbf{clusterable} concepts (with noise percentages below the threshold) from \textbf{non-clusterable} ones (with noise percentages above the threshold) under the current clustering approach.

\subsubsection*{Grid Components Clustering}
Because our primary goal is to detect prototypes or exemplars for each concept and we are able to do so on the clusterable concepts, we perform a complementary analysis on the non-clusterable ones.
This empirical method identifies connected components of high-density regions in the 2D UMAP embedding space by overlaying a grid and analyzing the density of points within each cell.
Specifically, for each concept, we: (i) impose a grid over the 2D UMAP representation of the sketches from the concept, (ii) compute the proportion of sketches falling into each grid cell.
Next, we examine the distribution of proportions across all cells and concepts to determine percentile-based thresholds for selecting high-density cells uniformly across concepts. We considered the 60th, 70th, 80th, 90th, and 95th percentile values.
For each concept and each percentile-based threshold, we then: (i) filter grid cells to keep only high-density ones according to the threshold, (ii) construct an adjacency matrix connecting high-density grid cells, and (iii) identify connected components within this adjacency matrix.
The process is visually summarized in SI.
These connected components represent contiguous regions of high density in the UMAP space, from which prototypes or exemplar drawings can be derived.
To identify the optimal value for the high-density threshold parameter, we run the pipeline on the clusterable categories, over all percentile values for the high-density threshold. 
For each sketch, we assign 1 if it belongs to any cluster according to the grid-based procedure and 0 if it is labeled as noise. We then use a binary cluster label based on the DBSCAN process (i.e., 1 if the sketch is classified as belonging to any DBSCAN cluster and 0 otherwise) as the ground truth.
We evaluate performance in terms of precision and find the optimal value for the threshold to be the 60th percentile value. 

\subsection*{Conceptual Properties of Concepts}
Psycholinguistic and neuroscience research has identified \textbf{concreteness} as a crucial factor influencing memory, word processing, affective connotation, and neural activation patterns.
\citet{brysbaert2014concreteness} collected concreteness ratings for about 40,000 English lemmas through a large-scale crowd-sourcing effort, providing a valuable resource to quantify the abstractness or concreteness of concepts. 

Beyond concreteness, \textbf{sensorimotor experience} plays a fundamental role in cognition. 
\citet{lynott2020lancaster} compiled crowdsourced scores for about 40,000 English concepts, assessing their association with perceptual modalities—touch, hearing, smell, taste, vision, and interoception—as well as their linkage to specific action effectors, including the mouth/throat, hand/arm, foot/leg, head (excluding mouth/throat), and torso.

We leverage the validated concreteness and sensorimotor experience ratings for the concepts in our dataset. 
Despite good coverage, 40 concepts out of 344 lack crowd-sourced annotations for concreteness and 38 out of 344 for sensorimotor experience.
We supplement these gaps by having three of the authors independently annotate the missing concepts, following the guidelines from the original papers. Inter-annotator agreement is moderate-to-high (see SI), and we use the average of the three annotations as the final score for each concept.

\subsubsection*{Correlation with clusterability}
We are interested in understanding whether certain properties of concepts correlate with a more clusterable visual representation. Specifically, we explore the properties of how concrete they are, which sense they are mostly related to and with which part of the body they are mainly experienced with.
We consider a concept to be \emph{clusterable} if the proportion of its sketches classified as noise by clustering is below a certain threshold.
Clusterability, for each concept, can then be defined as the proportion of non-noise sketches. 
We then measure the Spearman correlation between clusterability and each of the properties of concepts we are interested in analyzing. Given the presence of multiple comparisons, we adjust the significance level $\alpha=0.05$ with a Bonferroni correction that considers the total number of comparisons. 

\subsection*{Word vs. Image Semantics}
We compare image and word embedding representation of the concepts in our data.
For the computation of the former, each sketch is embedded using the DINOv2 image embedding model.
For the latter, we compare two word-embeddings approaches: a Word2Vec model pre-trained on Google News~\cite{word2vec-google-news} and a multilingual General Text Embedding model~\cite{gte-multilingual-base}.
For each of the 344 target concepts, we then identify a ranking of the most similar concepts, using cosine similarity between embeddings (image- or word-based).
Because our image data includes multiple sketches per concept and these sketches are organized into clusters (e.g., pizza as a slice and pizza as a wheel), we retain this information when computing similarity rankings. First, we compute the centroid of image embeddings for each concept cluster. Then, we aggregate clusters referring to the same concepts in the ranking. For example, the image-based ranking for \texttt{pizza\_1} (pizza depicted as a slice) might be: \texttt{[triangle\_1, triangle\_2, icecream\_1, diamond\_1, diamond\_2, stop\_sign\_1, star\_1, …]}. To make the rankings comparable with the ones based on word embeddings, we collapse each cluster back to its parent concept. In the example for pizza, the top-5 ranking becomes: \texttt{[triangle, icecream, diamond, stop\_sign, star]}.

We then measure the Rank-Biased Overlap (RBO) between the top-5, top-10, and top-20 most similar concepts identified by each pair of methods (Word2Vec vs. Image embeddings and Multilingual BERT vs. Image embeddings). We also compute the percentage of overlapped concepts between the top-10 ranked items and Kendall Tau correlation scores on the overall rankings.
We define a randomly shuffled version of the image-based ranking to provide a baseline for comparison.

\subsection*{Networks}
Our dataset includes metadata on the country of origin of users who produced sketches, allowing us to examine whether individuals from similar countries depict concepts in systematically similar ways compared to individuals from very different countries. To do so, we compare sketch-based similarities with similarities derived from cultural and linguistic information.

As a cultural benchmark, we use the Cultural Fixation (CFs) index developed by~\citet{muthukrishna2020beyond}, which provides distance scores between pairs of countries based on the World Values Survey of beliefs and behaviors. For the most represented countries in our dataset, we match available CFs scores using the most recent survey wave (2010--2014), which we adopt to ensure consistency across survey measures.
Some of the countries of interest (e.g., Denmark) are missing from the original World Values Survey data. This prevents us from matching the missing countries with those involved in the image-based and language-based comparisons. We decide to retain only the intersection between the countries involved in the culture-, image-, and language-based comparisons. 
As a result, when focusing on the top-100 countries in our dataset we retain 51 of them; for the top-50, we retain 27; and for the top-25, we retain 19.
For comparison, we also repeat the analysis using a more inclusive matching criterion, prioritizing country coverage over consistency of survey measures: we match all available CFs scores  across different waves and retain the most recent measure available for each country pair. This yields 79 of the top-100 countries, 43 of the top-50, and all 25 of the top-25.

We then transform the CFs distance matrix into a similarity matrix and construct a network in which nodes represent countries and edge weights capture their cultural similarity.
We experiment with two methods for network filtering:
\begin{itemize}
    \item \emph{Threshold on strongest edges.} We retain only the top $x\%$ highest-weight edges, producing a graph filtered to its strongest connections. \\
    \noindent We consider $x \in \{1, 5, 10, 20, 30, 40, 50, 60, 70, 80, 90\}\%$.
    \item \emph{Disparity filtering.} Based on the definition by~\citet{serrano2009extracting}, we evaluate the weight of each edge relative to a null model and assign it a significance score $\alpha$. We then retain only edges with $\alpha < \alpha_t$, where $\alpha_t$ is a threshold to be defined. For the choice of $\alpha_t$, we evaluate the proportion of the original Giant Connected Component (GCC) retained at different thresholds (0.01, 0.05, 0.1, 0.15, 0.2, 0.25, 0.275, 0.3, 0.4, 0.5, 0.9). A threshold of 0.25 preserves 98\% of the GCC, which we therefore adopt as our reference.
\end{itemize}
We also build a null model from this culture-based network for later comparisons. Specifically, we define a configuration model~\cite{newman2001random} that preserves the degree sequence of the culture-based network.

We compare this cultural-similarity network with two additional networks: one based on language and one based on images.
We start from the language-based network. From each of the top 100 countries by prevalence, we identify the primary national language and translate all concept names into that language. We then embed the translated names using multilingual word embeddings (built from the same BERT-based considered previously). The pairwise similarity between two languages is defined as the average cosine similarity across all pairs of concept embeddings. To link back to countries, we assign each country the similarity values corresponding to its primary language so that, for instance, the links Australia–Italy and United States–Italy are identical.
Similarly as for the culture-based network, we identify the complete graph and then compare two filtering methods.

For the image-based network, we quantify the association between each country and visual concepts using odds ratios. 
We treat each concept independently by considering its set of clusters. For every country–cluster pair, we compute the odds ratio
$$
OR_{c,k} = \frac{N_{c,k}}{N_{c, \neg k}} \cdot  \frac{N_{\neg c \neg k}}{N_{\neg c,k}}
$$ 
where $N_{c,k}$ is the number of images from country $c$ in cluster $k$, 
$N_{c, \neg k}$ is the number of images from country $c$ in all other clusters, 
$N_{\neg c \neg k}$ is the number of images from all other countries in all other clusters, and $N_{\neg c,k}$ is the number of images from all other countries in cluster $k$.
Odds ratios greater than 1 indicate over-representation of country $c$ in cluster $k$, values below 1 indicate under-representation, and values close to 1 imply no meaningful deviation.
For each country and concept, we define a vector of odds ratios across all clusters of that concept, capturing how strongly a country’s images concentrate in specific visual clusters relative to others.
We then compute the pairwise similarities between countries based on such odds ratios vectors.
If all entries of two countries’ vectors fall between 0.9 and 1.1, we assign their pairwise similarity a value of 0. 
Otherwise, we compute similarity using the Euclidean distance between their standardized (z-scored) log-odds ratios. This gives a country-by-country similarity matrix for each concept.
We then average similarities across concepts to obtain a global similarity matrix and construct a network in which nodes represent countries and edge weights encode similarity in visual-prototype usage. 
As with the culture- and language-based networks, we start from the complete graph and compare two filtering methods.

We compute three measures of network similarity between the cultural benchmark and the language- or image-based networks: 
\begin{enumerate}
    \item \emph{Edge similarity}: the Jaccard index of edge sets, i.e., the proportion of shared edges relative to the union of edges in both networks.
    \item \emph{Neighborhood similarity}: for each node, we compute the Jaccard index between its sets of neighbors in the two networks, and then average across all nodes. This captures local structural similarity even when the exact edges differ.
    \item \emph{Community similarity}: we compare Louvain communities~\cite{blondel2008fast} and Leiden communities~\cite{traag2019louvain}  detected in the two networks using the Normalized Mutual Information (NMI) score. This measure captures whether networks produce similar higher-level groupings of countries.
\end{enumerate}

We repeat the procedure by considering the two methods for network filtering (i.e., threshold on strongest edges and disparity filtering) and different numbers of nodes defining the network (namely, the top-100, top-50, and top-25 countries by prominence in our dataset). 

In the comparison between image- and culture-based networks and language- and culture-based networks, we evaluate the lift of each metric from the one computed between image-based and null network and language-based and null network, respectively. This means that the similarities between image (or language) and culture are defined as:
$$
sim_m(X,C) = \frac{sim_{m}(X,C)}{sim_m(X,null)},
$$
where $m$ is a metric of interest (for instance, edge similarity), $X$ is an image-based or language-based network, $C$ is the culture-based network and $null$ is the configuration model built from the culture-based network.

\subsection*{Data and Code Availability}
The dataset was shared with the authors under a non-disclosure agreement. The original data was collected by the QuickDraw team with informed consent, in accordance with \href{https://policies.google.com/terms}{Google's terms} and \href{https://policies.google.com/privacy}{privacy policy}. For reproducibility purposes, a sample of 50 million drawings across all categories is available from \href{https://github.com/googlecreativelab/quickdraw-dataset}{Google Creative Lab}. The code is available on \href{https://github.com/ariannap13/billions_sketches-cultural-diversity}{Github}.

\clearpage

\section*{References}
\printbibliography[heading=none]

@book{berlin1991basic,
  title={Basic color terms: Their universality and evolution},
  author={Berlin, Brent and Kay, Paul},
  year={1969},
  publisher={Univ of California Press}
}

@book{smith1981categories,
  title={Categories and concepts},
  author={Smith, Edward E and Medin, Douglas L},
  year={1981},
  publisher={Harvard University Press}
}

@article{medin1978context,
  title={Context theory of classification learning.},
  author={Medin, Douglas L and Schaffer, Marguerite M},
  journal={Psychological review},
  volume={85},
  number={3},
  pages={207},
  year={1978},
  publisher={American Psychological Association}
}

@article{kemp2012kinship,
  title={Kinship categories across languages reflect general communicative principles},
  author={Kemp, Charles and Regier, Terry},
  journal={Science},
  volume={336},
  number={6084},
  pages={1049--1054},
  year={2012},
  publisher={American Association for the Advancement of Science}
}

@article{fedorenko2024language,
  title={Language is primarily a tool for communication rather than thought},
  author={Fedorenko, Evelina and Piantadosi, Steven T and Gibson, Edward AF},
  journal={Nature},
  volume={630},
  number={8017},
  pages={575--586},
  year={2024},
  publisher={Nature Publishing Group UK London}
}

@article{mukherjee2025drawings,
  title={Drawings of {THINGS}: A large-scale drawing dataset of 1,854 object concepts},
  author={Mukherjee, Kushin and Huey, Holly and Stoinski, Laura M. and Hebart, Martin N. and Fan, Judith and Bainbridge, Wilma A.},
  journal={Behavior Research Methods},
  volume={58, 57},
  year={2025},
  doi={10.3758/s13428-025-02887-w}
}

@article{zhu2025crosscontextual,
  title={Cross-Contextual Variability in Children's Early Understanding of Visual Media},
  author={Zhu, Rebecca and Kilonzo, Tabitha N. and Zhu, Lily Z. and Fan, Judith E. and Frank, Michael C.},
  journal={Topics in Cognitive Science},
  year={2025}
}

@book{bergen2012louder,
  title={Louder than words: The new science of how the mind makes meaning},
  author={Bergen, Benjamin K},
  year={2012},
  publisher={Basic Books},
  address={New York}
}

@article{guilbeault2020color,
  title={Color associations in abstract semantic domains},
  author={Guilbeault, Douglas and Nadler, Ethan O and Chu, Mark and Lo Sardo, Donald Ruggiero and Kar, Aabir Abubaker and Desikan, Bhargav Srinivasa},
  journal={Cognition},
  volume={201},
  pages={104306},
  year={2020},
  publisher={Elsevier}
}

@article{long2023developmental,
  title={Developmental changes in drawing production under different memory demands in a {US} and {Chinese} sample},
  author={Long, Bria and Wang, Ying and Christie, Stella and Frank, Michael C. and Fan, Judith E.},
  journal={Developmental Psychology},
  volume={59},
  number={10},
  pages={1784},
  year={2023}
}

@article{long2024parallel,
  title={Parallel developmental changes in children's production and recognition of line drawings of visual concepts},
  author={Long, Bria and Fan, Judith E. and Huey, Holly and Chai, Zixian and Frank, Michael C.},
  journal={Nature Communications},
  volume={15},
  number={1},
  pages={1191},
  year={2024}
}

@inproceedings{yu2016sketch,
  title={Sketch me that shoe},
  author={Yu, Qian and Liu, Feng and Song, Yi-Zhe and Xiang, Tao and Hospedales, Timothy M. and Loy, Chen Change},
  booktitle={Proceedings of the IEEE Conference on Computer Vision and Pattern Recognition},
  pages={799--807},
  year={2016}
}

@book{mcneill1992hand,
  title={Hand and Mind: What Gestures Reveal about Thought},
  author={McNeill, David},
  year={1992},
  publisher={University of Chicago Press},
  address={Chicago}
}

@article{xu2009symbolic,
  title={Symbolic gestures and spoken language are processed by a common neural system},
  author={Xu, Jiang and Gannon, Patrick J. and Emmorey, Karen and Smith, Jason F. and Braun, Allen R.},
  journal={Proceedings of the National Academy of Sciences},
  volume={106},
  number={49},
  pages={20664--20669},
  year={2009},
  doi={10.1073/pnas.0909197106}
}

@article{willems2007language,
  title={When language meets action: The neural integration of gesture and speech},
  author={Willems, Roel M. and {\"O}zy{\"u}rek, Asl{\i} and Hagoort, Peter},
  journal={Cerebral Cortex},
  volume={17},
  number={10},
  pages={2322--2333},
  year={2007},
  doi={10.1093/cercor/bhl141}
}

@article{ozyurek2007online,
  title={On-line integration of semantic information from speech and gesture: Insights from event-related brain potentials},
  author={{\"O}zy{\"u}rek, Asl{\i} and Willems, Roel M. and Kita, Sotaro and Hagoort, Peter},
  journal={Journal of Cognitive Neuroscience},
  volume={19},
  number={4},
  pages={605--616},
  year={2007},
  doi={10.1162/jocn.2007.19.4.605}
}

@article{xu2022deep,
  title={Deep learning for free-hand sketch: A survey},
  author={Xu, Peng and Hospedales, Timothy M. and Yin, Qiyue and Song, Yi-Zhe and Xiang, Tao and Wang, Liang},
  journal={IEEE Transactions on Pattern Analysis and Machine Intelligence},
  volume={45},
  number={1},
  pages={285--312},
  year={2022}
}

@article{lynott2020lancaster,
  title={The Lancaster Sensorimotor Norms: multidimensional measures of perceptual and action strength for 40,000 English words},
  author={Lynott, Dermot and Connell, Louise and Brysbaert, Marc and Brand, James and Carney, James},
  journal={Behavior research methods},
  volume={52},
  number={3},
  pages={1271--1291},
  year={2020},
  publisher={Springer}
}

@article{guilbeault2021experimental,
  title={Experimental evidence for scale-induced category convergence across populations},
  author={Guilbeault, Douglas and Baronchelli, Andrea and Centola, Damon},
  journal={Nature communications},
  volume={12},
  number={1},
  pages={327},
  year={2021},
  publisher={Nature Publishing Group UK London}
}

@article{rogers2004structure,
  title={Structure and deterioration of semantic memory: a neuropsychological and computational investigation.},
  author={Rogers, Timothy T and Lambon Ralph, Matthew A and Garrard, Peter and Bozeat, Sasha and McClelland, James L and Hodges, John R and Patterson, Karalyn},
  journal={Psychological review},
  volume={111},
  number={1},
  pages={205},
  year={2004},
  publisher={American Psychological Association}
}

@article{rosch1975family,
  title={Family resemblances: Studies in the internal structure of categories},
  author={Rosch, Eleanor and Mervis, Carolyn B},
  journal={Cognitive psychology},
  volume={7},
  number={4},
  pages={573--605},
  year={1975},
  publisher={Elsevier}
}

@article{murphy2016there,
  title={Is there an exemplar theory of concepts?},
  author={Murphy, Gregory L},
  journal={Psychonomic bulletin \& review},
  volume={23},
  number={4},
  pages={1035--1042},
  year={2016},
  publisher={Springer}
}

@book{goldstein2015cognitive,
  title={Cognitive psychology: Connecting mind, research, and everyday experience},
  author={Goldstein, E Bruce},
  year={2015},
  publisher={Cengage learning Stamford, CT}
}

@article{michel2011quantitative,
  title={Quantitative analysis of culture using millions of digitized books},
  author={Michel, Jean-Baptiste and Shen, Yuan Kui and Aiden, Aviva Presser and Veres, Adrian and Gray, Matthew K and Google Books Team and Pickett, Joseph P and Hoiberg, Dale and Clancy, Dan and Norvig, Peter and others},
  journal={science},
  volume={331},
  number={6014},
  pages={176--182},
  year={2011},
  publisher={American Association for the Advancement of Science}
}

@article{muthukrishna2020beyond,
  title={Beyond Western, Educated, Industrial, Rich, and Democratic (WEIRD) psychology: Measuring and mapping scales of cultural and psychological distance},
  author={Muthukrishna, Michael and Bell, Adrian V and Henrich, Joseph and Curtin, Cameron M and Gedranovich, Alexander and McInerney, Jason and Thue, Braden},
  journal={Psychological science},
  volume={31},
  number={6},
  pages={678--701},
  year={2020},
  publisher={Sage Publications Sage CA: Los Angeles, CA}
}

@article{brysbaert2014concreteness,
  title={Concreteness ratings for 40 thousand generally known English word lemmas},
  author={Brysbaert, Marc and Warriner, Amy Beth and Kuperman, Victor},
  journal={Behavior research methods},
  volume={46},
  number={3},
  pages={904--911},
  year={2014},
  publisher={Springer}
}

@article{ha2017neural,
  title={A neural representation of sketch drawings},
  author={Ha, David and Eck, Douglas},
  journal={arXiv preprint arXiv:1704.03477},
  year={2017}
}

@inproceedings{xu2018sketchmate,
  title={Sketchmate: Deep hashing for million-scale human sketch retrieval},
  author={Xu, Peng and Huang, Yongye and Yuan, Tongtong and Pang, Kaiyue and Song, Yi-Zhe and Xiang, Tao and Hospedales, Timothy M and Ma, Zhanyu and Guo, Jun},
  booktitle={Proceedings of the IEEE conference on computer vision and pattern recognition},
  pages={8090--8098},
  year={2018}
}

@article{xu2021multigraph,
  title={Multigraph transformer for free-hand sketch recognition},
  author={Xu, Peng and Joshi, Chaitanya K and Bresson, Xavier},
  journal={IEEE Transactions on Neural Networks and Learning Systems},
  volume={33},
  number={10},
  pages={5150--5161},
  year={2021},
  publisher={IEEE}
}

@inproceedings{lamb2020sketchtransfer,
  title={Sketchtransfer: A new dataset for exploring detail-invariance and the abstractions learned by deep networks},
  author={Lamb, Alex and Ozair, Sherjil and Verma, Vikas and Ha, David},
  booktitle={Proceedings of the IEEE/CVF Winter Conference on Applications of Computer Vision},
  pages={963--972},
  year={2020}
}

@article{fernandez2019quick,
  title={Quick, stat!: A statistical analysis of the quick, draw! dataset},
  author={Fernandez-Fernandez, Raul and Victores, Juan G and Estevez, David and Balaguer, Carlos},
  journal={arXiv preprint arXiv:1907.06417},
  year={2019}
}

@article{serrano2009extracting,
  title={Extracting the multiscale backbone of complex weighted networks},
  author={Serrano, M {\'A}ngeles and Bogun{\'a}, Mari{\'a}n and Vespignani, Alessandro}, 
  journal={Proceedings of the national academy of sciences},
  volume={106},
  number={16},
  pages={6483--6488},
  year={2009},
  publisher={National Academy of Sciences}
}

@article{blondel2008fast,
  title={Fast unfolding of communities in large networks},
  author={Blondel, Vincent D and Guillaume, Jean-Loup and Lambiotte, Renaud and Lefebvre, Etienne},
  journal={Journal of statistical mechanics: theory and experiment},
  volume={2008},
  number={10},
  pages={P10008},
  year={2008},
  publisher={IOP Publishing}
}

@article{wilson2002six,
  title={Six views of embodied cognition},
  author={Wilson, Margaret},
  journal={Psychonomic bulletin \& review},
  volume={9},
  number={4},
  pages={625--636},
  year={2002},
  publisher={Springer}
}

@article{zaslavsky2018efficient,
  title={Efficient compression in color naming and its evolution},
  author={Zaslavsky, Noga and Kemp, Charles and Regier, Terry and Tishby, Naftali},
  journal={Proceedings of the National Academy of Sciences},
  volume={115},
  number={31},
  pages={7937--7942},
  year={2018},
  publisher={National Academy of Sciences}
}

@article{san2018universal,
  title={Universal meaning extensions of perception verbs are grounded in interaction},
  author={San Roque, Lila and Kendrick, Kobin H and Norcliffe, Elisabeth and Majid, Asifa},
  journal={Cognitive Linguistics},
  volume={29},
  number={3},
  pages={371--406},
  year={2018},
  publisher={De Gruyter}
}

@article{jackson2019emotion,
  title={Emotion semantics show both cultural variation and universal structure},
  author={Jackson, Joshua Conrad and Watts, Joseph and Henry, Teague R and List, Johann-Mattis and Forkel, Robert and Mucha, Peter J and Greenhill, Simon J and Gray, Russell D and Lindquist, Kristen A},
  journal={Science},
  volume={366},
  number={6472},
  pages={1517--1522},
  year={2019},
  publisher={American Association for the Advancement of Science}
}

@article{tjuka2024universal,
  title={Universal and cultural factors shape body part vocabularies},
  author={Tjuka, Annika and Forkel, Robert and List, Johann-Mattis},
  journal={Scientific Reports},
  volume={14},
  number={1},
  pages={10486},
  year={2024},
  publisher={Nature Publishing Group UK London}
}

@article{thompson2020cultural,
  title={Cultural influences on word meanings revealed through large-scale semantic alignment},
  author={Thompson, Bill and Roberts, Se{\'a}n G and Lupyan, Gary},
  journal={Nature Human Behaviour},
  volume={4},
  number={10},
  pages={1029--1038},
  year={2020},
  publisher={Nature Publishing Group UK London}
}

@inproceedings{lewis2021characterizing,
  title={Characterizing variability in shared meaning through millions of sketches},
  author={Lewis, Molly and Balamurugan, Anjali and Zheng, Bin and Lupyan, Gary},
  booktitle={Proceedings of the Annual Meeting of the Cognitive Science Society},
  volume={43},
  year={2021}
}

@article{lewis2023local,
  title={Local similarity and global variability characterize the semantic space of human languages},
  author={Lewis, Molly and Cahill, Aoife and Madnani, Nitin and Evans, James},
  journal={Proceedings of the National Academy of Sciences},
  volume={120},
  number={51},
  pages={e2300986120},
  year={2023},
  publisher={National Academy of Sciences}
}

@article{malt2024representing,
  title={Representing the world in language and thought},
  author={Malt, Barbara C},
  journal={Topics in Cognitive Science},
  volume={16},
  number={1},
  pages={6--24},
  year={2024},
  publisher={Wiley Online Library}
}

@article{xu2020conceptual,
  title={Conceptual relations predict colexification across languages},
  author={Xu, Yang and Duong, Khang and Malt, Barbara C and Jiang, Serena and Srinivasan, Mahesh},
  journal={Cognition},
  volume={201},
  pages={104280},
  year={2020},
  publisher={Elsevier}
}

@article{kemp2018semantic,
  title={Semantic typology and efficient communication},
  author={Kemp, Charles and Xu, Yang and Regier, Terry},
  journal={Annual Review of Linguistics},
  volume={4},
  number={1},
  pages={109--128},
  year={2018},
  publisher={Annual Reviews}
}

@article{liang2024shared,
  title={Shared structure of fundamental human experience revealed by polysemy network of basic vocabularies across languages},
  author={Liang, Yuzhu and Xu, Ke and Ran, Qibin},
  journal={Scientific Reports},
  volume={14},
  number={1},
  pages={5877},
  year={2024},
  publisher={Nature Publishing Group UK London}
}

@inproceedings{thompson2018quantifying,
  title={Quantifying semantic similarity across languages},
  author={Thompson, Bill and Roberts, Sean G and Lupyan, Gary},
  booktitle={Annual Meeting of the Cognitive Science Society},
  pages={2554--2559},
  year={2018},
  organization={Cognitive Science Society}
}

@article{fernandino2022decoding,
  title={Decoding the information structure underlying the neural representation of concepts},
  author={Fernandino, Leonardo and Tong, Jia-Qing and Conant, Lisa L and Humphries, Colin J and Binder, Jeffrey R},
  journal={Proceedings of the National Academy of Sciences},
  volume={119},
  number={6},
  pages={e2108091119},
  year={2022},
  publisher={National Academy of Sciences}
}

@article{barsalou2010grounded,
  title={Grounded cognition: Past, present, and future},
  author={Barsalou, Lawrence W},
  journal={Topics in cognitive science},
  volume={2},
  number={4},
  pages={716--724},
  year={2010},
  publisher={Wiley Online Library}
}

@article{bechtold2023brain,
  title={Brain signatures of embodied semantics and language: A consensus paper},
  author={Bechtold, Laura and Cosper, Samuel H and Malyshevskaya, Anastasia and Montefinese, Maria and Morucci, Piermatteo and Niccolai, Valentina and Repetto, Claudia and Zappa, Ana and Shtyrov, Yury},
  journal={Journal of cognition},
  volume={6},
  number={1},
  pages={61},
  year={2023}
}

@misc{atari2023humans,
    author = {Atari, Mohammad and Xue, Mona J and Park, Peter S and Blasi, Damián E and Henrich, Joseph},
    title={Which Humans?},
    publisher={PsyArXiv},
    year={2023}
}

@article{marjieh2024large,
  title={Large language models predict human sensory judgments across six modalities},
  author={Marjieh, Raja and Sucholutsky, Ilia and van Rijn, Pol and Jacoby, Nori and Griffiths, Thomas L},
  journal={Scientific Reports},
  volume={14},
  number={1},
  pages={21445},
  year={2024},
  publisher={Nature Publishing Group UK London}
}

@article{nadler2025statistical,
  title={Statistical or embodied? Comparing colorseeing, colorblind, painters, and Large Language Models in their processing of color metaphors},
  author={Nadler, Ethan O and Guilbeault, Douglas and Ringold, Sofronia M and Williamson, TR and Bellemare-Pepin, Antoine and Comșa, Iulia M and Jerbi, Karim and Narayanan, Srini and Aziz-Zadeh, Lisa},
  journal={Cognitive Science},
  volume={49},
  number={7},
  pages={e70083},
  year={2025},
  publisher={Wiley Online Library}
}

@article{piantadosi2024concepts,
  title={Why concepts are (probably) vectors},
  author={Piantadosi, Steven T and Muller, Dyana CY and Rule, Joshua S and Kaushik, Karthikeya and Gorenstein, Mark and Leib, Elena R and Sanford, Emily},
  journal={Trends in Cognitive Sciences},
  volume={28},
  number={9},
  pages={844--856},
  year={2024},
  publisher={Elsevier}
}

@article{frank2025cognitive,
  title={Cognitive modeling using artificial intelligence},
  author={Frank, Michael C and Goodman, Noah D},
  journal={Annual Review of Psychology},
  volume={777:543-566},
  year={2026},
  publisher={Annual Reviews},
  doi={10.1146/annurev-psych-030625-040748}
}

@inproceedings{stockl2021watching,
  title={Watching a language model learning chess},
  author={St{\"o}ckl, Andreas},
  booktitle={Proceedings of the International Conference on Recent Advances in Natural Language Processing (RANLP 2021)},
  pages={1369--1379},
  year={2021}
}

@inproceedings{loyola2023perceptual,
  title={Perceptual structure in the absence of grounding: the impact of abstractedness and subjectivity in color language for LLMs},
  author={Loyola, Pablo and Marrese-Taylor, Edison and Hoyos-Idrobo, Andres},
  booktitle={Findings of the Association for Computational Linguistics: EMNLP 2023},
  pages={1536--1542},
  year={2023}
}

@article{oquab2023dinov2,
  title={Dinov2: Learning robust visual features without supervision},
  author={Oquab, Maxime and Darcet, Timoth{\'e}e and Moutakanni, Th{\'e}o and Vo, Huy and Szafraniec, Marc and Khalidov, Vasil and Fernandez, Pierre and Haziza, Daniel and Massa, Francisco and El-Nouby, Alaaeldin and others},
  journal={arXiv preprint arXiv:2304.07193},
  year={2023}
}

@inproceedings{ester1996density,
  title={A density-based algorithm for discovering clusters in large spatial databases with noise},
  author={Ester, Martin and Kriegel, Hans-Peter and Sander, J{\"o}rg and Xu, Xiaowei and others},
  booktitle={kdd},
  volume={96},
  number={34},
  pages={226--231},
  year={1996}
}

@inproceedings{moulavi2014density,
  title={Density-based clustering validation},
  author={Moulavi, Davoud and Jaskowiak, Pablo A and Campello, Ricardo JGB and Zimek, Arthur and Sander, J{\"o}rg},
  booktitle={Proceedings of the 2014 SIAM international conference on data mining},
  pages={839--847},
  year={2014},
  organization={SIAM}
}

@article{mcinnes2018umap,
  title={Umap: Uniform manifold approximation and projection for dimension reduction},
  author={McInnes, Leland and Healy, John and Melville, James},
  journal={arXiv preprint arXiv:1802.03426},
  year={2018}
}

@article{lakoff2012explaining,
  title={Explaining embodied cognition results},
  author={Lakoff, George},
  journal={Topics in cognitive science},
  volume={4},
  number={4},
  pages={773--785},
  year={2012},
  publisher={Wiley Online Library}
}

@article{newman2001random,
  title={Random graphs with arbitrary degree distributions and their applications},
  author={Newman, Mark EJ and Strogatz, Steven H and Watts, Duncan J},
  journal={Physical review E},
  volume={64},
  number={2},
  pages={026118},
  year={2001},
  publisher={APS}
}

@online{word2vec-google-news,
  author       = {Mikolov, Tomas and Chen, Kai and Corrado, Greg and Dean, Jeffrey},
  title        = {Word2Vec Google News 300-dimensional embeddings},
  year         = {n.d.},
  url          = {https://huggingface.co/fse/word2vec-google-news-300},
  note         = {Pre-trained model, accessed 2025-03-01}
}

@online{gte-multilingual-base,
  author       = {{Alibaba-NLP}},
  title        = {GTE Multilingual Base},
  year         = {n.d.},
  url          = {https://huggingface.co/Alibaba-NLP/gte-multilingual-base},
  note         = {Pre-trained BERT-based model, accessed 2025-03-01}
}

@article{traag2019louvain,
  title={From Louvain to Leiden: guaranteeing well-connected communities},
  author={Traag, Vincent A and Waltman, Ludo and Van Eck, Nees Jan},
  journal={Scientific reports},
  volume={9},
  number={1},
  pages={5233},
  year={2019},
  publisher={Nature Publishing Group UK London}
}

@inproceedings{radford2021learning,
  title={Learning transferable visual models from natural language supervision},
  author={Radford, Alec and Kim, Jong Wook and Hallacy, Chris and Ramesh, Aditya and Goh, Gabriel and Agarwal, Sandhini and Sastry, Girish and Askell, Amanda and Mishkin, Pamela and Clark, Jack and others},
  booktitle={International conference on machine learning},
  pages={8748--8763},
  year={2021},
  organization={PmLR}
}
\end{refsection}

\section*{Acknowledgements}

A.P. and L.M.A. acknowledge funding from Carlsberg Foundation Project COCOONS (Grant ID: CF21-0432).

\section*{Author Contributions}
A.P., M.M., D.G., L.M.A, and A.B. designed the project. A.P. and M.M. analysed the data. A.P., M.M., and N.D. developed the algorithmic methods. A.P., M.M., D.G., L.M.A, and A.B. wrote the paper. 

\section*{Competing Interests}
The authors declare no competing interests.

\section*{Additional Information}
\noindent\textbf{Supplementary Information} is available for this paper.\\
\noindent\textbf{Correspondence} should be addressed to Luca Maria Aiello and Andrea Baronchelli.

\section*{Extended Data and Supplementary Information}

\subsection{Data Exploration}

We report a distribution of drawing duration times in Supplementary Figure~\ref{fig:drawing_duration}. Although the game limits drawing time to 20 seconds, we observe a notable number of instances exceeding this threshold, reaching up to 30 seconds. This discrepancy might reflect latency time or data recorded from an older version of the game in which the time cap had not yet been enforced.

\begin{figure}[htbp]
    \centering
    \includegraphics[width=0.5\textwidth]{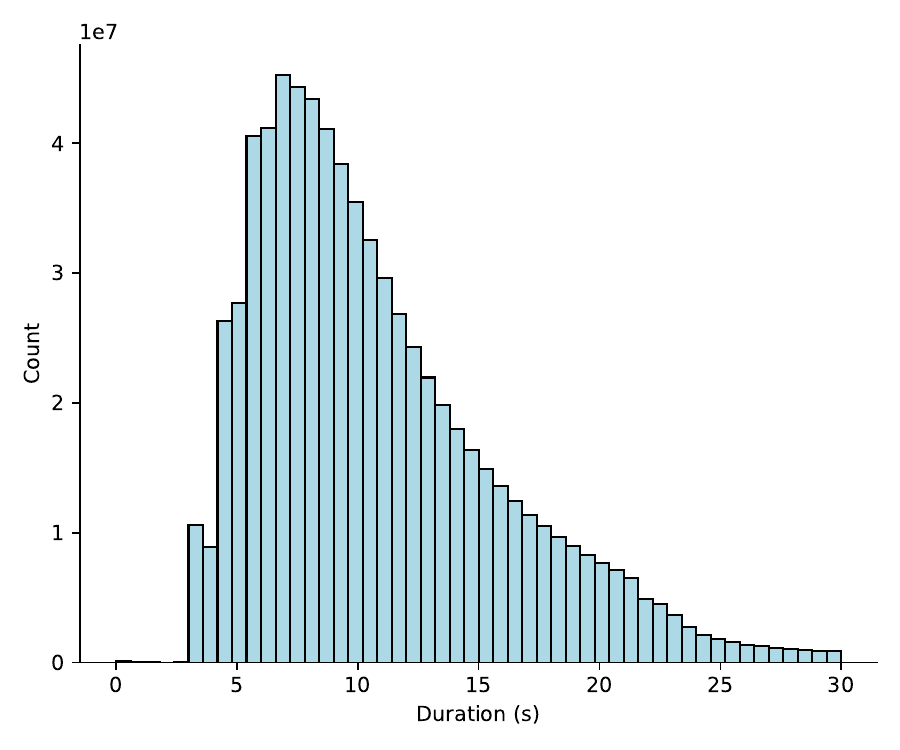} 
    \caption{Distribution of number of seconds passed before the sketched was recognized (drawing duration) by the neural network in the QuickDraw game.}
    \label{fig:drawing_duration}
\end{figure}

Supplementary Table \ref{tab:countries_distribution} contains the percentage of sketches in our dataset from the top-100 most prominent countries. The majority of data comes from the US (41.30\%), Great Britain (6.28\%), and Brazil (4.44\%). Location is unknown (i.e., country code reported as ZZ) for 0.08\% of the data .

\begin{table}[htbp]
\centering
\small
\begin{tabular}{l r  l r  l r}
\hline
Country & \% & Country & \% & Country & \% \\
\hline
US & 41.30 & AE & 0.42 & QA & 0.07 \\
GB & 6.28  & PT & 0.40 & MD & 0.07 \\
BR & 4.44  & HR & 0.40 & MA & 0.06 \\
RU & 3.47  & BG & 0.39 & BD & 0.06 \\
CA & 3.23  & SK & 0.35 & MK & 0.06 \\
AU & 2.51  & SG & 0.34 & UY & 0.06 \\
IN & 2.30  & RS & 0.34 & CR & 0.05 \\
TW & 2.29  & CO & 0.31 & LB & 0.04 \\
DE & 1.97  & BE & 0.30 & PR & 0.04 \\
KR & 1.86  & CH & 0.29 & BH & 0.04 \\
JP & 1.53  & GR & 0.27 & GT & 0.04 \\
FR & 1.46  & AT & 0.27 & OM & 0.04 \\
SE & 1.40  & CL & 0.25 & TN & 0.04 \\
FI & 1.36  & ZA & 0.24 & IS & 0.04 \\
IT & 1.19  & KZ & 0.23 & SV & 0.04 \\
TR & 1.03  & IL & 0.23 & NP & 0.03 \\
PH & 1.03  & BY & 0.22 & PS & 0.03 \\
ID & 0.99  & EG & 0.19 & DO & 0.03 \\
NL & 0.90  & PE & 0.18 & AZ & 0.03 \\
UA & 0.90  & EE & 0.15 & AM & 0.03 \\
PL & 0.86  & IQ & 0.14 & CY & 0.03 \\
MX & 0.80  & LT & 0.13 & TT & 0.03 \\
TH & 0.78  & BA & 0.13 & MT & 0.03 \\
RO & 0.77  & PK & 0.13 & HN & 0.03 \\
ES & 0.71  & LV & 0.10 & PA & 0.03 \\
CZ & 0.68  & DZ & 0.10 & BO & 0.03 \\
NZ & 0.58  & EC & 0.09 & KE & 0.03 \\
DK & 0.57  & KH & 0.09 & MN & 0.03 \\
SA & 0.56  & GE & 0.09 & KG & 0.03 \\
MY & 0.56  & KW & 0.08 & PY & 0.03 \\
HK & 0.53  & ZZ & 0.08 &    &      \\
AR & 0.52  & SI & 0.08 &    &      \\
VN & 0.48  & VE & 0.08 &    &      \\ 
HU & 0.45  & JO & 0.07 &    &      \\ 
\hline
\end{tabular}
\caption{Distribution of sketches by country. Values are percentages with two decimal places.}
\label{tab:countries_distribution}
\end{table}

\subsection{Concreteness and Sensorimotor Data}
To supplement existing concreteness and sensorimotor experience datasets with the missing concepts needed for our analyses, we conducted an annotation round involving three of the authors. Table~\ref{tab:krippendorff} reports Krippendorff's $\alpha$  for each conceptual property, showing moderate-to-high agreement. Agreement for the \emph{visual} property could not be reliably estimated with Krippendorff's $\alpha$ due to low variability in annotation scores (almost all 5). We therefore report percentage agreement instead, which was 81\%.

\begin{table}[t!]
\centering
\begin{tabular}{lc}
\hline
\textbf{Conceptual property} & \textbf{Krippendorff's $\alpha$} \\
\hline
Concreteness    & 0.57 \\
Auditory        & 0.59 \\
Gustatory       & 0.90 \\
Haptic          & 0.59 \\
Interoceptive   & 0.64 \\
Olfactory       & 0.60 \\
Foot/leg        & 0.61 \\
Hand/arm        & 0.60 \\
Head            & 0.84 \\
Mouth           & 0.68 \\
Torso           & 0.58 \\
\hline
\end{tabular}
\caption{Inter-annotator agreement (Krippendorff's $\alpha$) for each conceptual property.}
\label{tab:krippendorff}
\end{table}

\subsection{Clustering}
The distribution of the percentage of noise resulting from the clustering of concepts is characterized by bimodality (see Supplementary Figure~\ref{fig:clustering_percnoise_nclusters}, left). 
We define a threshold of noise as the local minimum between the two peaks of this distribution. This threshold represents the cutoff above which a concept is considered too noisy to form meaningful clusters: if the percentage of sketches classified as noise for a category exceeds this value, the category is deemed non-clusterable.
The number of clusters per concept ranges from 1 to 9, with an outlier at 21. The median value is 2, and the distribution is right-skewed (see Supplementary Figure~\ref{fig:clustering_percnoise_nclusters}, right).

\begin{figure}[htbp]
    \centering

    \begin{subfigure}{0.4\textwidth}
        \includegraphics[width=\linewidth]{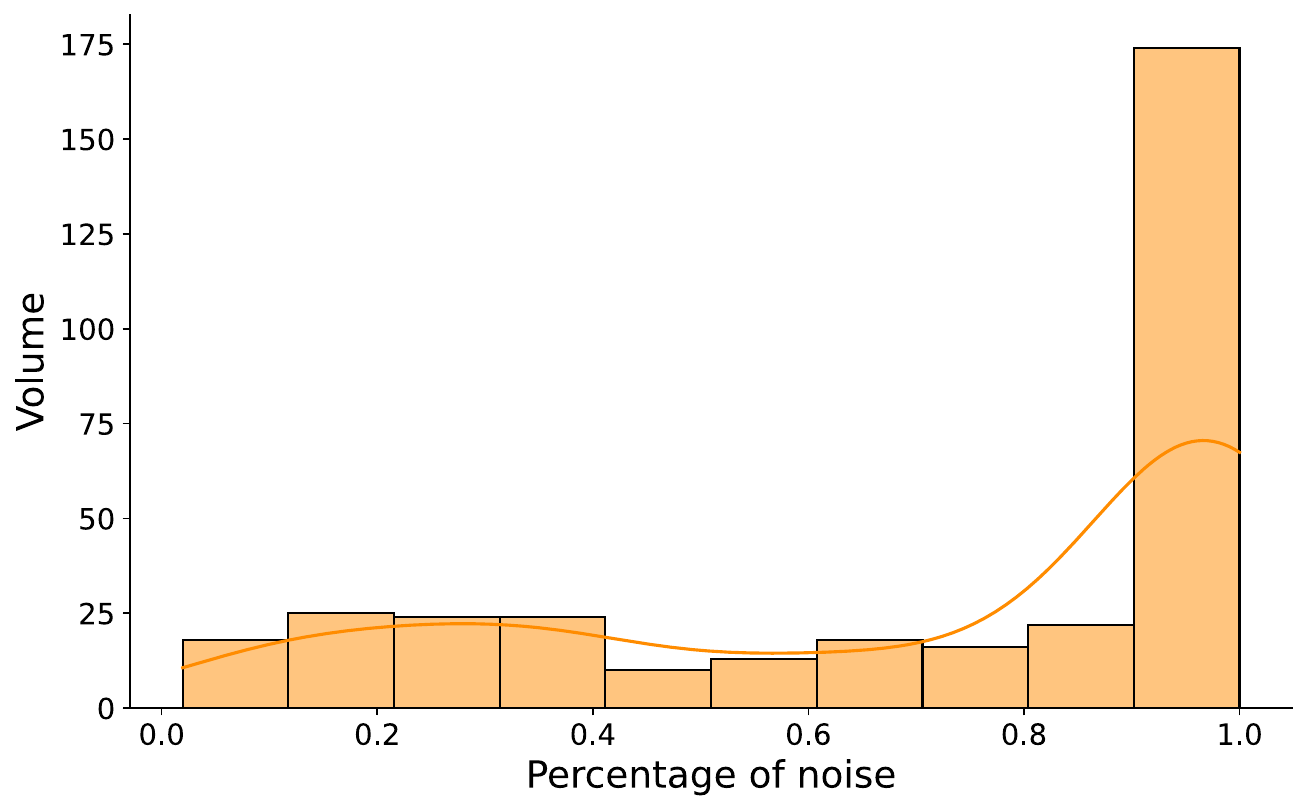} 
        \caption{Distribution of the percentage of noise in clustering per concept.}
    \end{subfigure}
    \hspace{0.7cm}
    \begin{subfigure}{0.5\textwidth}
        \includegraphics[width=\linewidth]{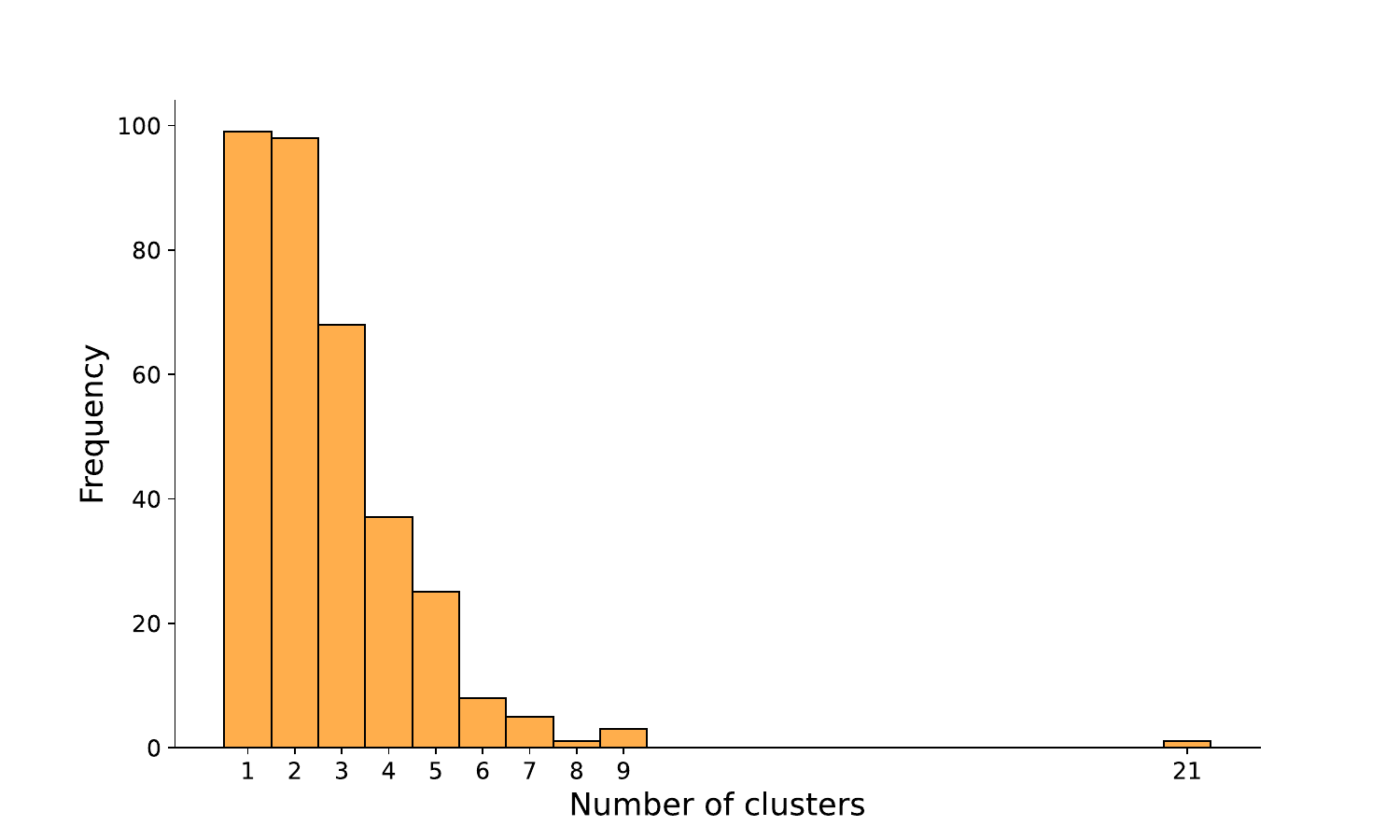}
        \caption{Distribution of the number of clusters per concept.}
    \end{subfigure}

    \caption{Distributions of the percentage of noise in clustering (left) and the number of clusters per concept (right).}
    \label{fig:clustering_percnoise_nclusters}
\end{figure}

\noindent To test whether \emph{clusterability} is robust to the choice of image embedding method, we sample concepts across the distribution of noise percentage from clustering based on DINOv2 embeddings. We compute the deciles of this distribution (0.0195, 0.171, 0.333, 0.535, 0.776, 0.909, 0.960, 0.977, 0.988, 1) and randomly sample one concept from each: clarinet (0.144), bear (0.258), balloon (0.378), hot dog (0.665), cloud (0.828), candle (0.940), skyscraper (0.976), wine bottle (0.983), and dragon (0.989). We embed the sketches for these concepts using CLIP-ViT-B-32, following the same pre-processing and clustering pipeline used for DINOv2. The rank correlation between clusterability of sampled concepts under DINOv2 and under CLIP-ViT-B-32 embeddings is high (Spearman's $\rho = 0.833$, $p = 0.005$), indicating that clusterability is robust to the choice of embedding method.

\noindent In Supplementary Figure \ref{fig:overlap_clusters}, we show the superposition of randomly sampled drawings within each cluster for the six hand-selected concepts reported in the main text, revealing the typical visual forms that clusters represent.

\begin{figure}[htbp]
    \centering
    \begin{tabular}{>{\centering\arraybackslash}m{2cm} *{10}{c}}
        Donut & 
        \raisebox{-.5\height}{\includegraphics[width=0.1\textwidth]{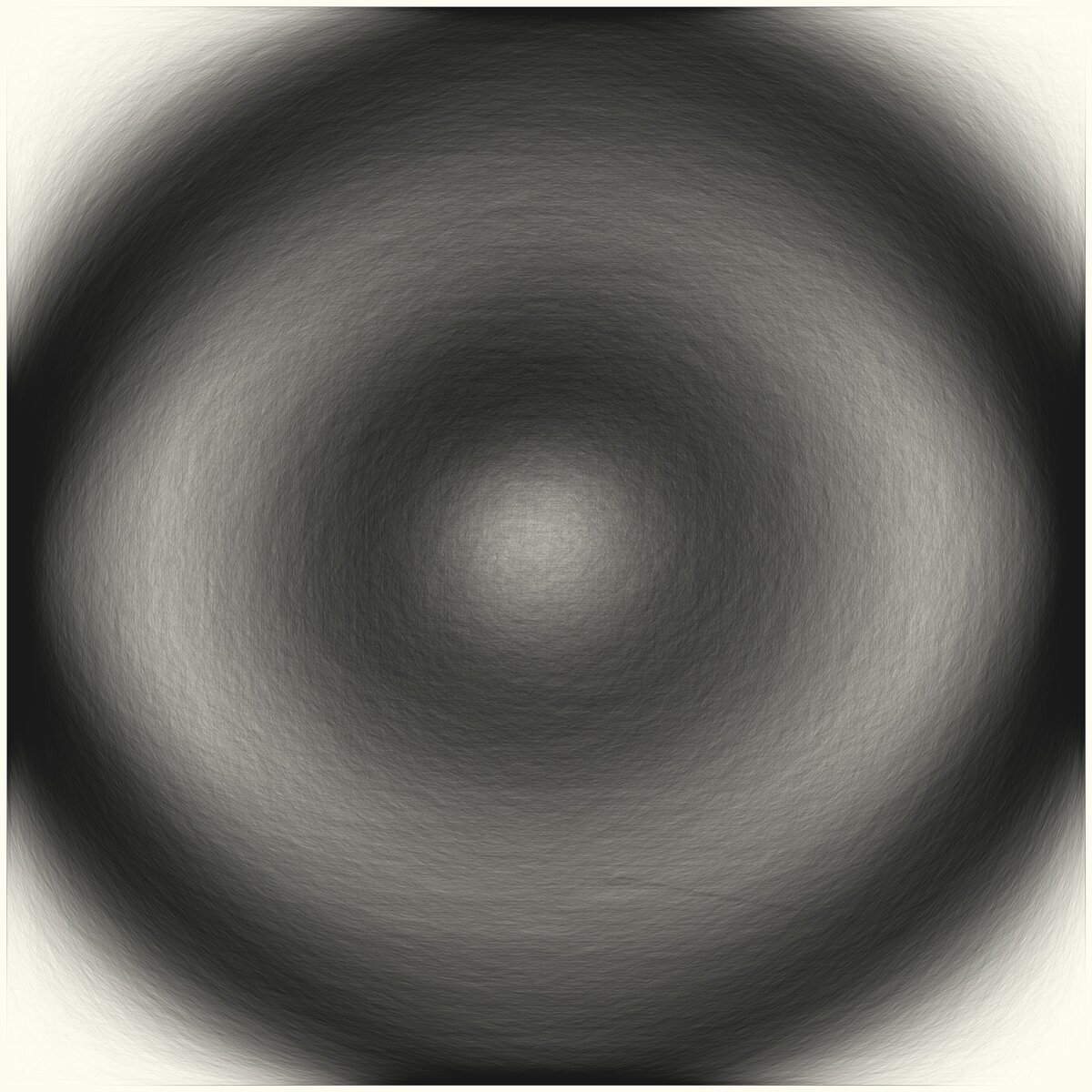}} \\[1.2cm]

        Fish &
        \raisebox{-.5\height}{\includegraphics[width=0.1\textwidth]{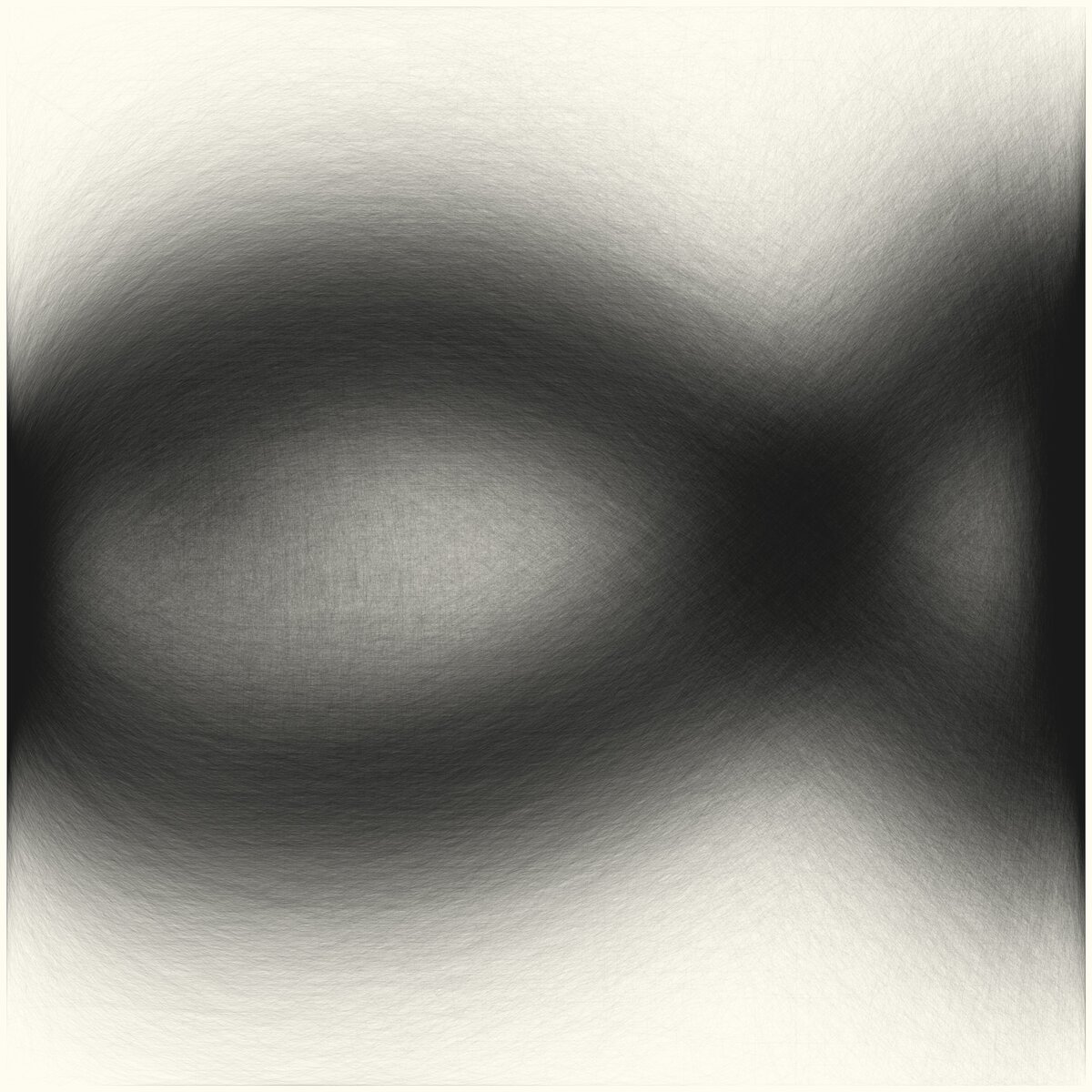}} &
        \raisebox{-.5\height}{\includegraphics[width=0.1\textwidth]{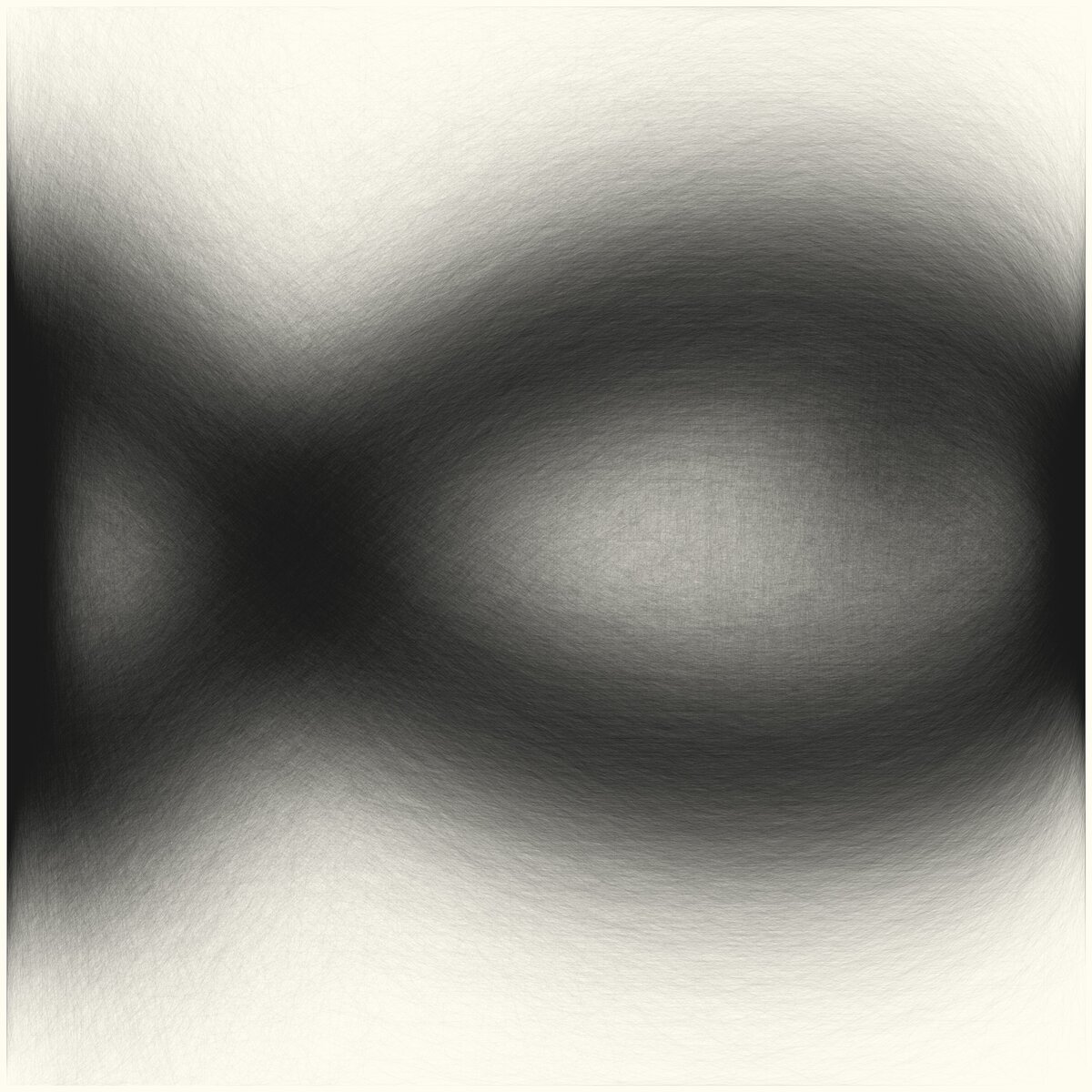}} \\[1.2cm]

        Pizza &
        \raisebox{-.5\height}{\includegraphics[width=0.1\textwidth]{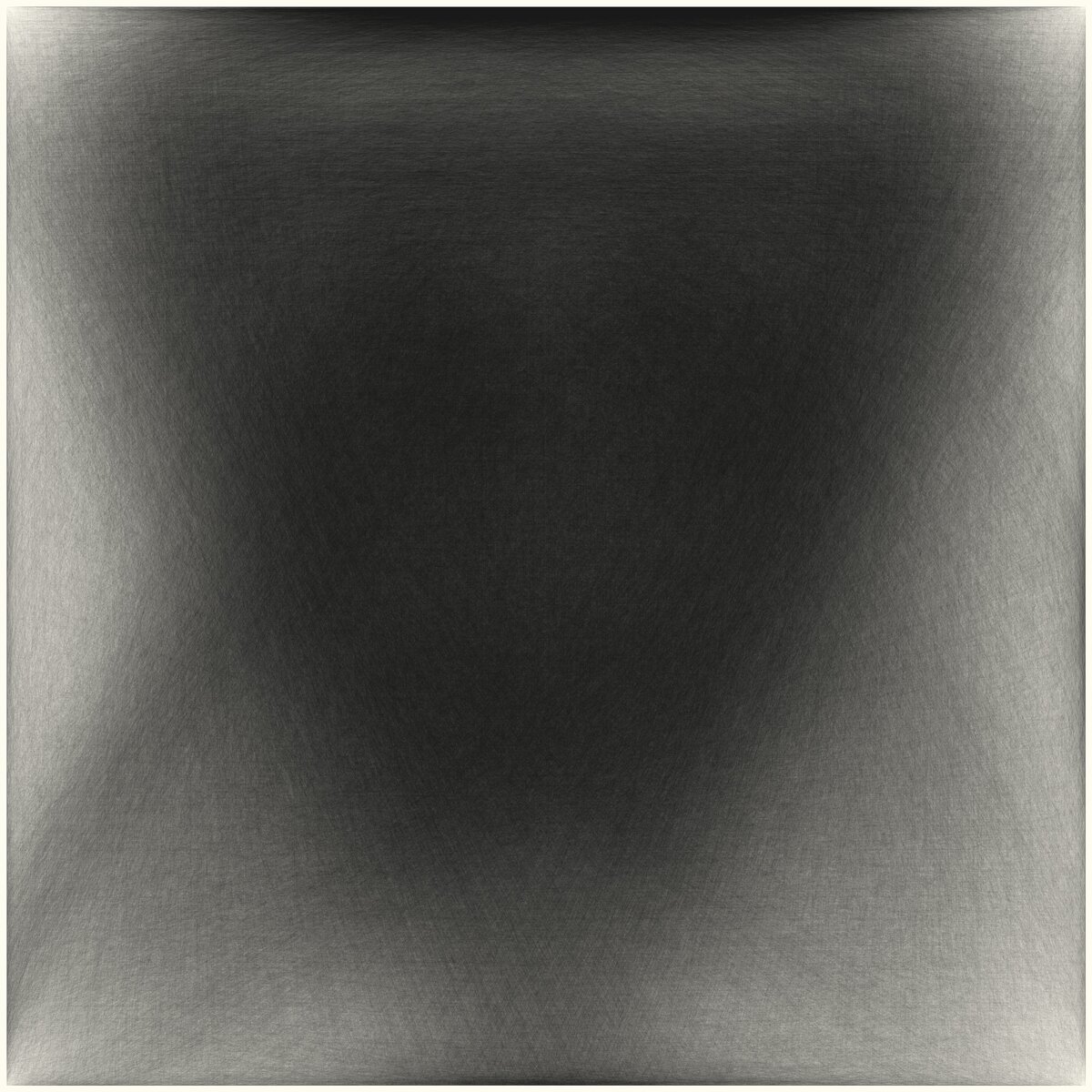}} &
        \raisebox{-.5\height}{\includegraphics[width=0.1\textwidth]{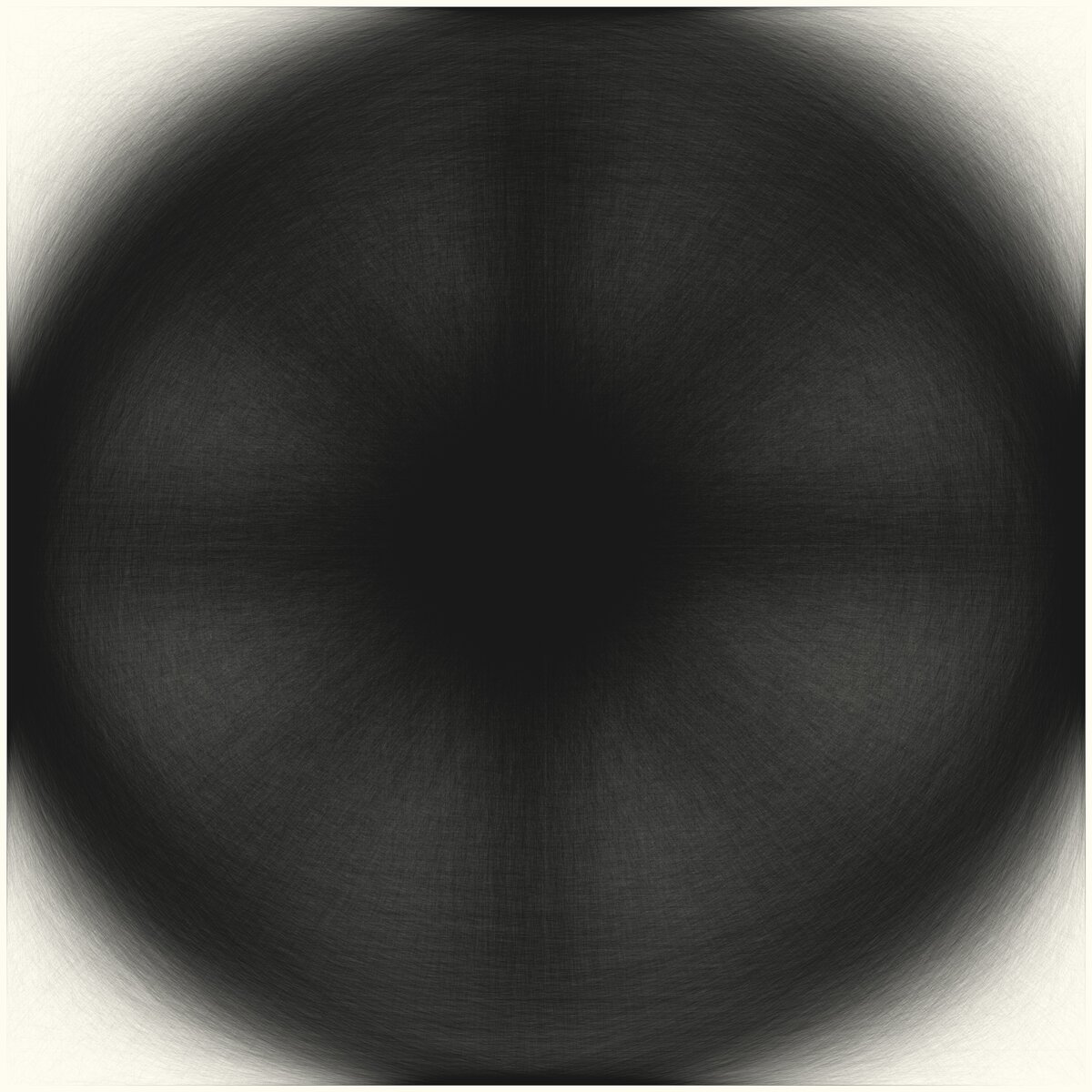}} \\[1.2cm]

        Moon &
        \raisebox{-.5\height}{\includegraphics[width=0.1\textwidth]{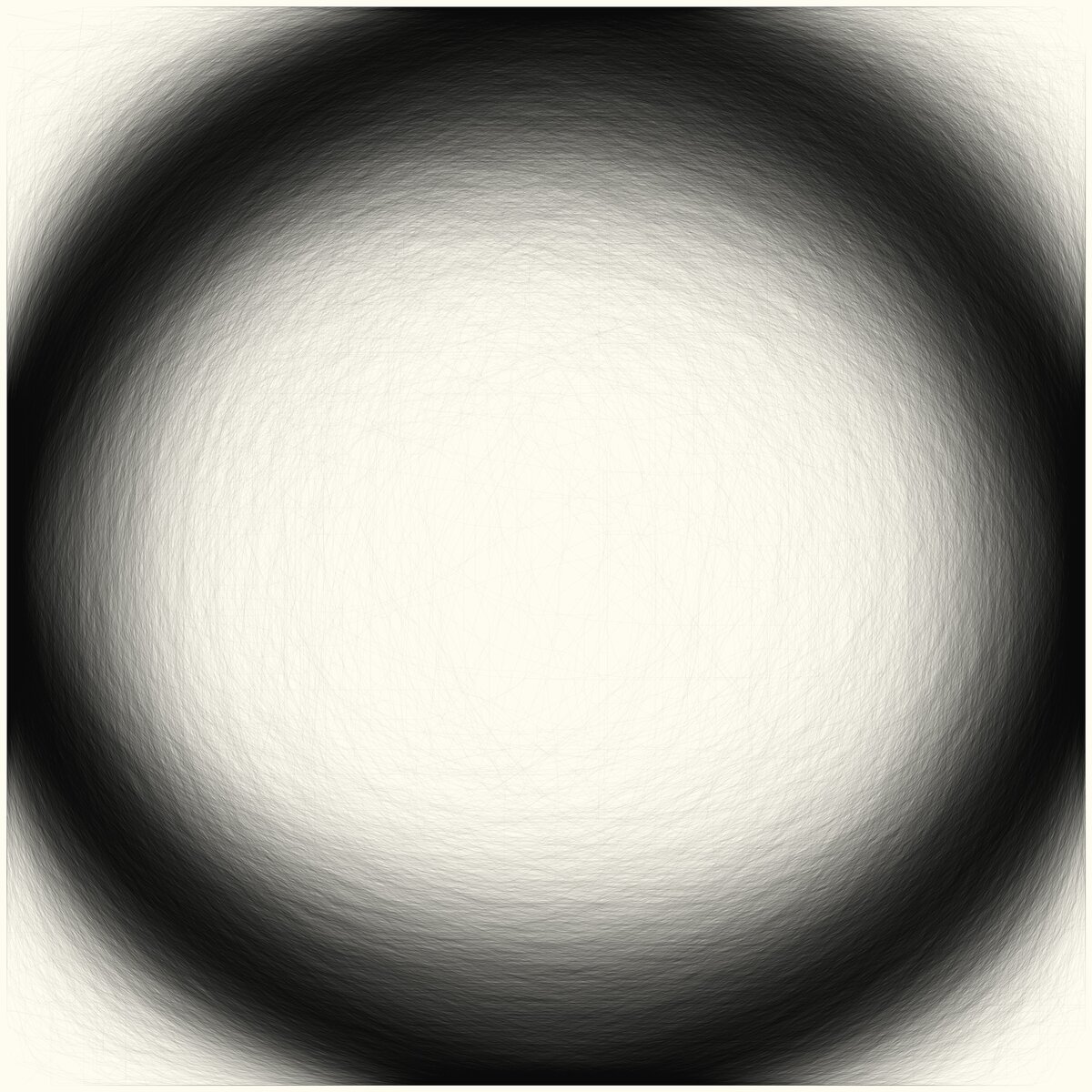}} &
        \raisebox{-.5\height}{\includegraphics[width=0.1\textwidth]{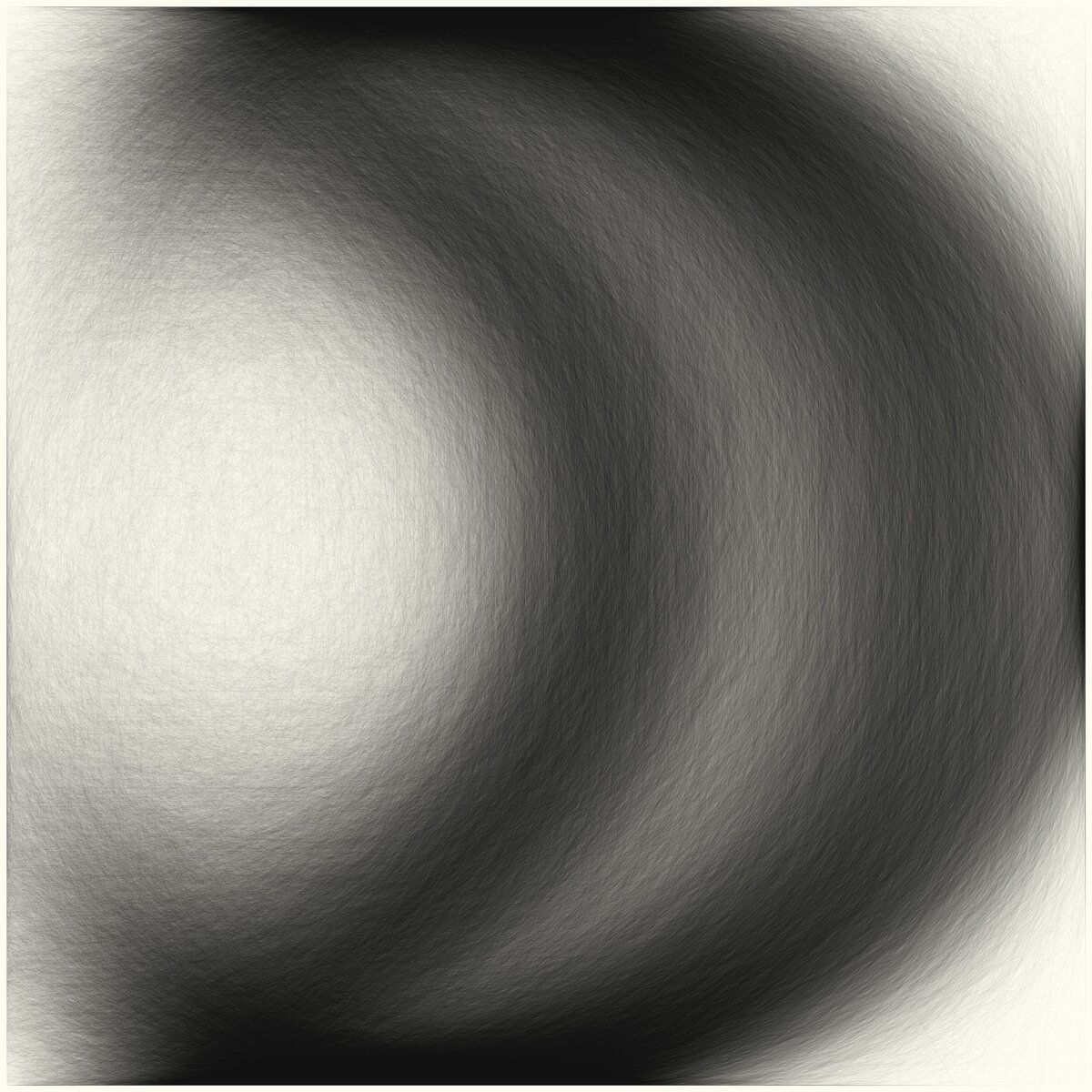}} &
        \raisebox{-.5\height}{\includegraphics[width=0.1\textwidth]{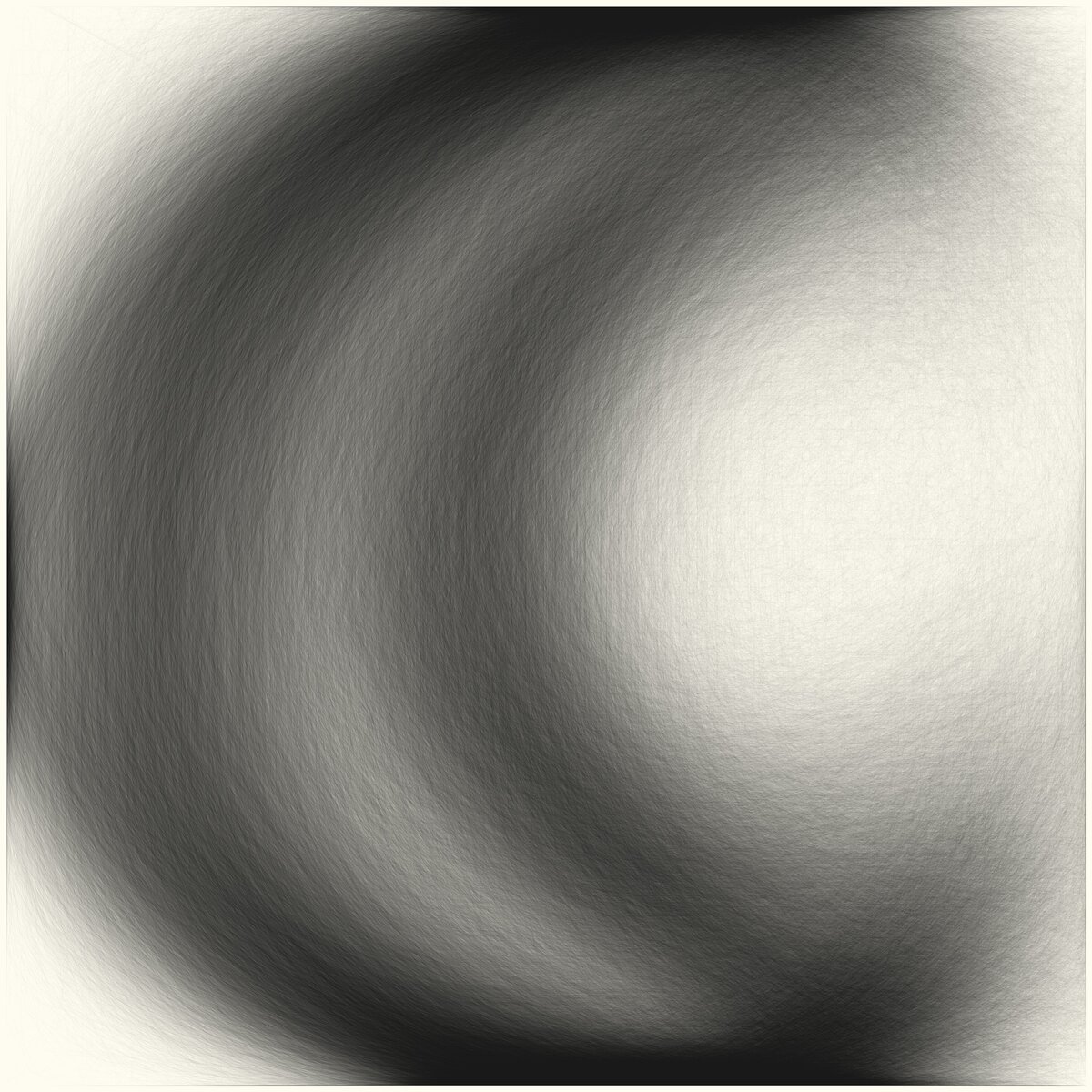}} &
        \raisebox{-.5\height}{\includegraphics[width=0.1\textwidth]{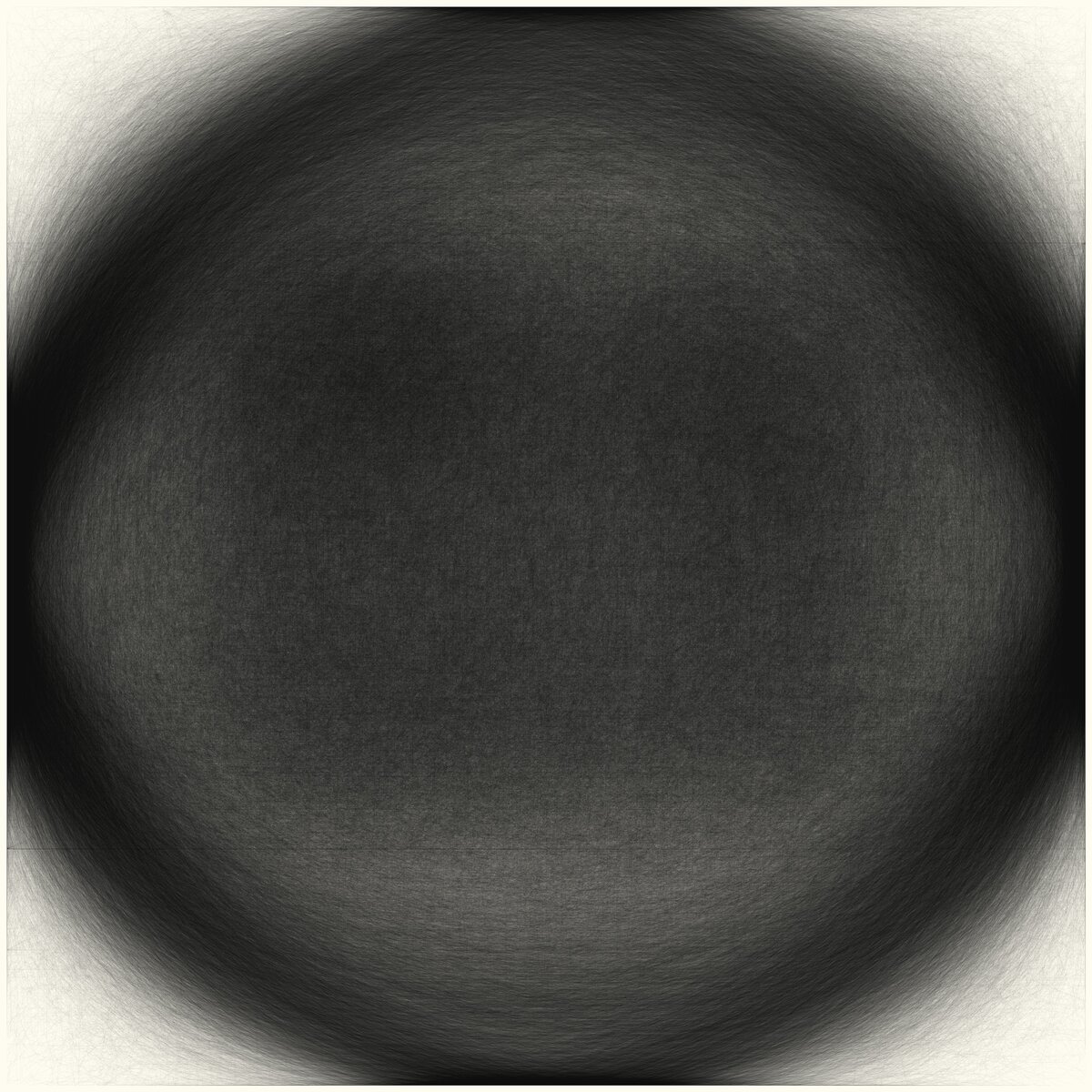}} &
        \raisebox{-.5\height}{\includegraphics[width=0.1\textwidth]{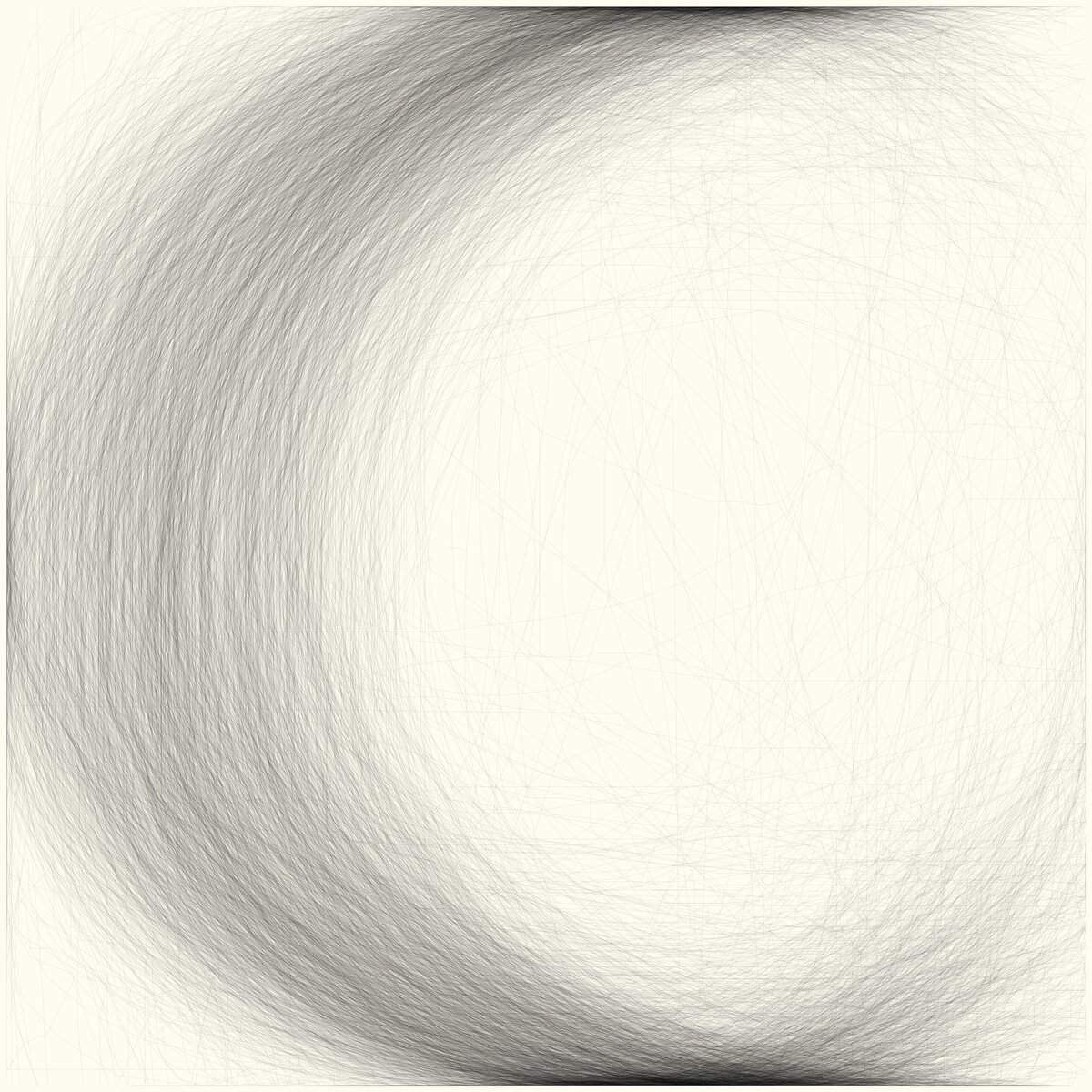}} \\[1.2cm]

        Smiley Face &
        \raisebox{-.5\height}{\includegraphics[width=0.1\textwidth]{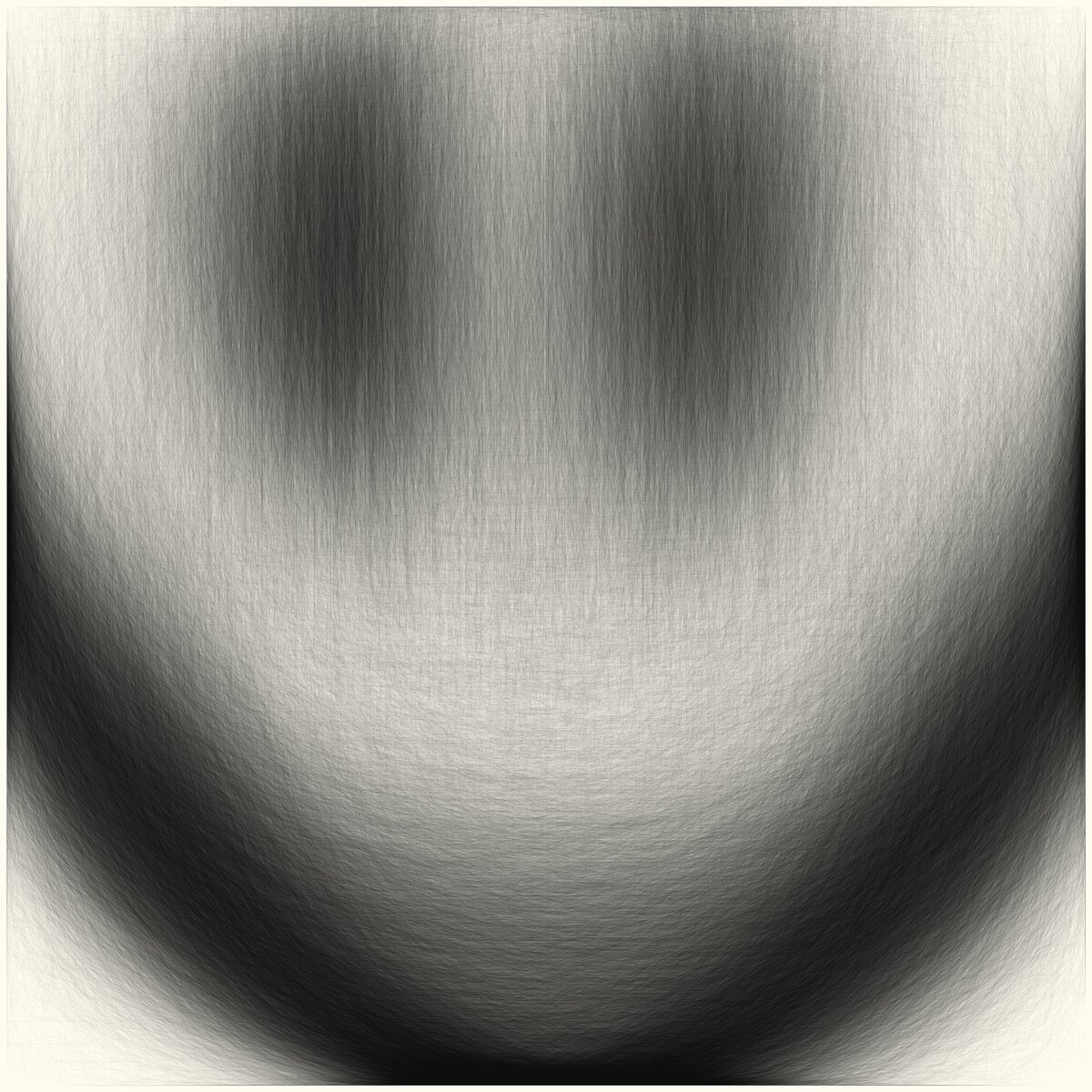}} &
        \raisebox{-.5\height}{\includegraphics[width=0.1\textwidth]{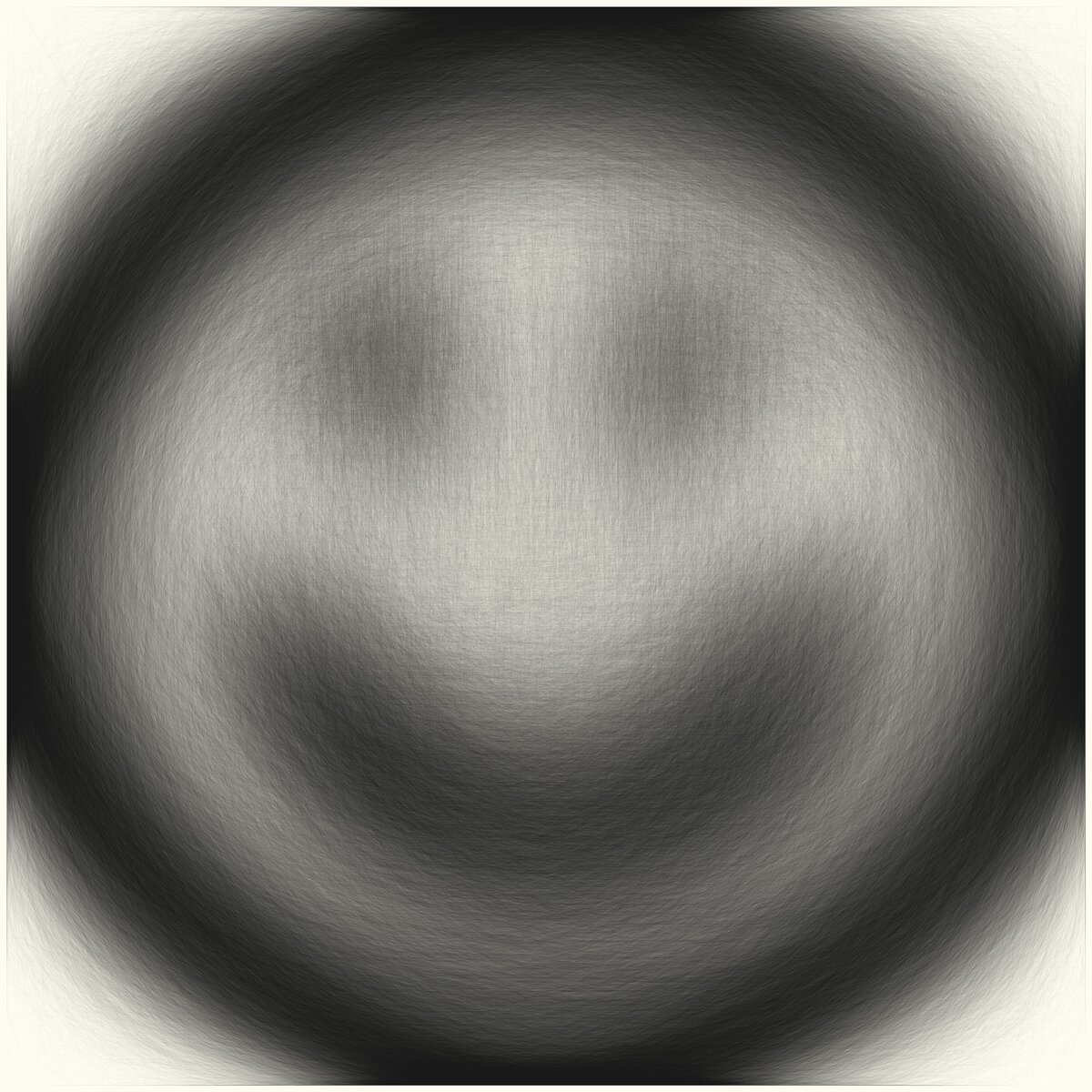}} &
        \raisebox{-.5\height}{\includegraphics[width=0.1\textwidth]{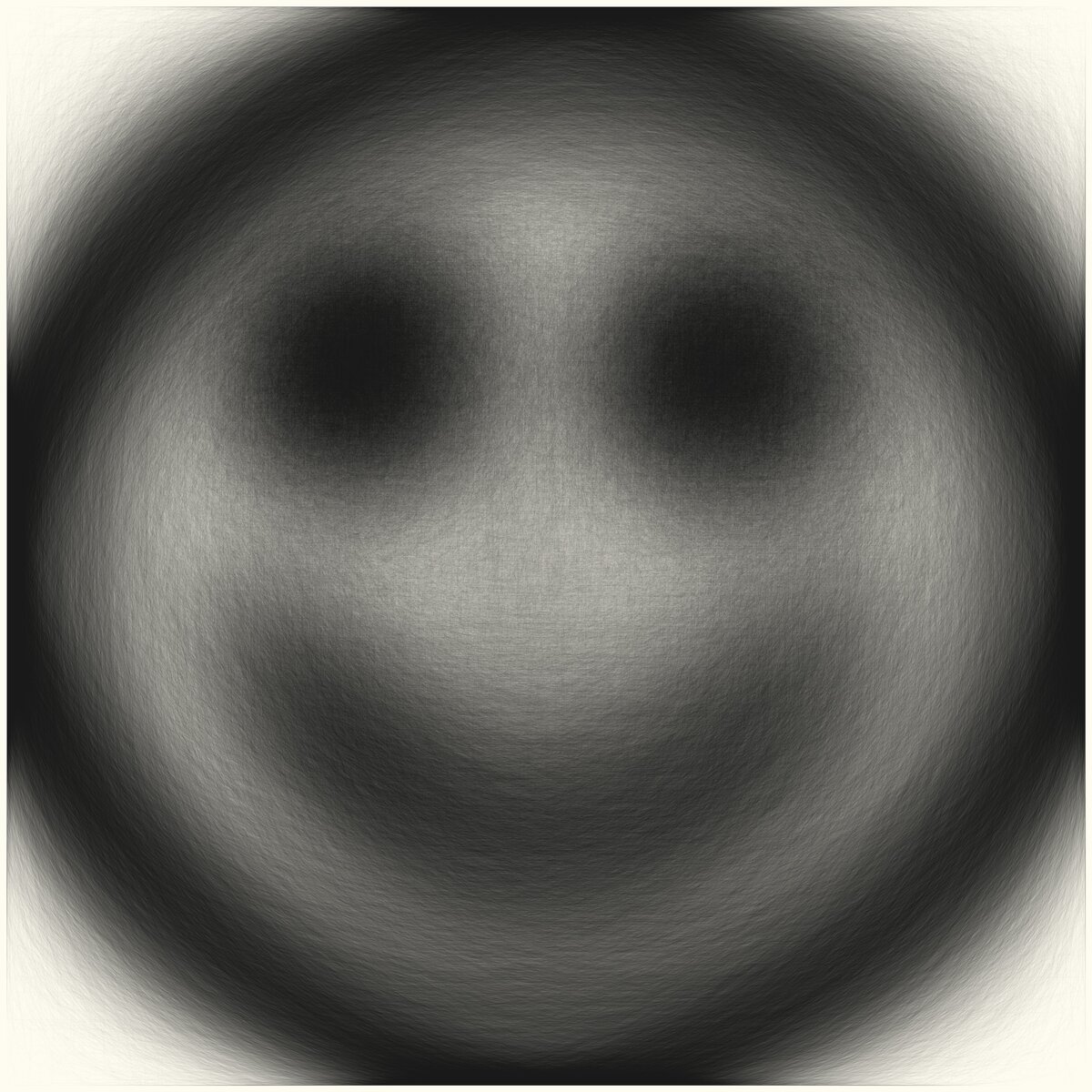}} &
        \raisebox{-.5\height}{\includegraphics[width=0.1\textwidth]{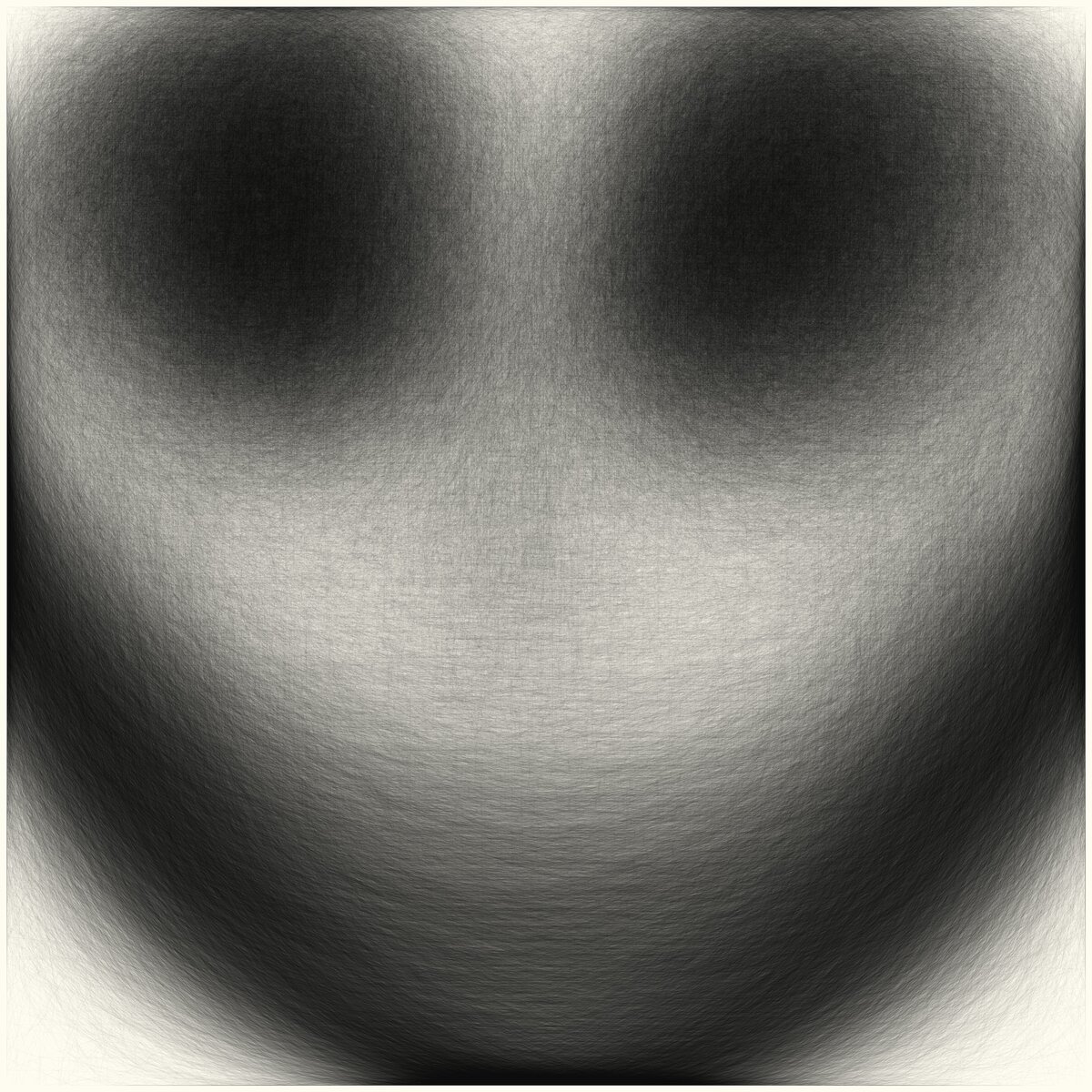}} &
        \raisebox{-.5\height}{\includegraphics[width=0.1\textwidth]{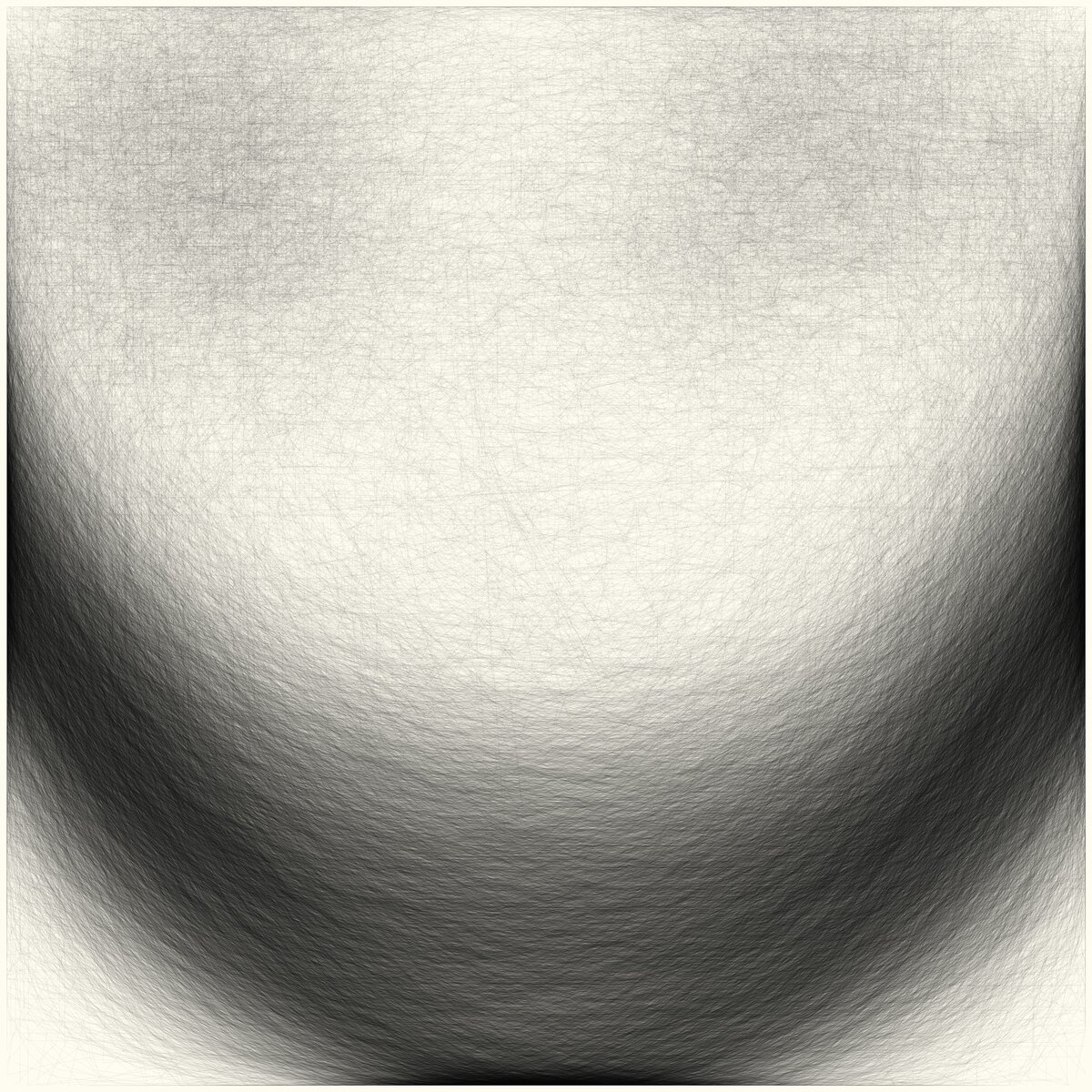}} \\[1.2cm]

        Phone &
        \raisebox{-.5\height}{\includegraphics[width=0.1\textwidth]{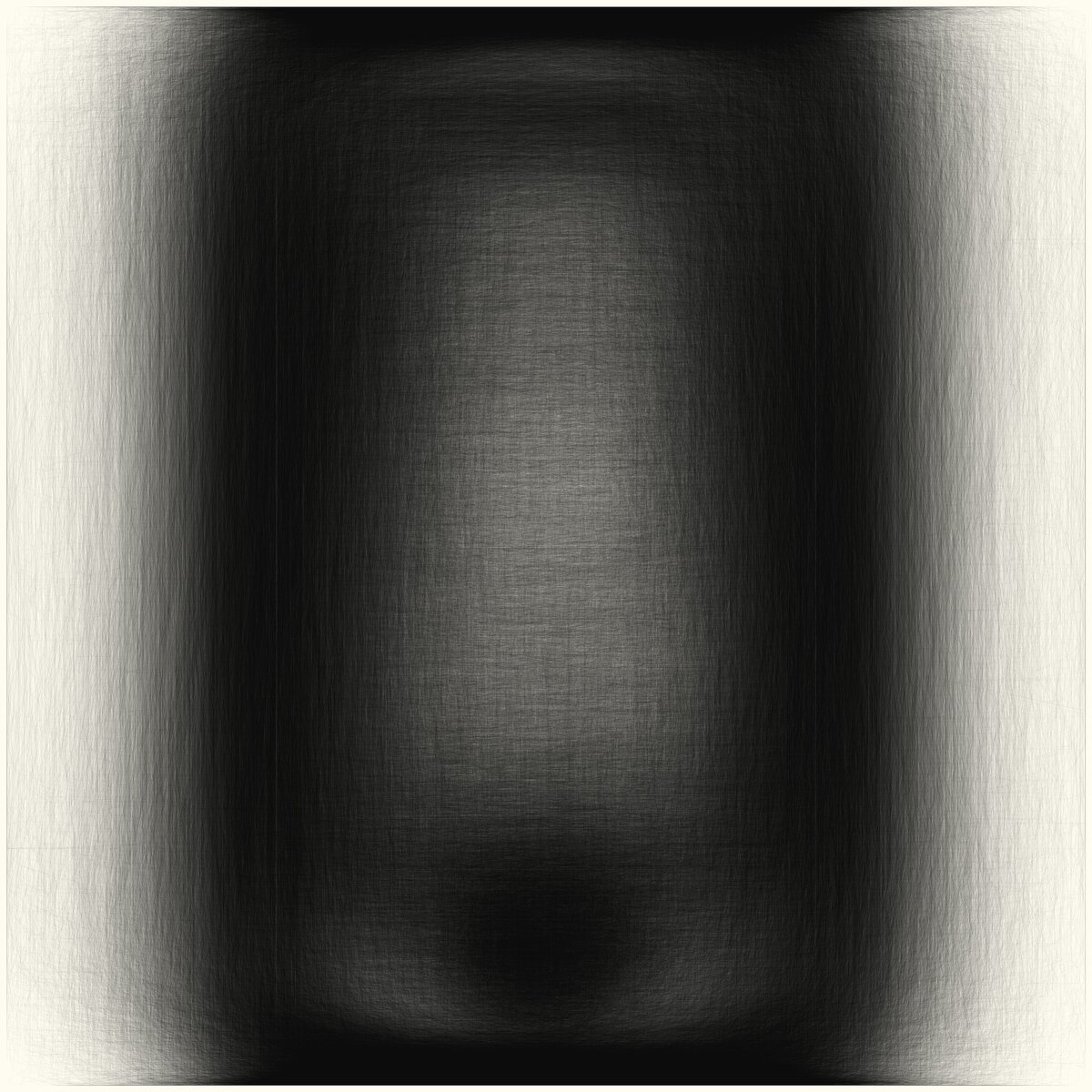}} &
        \raisebox{-.5\height}{\includegraphics[width=0.1\textwidth]{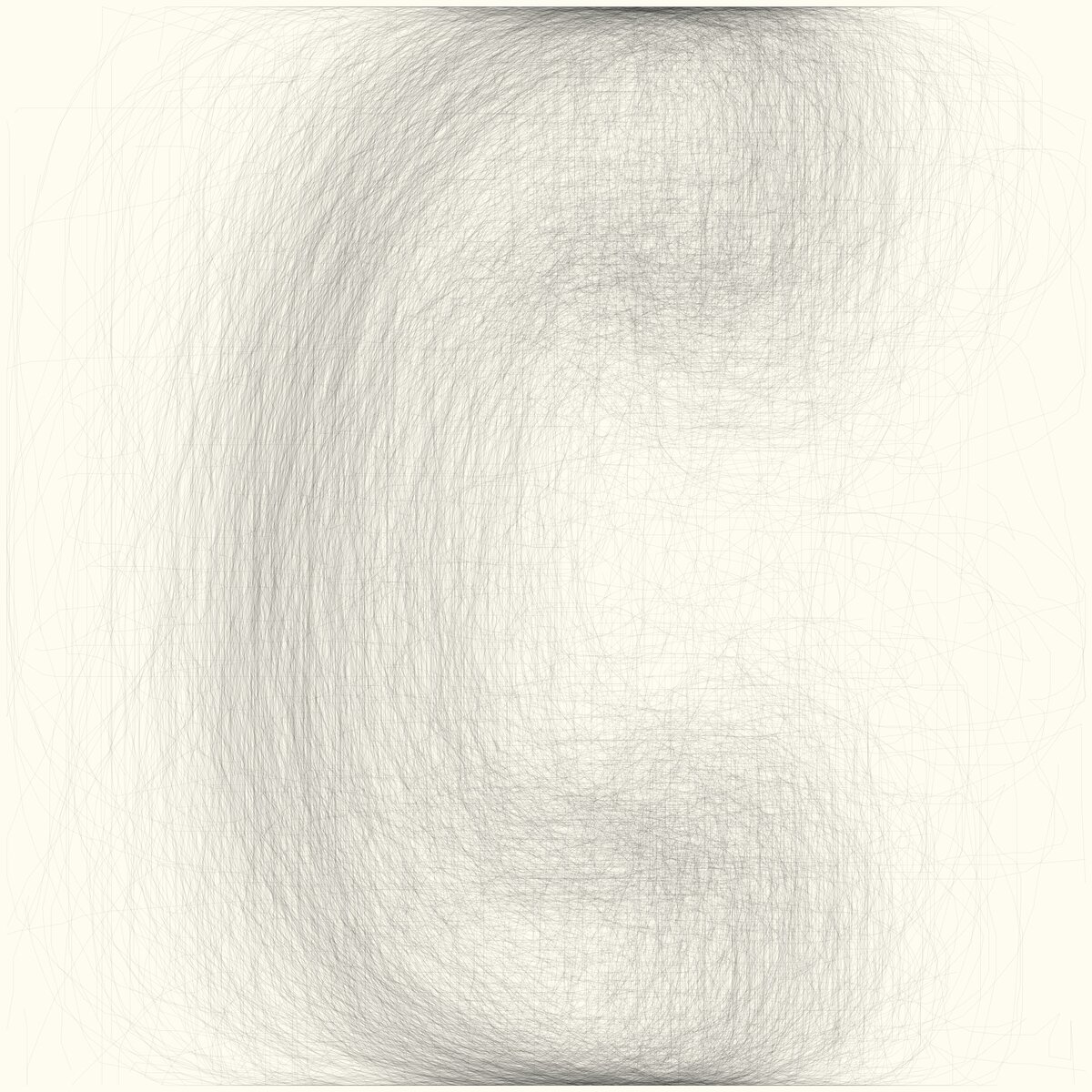}} &
        \raisebox{-.5\height}{\includegraphics[width=0.1\textwidth]{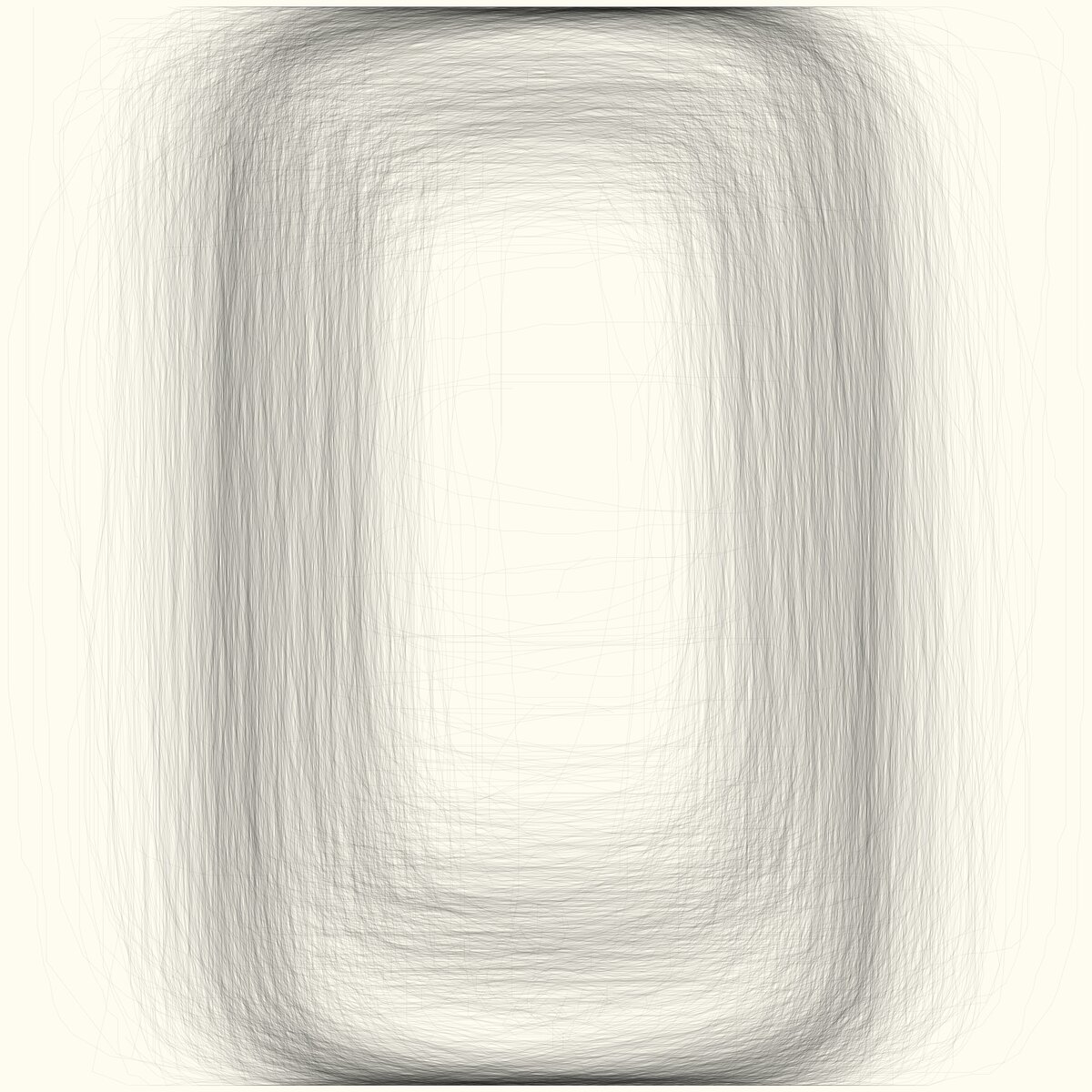}} &
        \raisebox{-.5\height}{\includegraphics[width=0.1\textwidth]{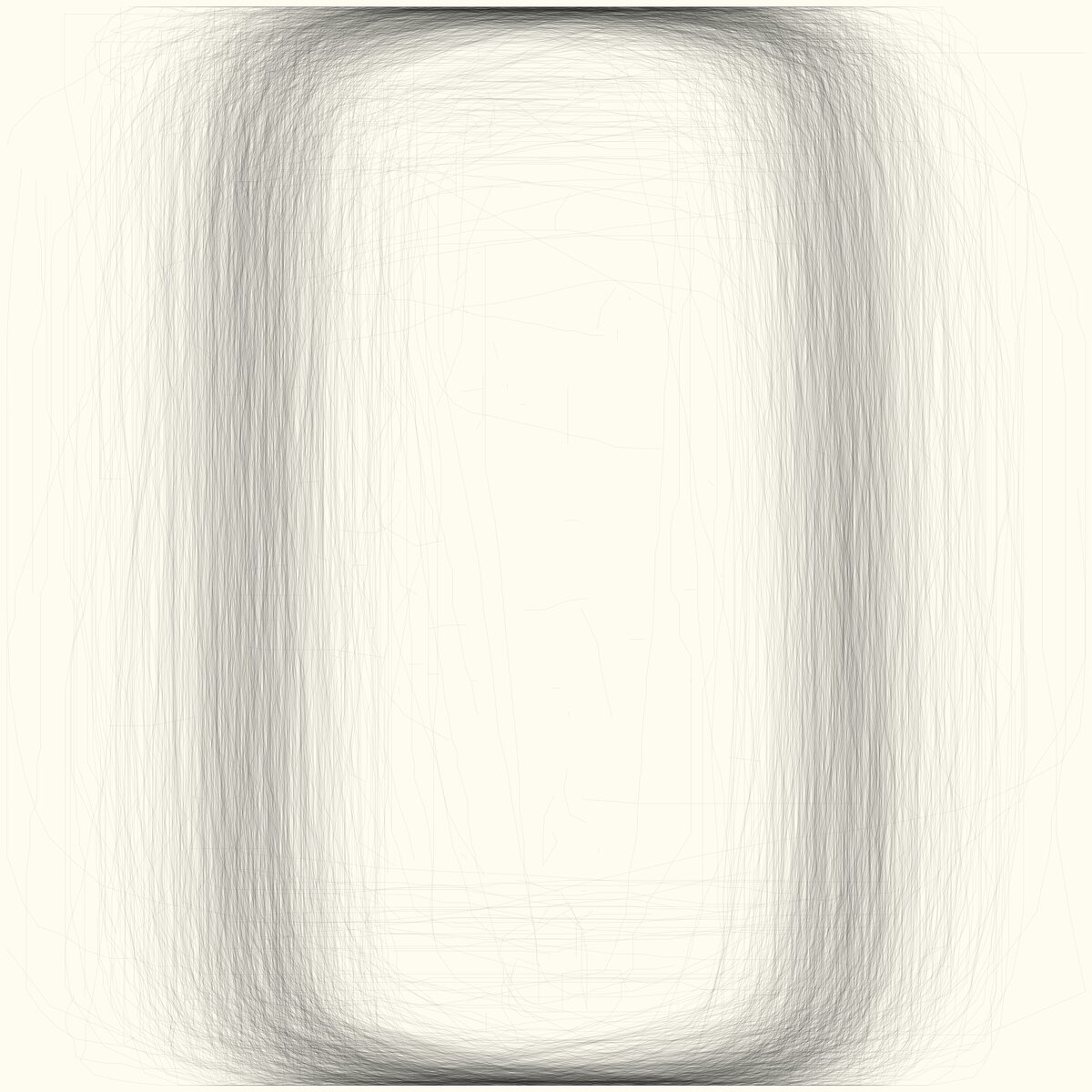}} &
        \raisebox{-.5\height}{\includegraphics[width=0.1\textwidth]{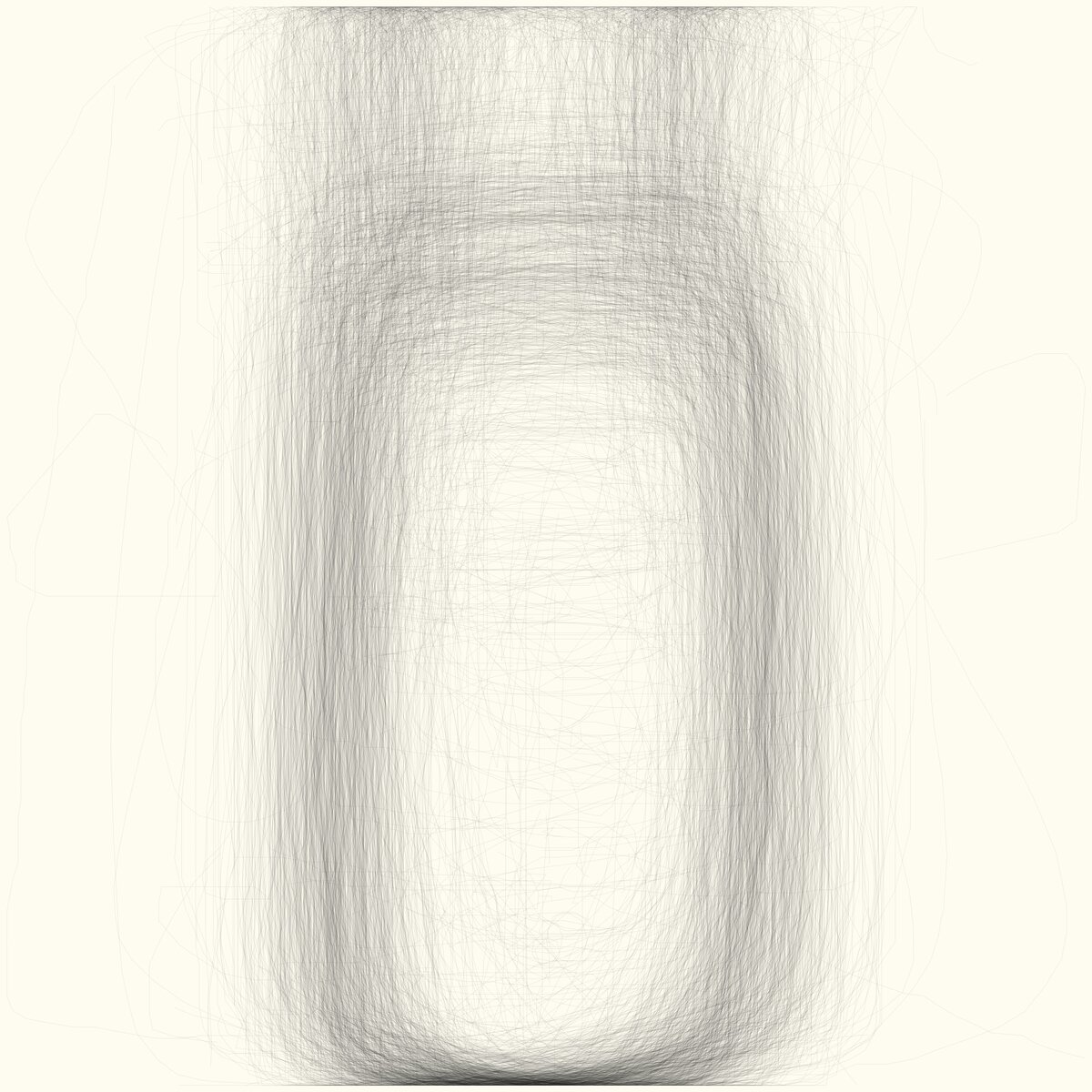}} &
        \raisebox{-.5\height}{\includegraphics[width=0.1\textwidth]{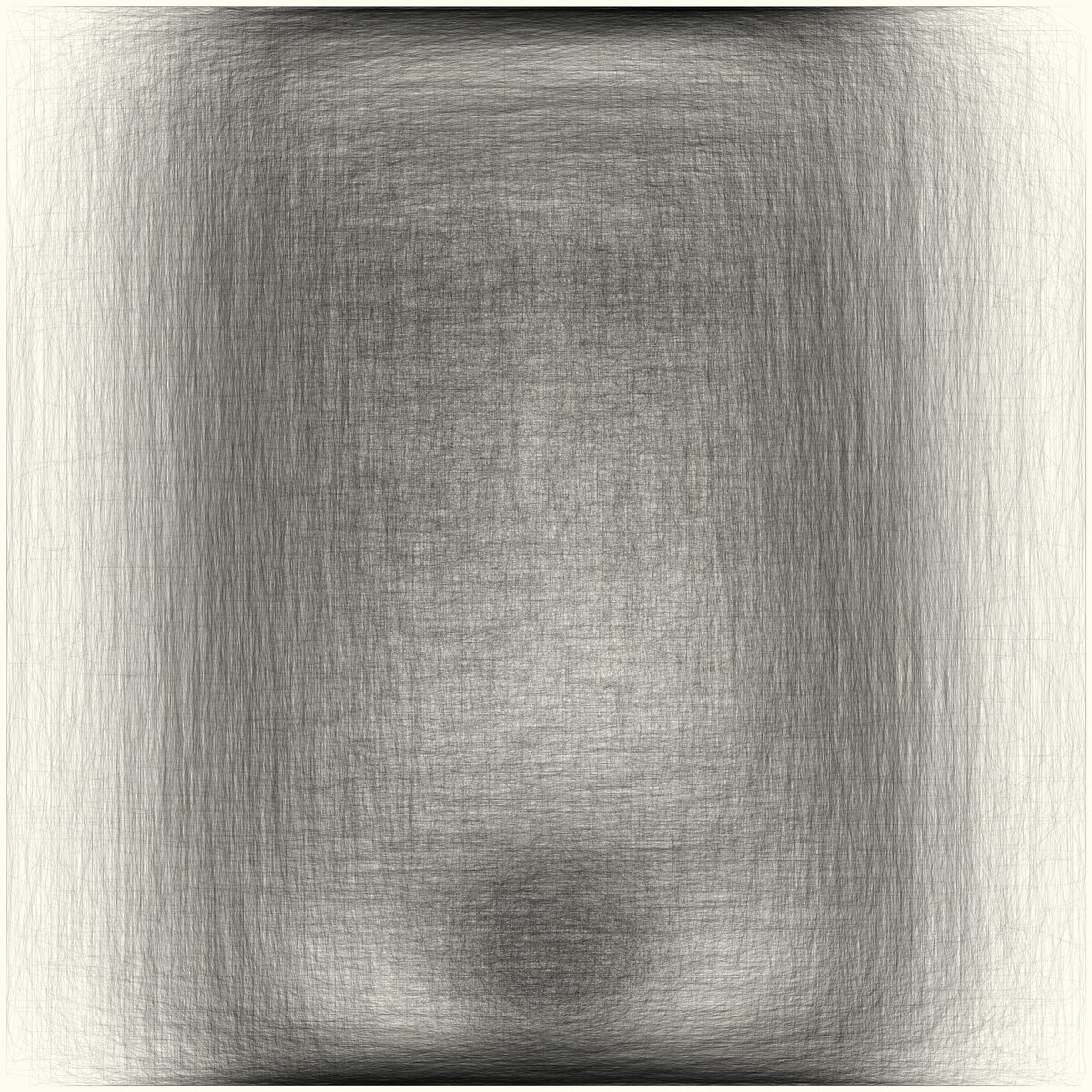}} &
        \raisebox{-.5\height}{\includegraphics[width=0.1\textwidth]{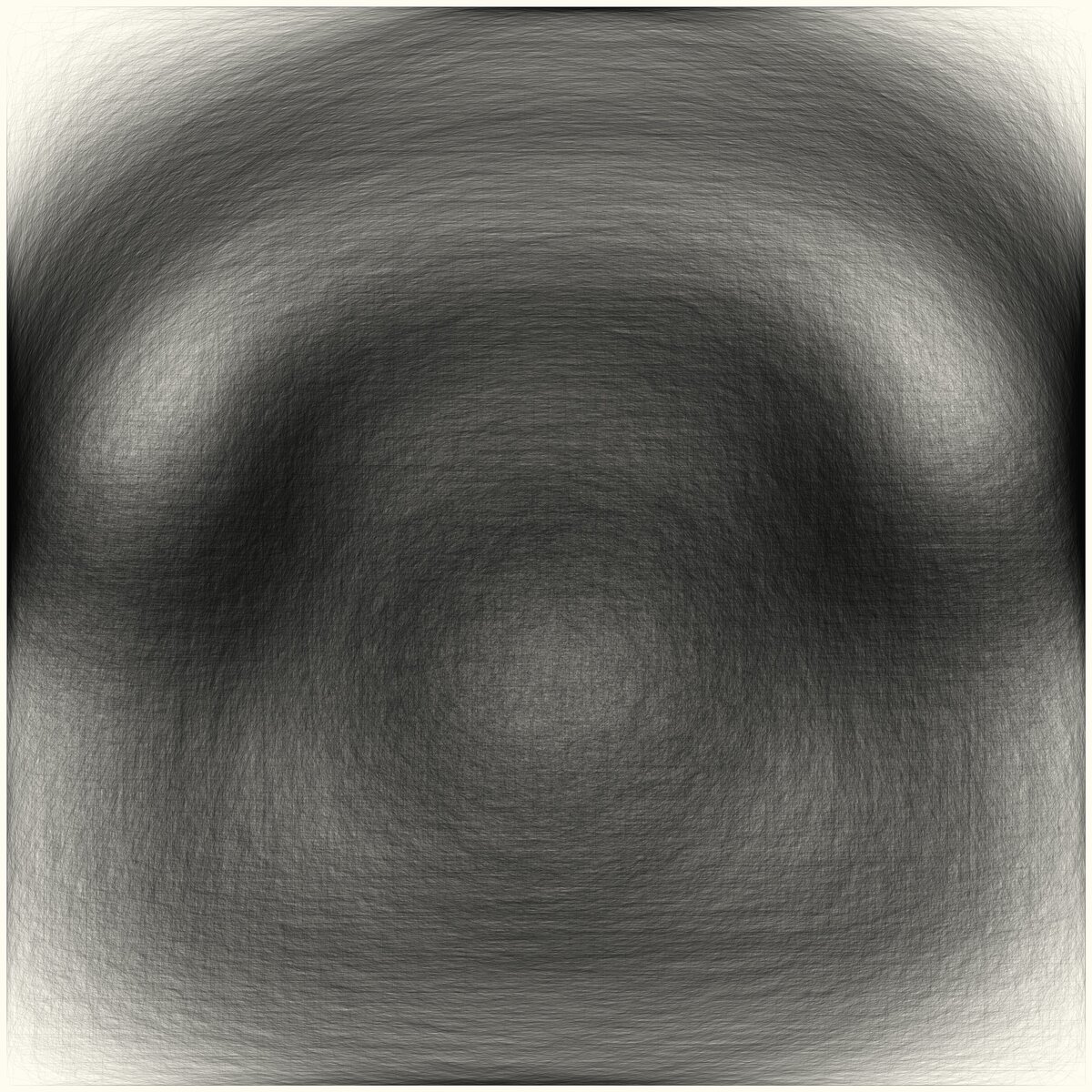}} \\
    \end{tabular}
    \caption{Superposition of randomly sampled drawings within each cluster, for six representative concepts.}
    \label{fig:overlap_clusters}
\end{figure}

\begin{figure}[htbp]
    \centering

    \begin{subfigure}{0.31\textwidth}
        \includegraphics[width=\linewidth]{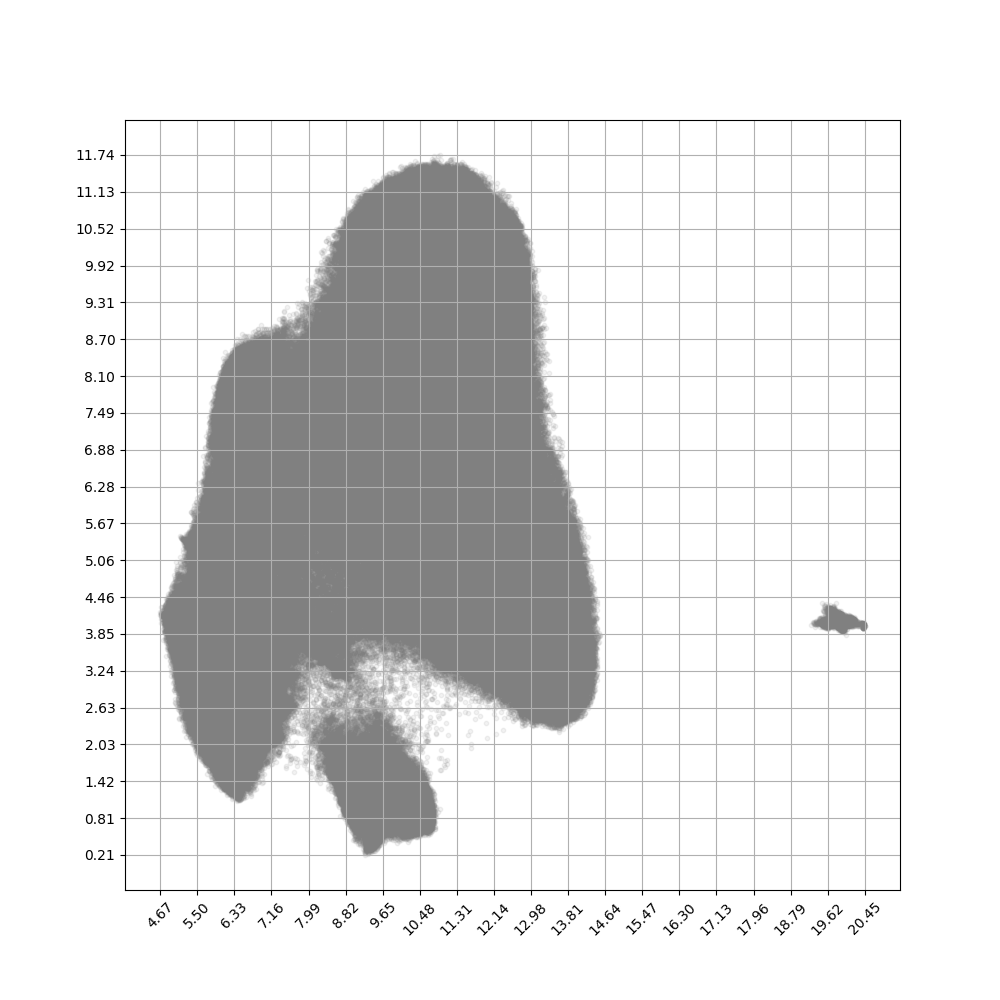} 
        \caption{Grid over UMAP.}
    \end{subfigure}
    \hspace{0.1cm}
    \begin{subfigure}{0.31\textwidth}
        \includegraphics[width=\linewidth]{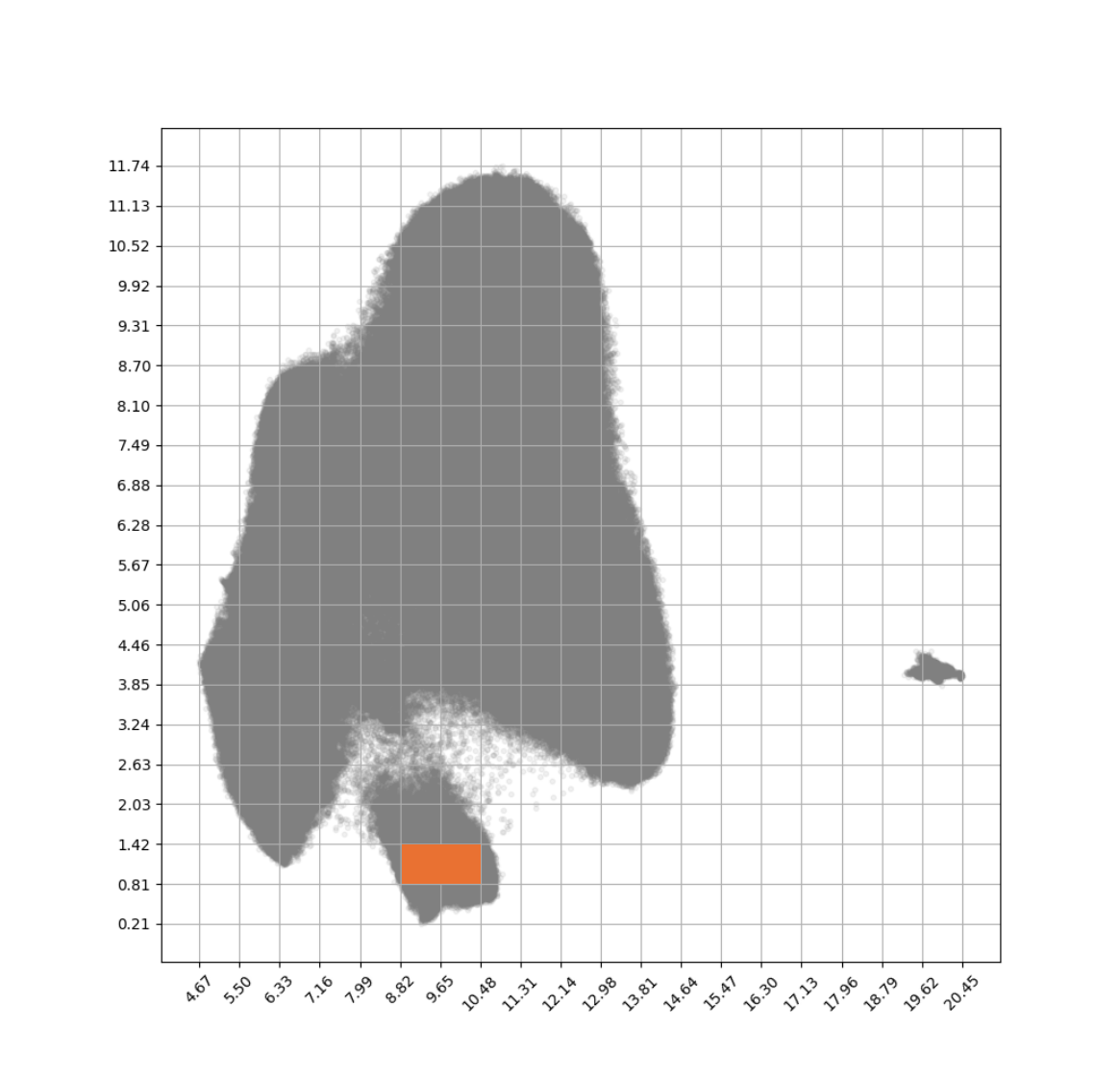}
        \caption{High-density cells.}
    \end{subfigure}
    \hspace{0.1cm}
    \begin{subfigure}{0.31\textwidth}
        \includegraphics[width=\linewidth]{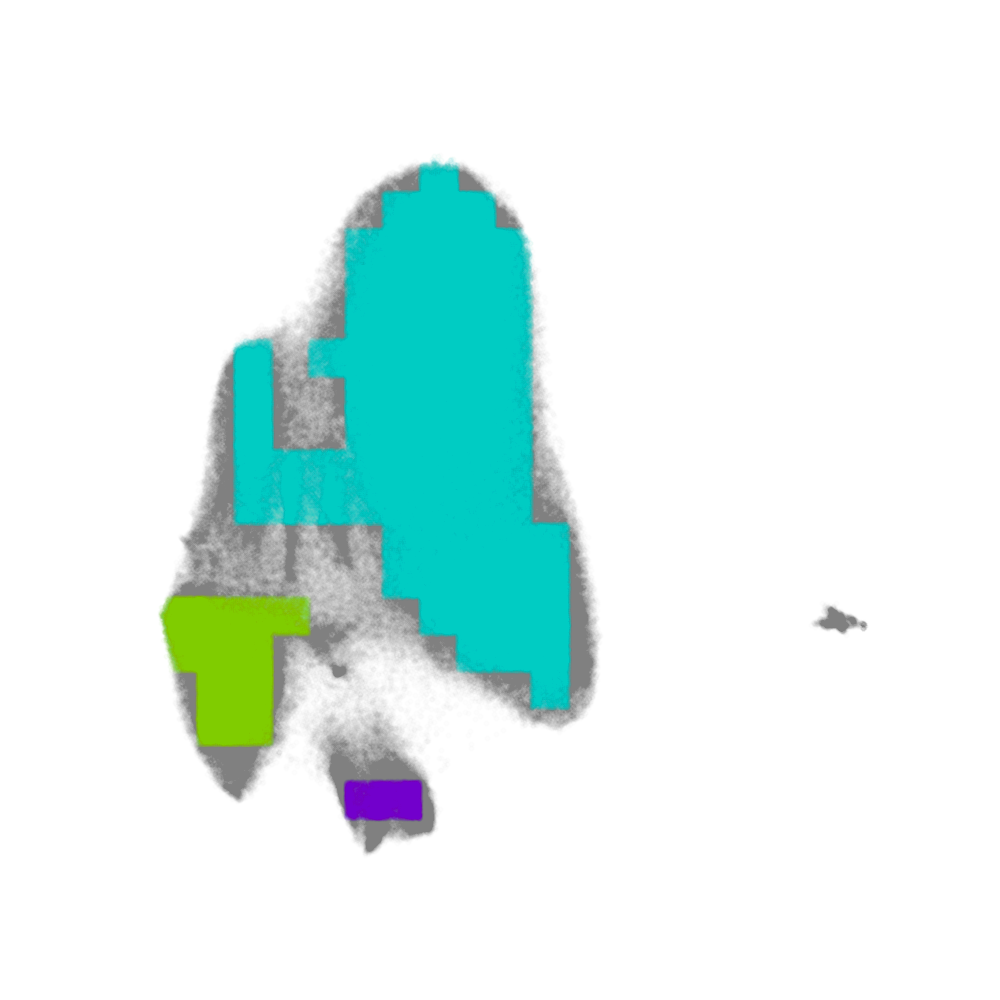}
        \caption{Cluster-like groups.}
    \end{subfigure}

    \caption{Steps of the grid-based pipeline on a selected example concepts (\emph{avocado}). (a) Imposition of a 20x20 cells grid over the UMAP representation of image embeddings. (b) Definition of high-density grid cells for the concept. (c) Identification of connected components of high-density grid cells.}
    \label{fig:grid_pipeline}
\end{figure}

\noindent We implement a grid-based pipeline to extract cluster-like groups of similar sketches from categories identified as non-clusterable using DBSCAN (cf. Supplementary Figure \ref{fig:grid_pipeline}).
In the pipeline, there is one parameter to be optimized: the proportions of points in each grid-cell for the cell to be considered \emph{high-density}. 
We evaluate the 60th, 70th, 80th, 90th, and 95th percentile values of the distribution of proportions across concepts.
To identify the optimal parameter value, we run the alternative clustering process on the categories that were instead classified as clusterable and report the results in Supplementary Figure~\ref{fig:precision_gridcomponents}, in terms of precision.

\begin{figure}[t!]
    \centering
    \includegraphics[width=0.6\textwidth]{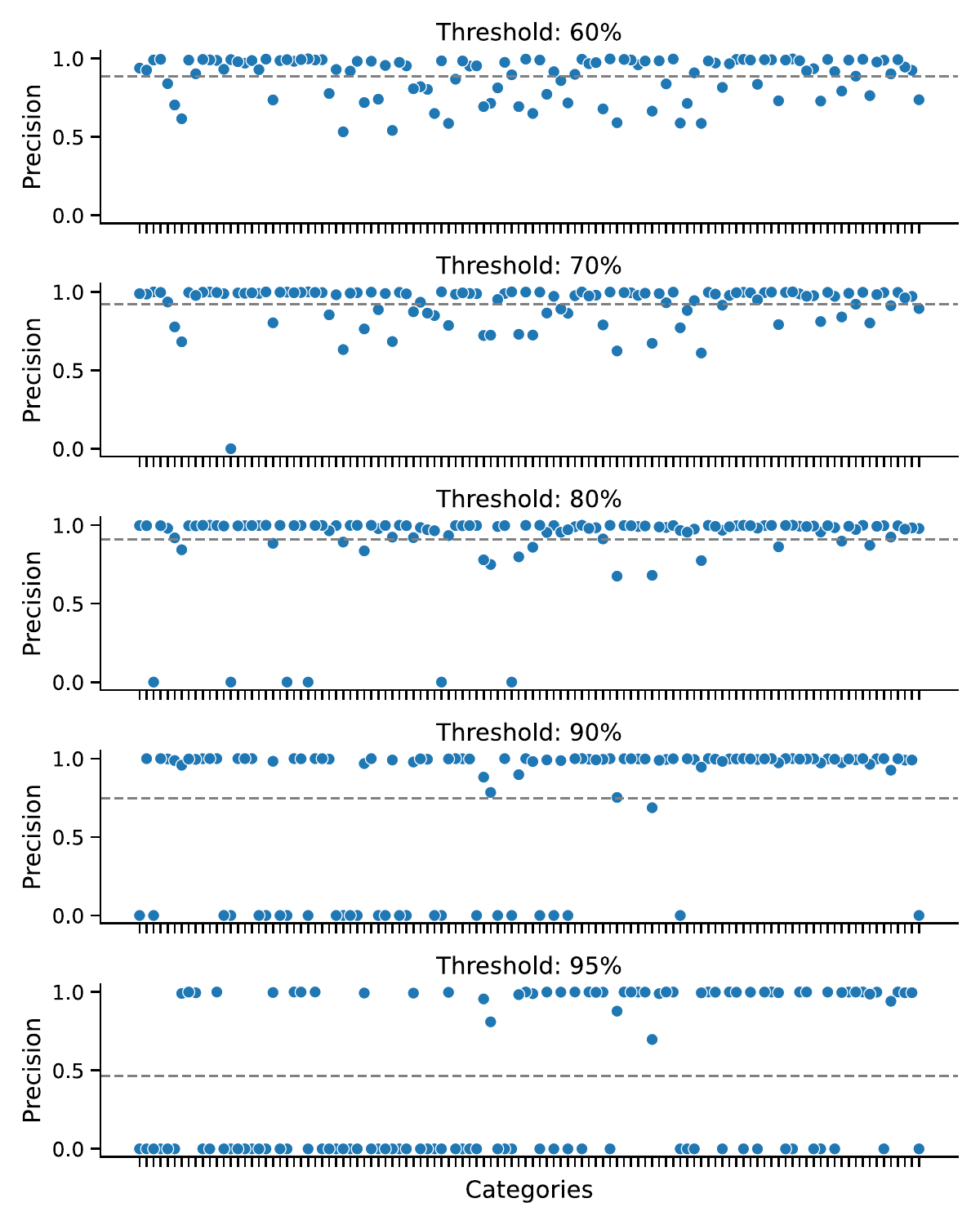} 
    \caption{Performance of the alternative clustering methodology on clusterable concept categories in terms of precision, considering different percentile threshold values. The grey dashed line represents, for each percentile threshold value, the average precision across categories.}
    \label{fig:precision_gridcomponents}
\end{figure}

\subsection{Image-based Network of Countries}

\begin{figure}[htbp]
    \centering
    \includegraphics[width=0.7\textwidth]{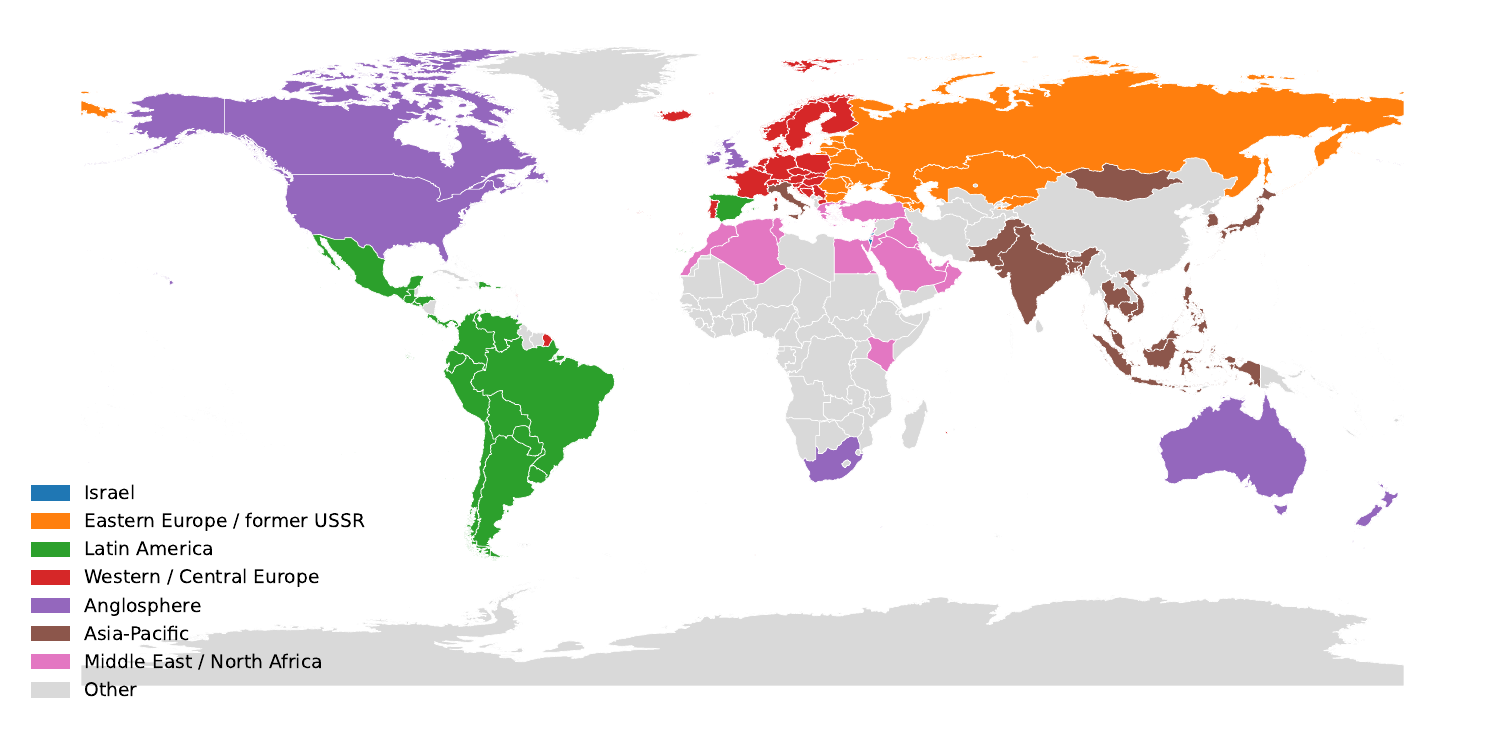} 
    \caption{Communities of countries in the image embeddings-based network, plotted on a world map.}
    \label{fig:communities_map}
\end{figure}

Supplementary Figure \ref{fig:communities_map} shows the Louvain-based communities of countries identified in the image-embedding-based network, plotted on a world map. Countries within the same community tend to form clear regional clusters, such as English-speaking countries, Latin America, Western-Central Europe, Eastern Europe, Africa and the Middle East, and Asia.

\subsection{Image vs. Language Comparison}
Given a target concept, we compare the ranking of most similar concepts based on image and word similarity. We consider Rank-Biased Overlap at top-5, top-10, and top-20, percentage overlap of top-10, and Kendall-Tau on full ranking as rank correlation metrics.
The correlation of word- and image-based rankings are very low, as seen in Supplementary Figure \ref{fig:rank_emb}, both when considering Word2Vec and a BERT-based multilingual model for word embeddings.

\begin{figure}[htbp]
    \centering

    \begin{subfigure}{0.3\textwidth}
        \includegraphics[width=\linewidth]{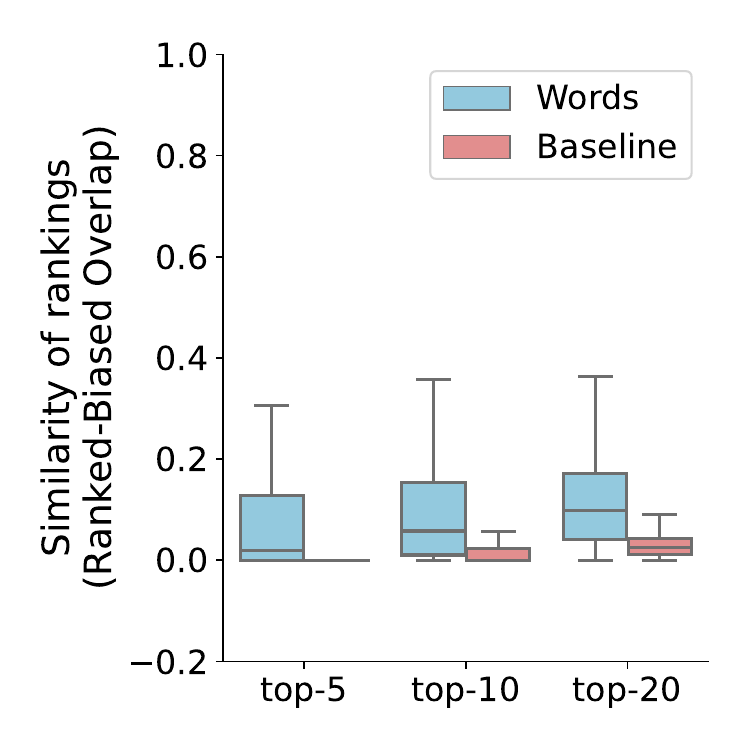}
        \caption{Rank-Biased Overlap at different top-$k$ (Word2Vec).}
    \end{subfigure}
    \hspace{0.2cm}
    \begin{subfigure}{0.3\textwidth}
        \includegraphics[width=\linewidth]{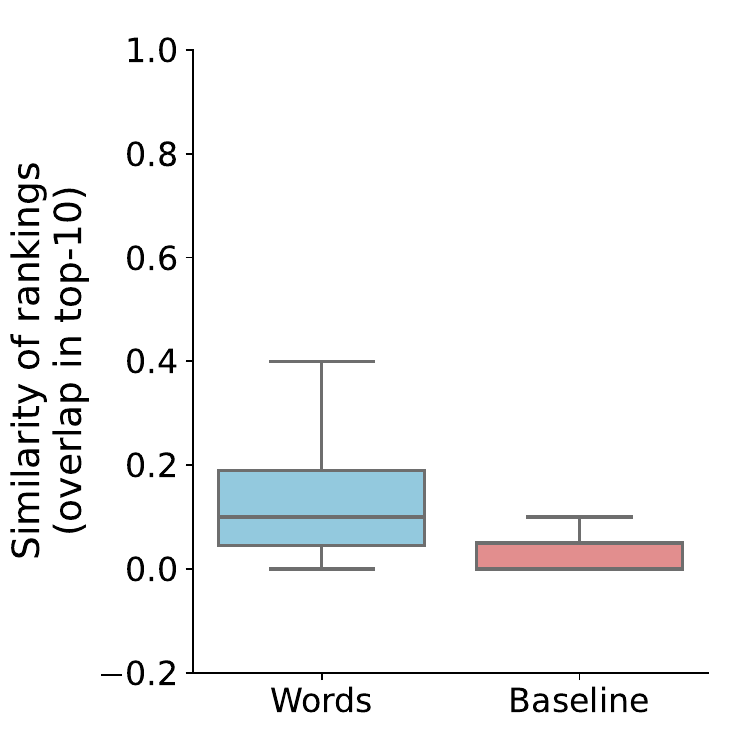}
        \caption{\% overlap of top-10 most similar concepts (Word2Vec).}
    \end{subfigure}
    \hspace{0.2cm}
    \begin{subfigure}{0.3\textwidth}
        \includegraphics[width=\linewidth]{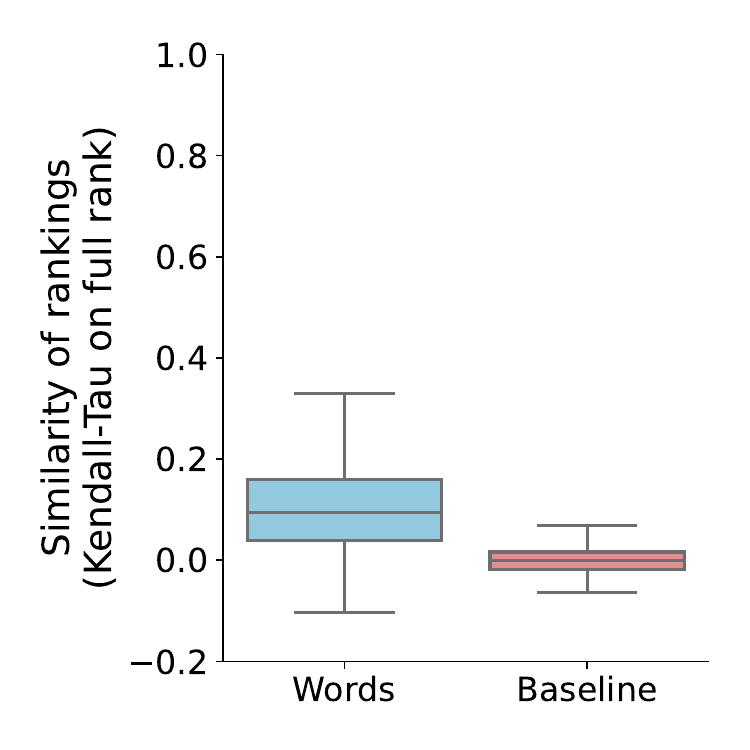}
        \caption{Kendall-Tau on full ranking (Word2Vec).}
    \end{subfigure}

    \vspace{0.5cm}
    \begin{subfigure}{0.3\textwidth}
        \includegraphics[width=\linewidth]{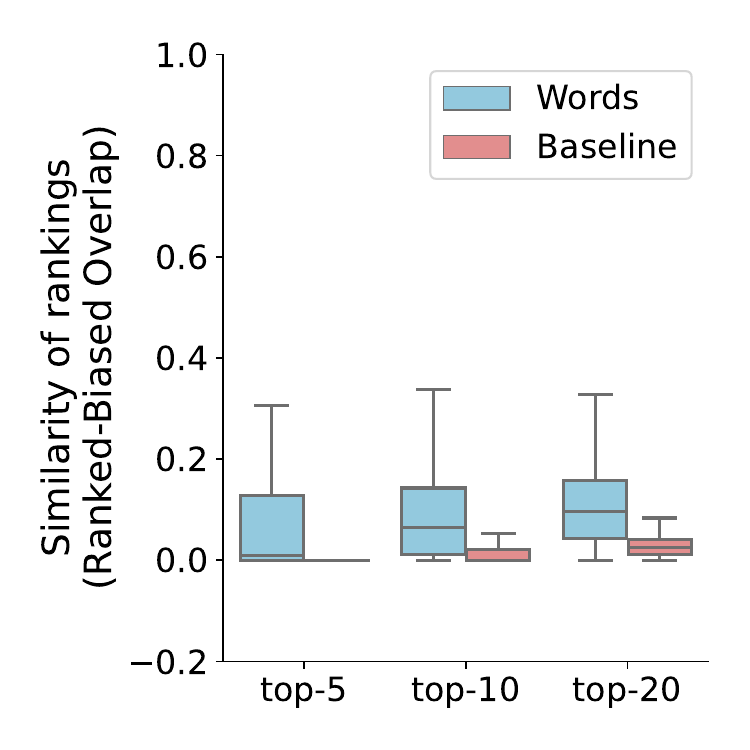} 
        \caption{Rank-Biased Overlap at different top-$k$ (Multilingual).}
    \end{subfigure}
    \hspace{0.2cm}
    \begin{subfigure}{0.3\textwidth}
        \includegraphics[width=\linewidth]{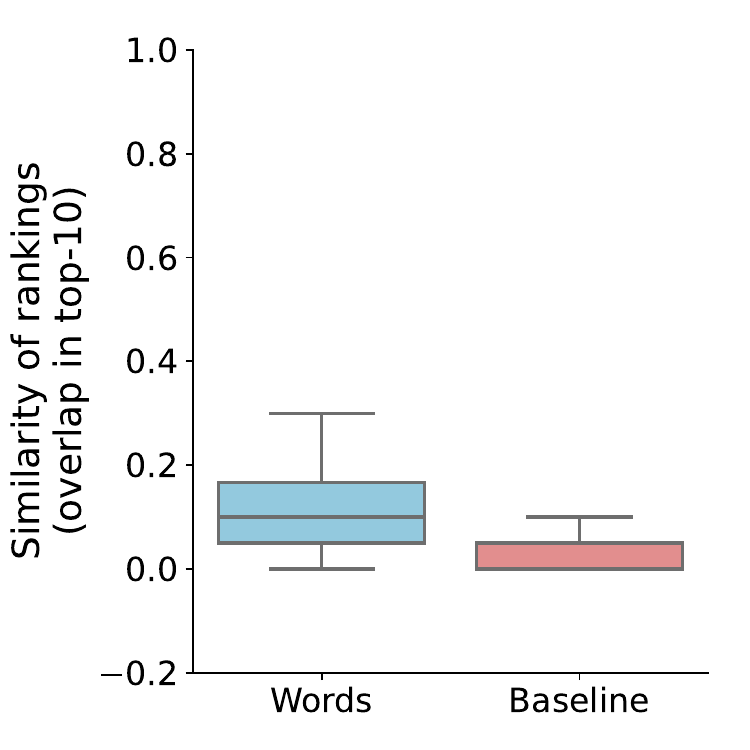}
        \caption{\% overlap of top-10 most similar concepts (Multil.).}
    \end{subfigure}
    \hspace{0.2cm}
    \begin{subfigure}{0.3\textwidth}
        \includegraphics[width=\linewidth]{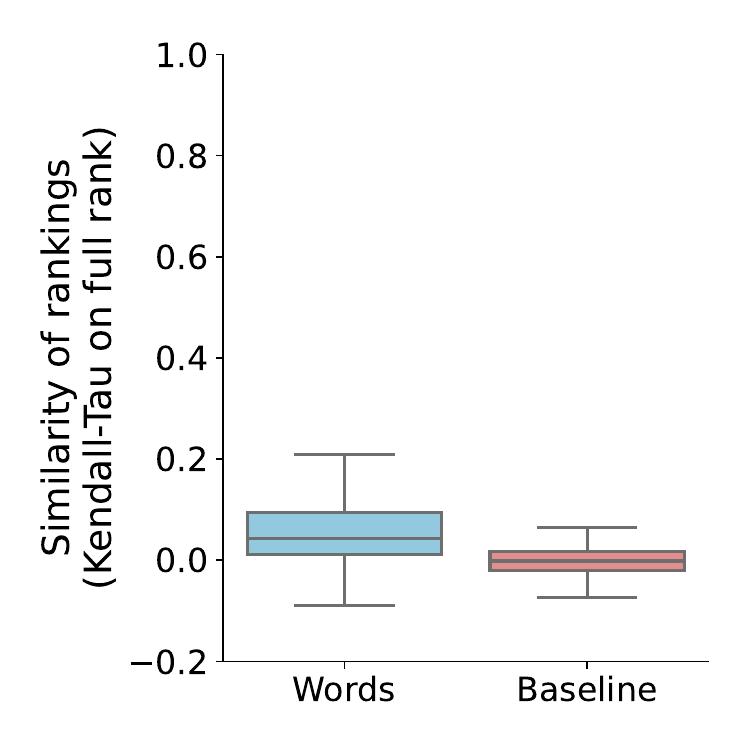}
        \caption{Kendall-Tau on full ranking (Multilingual).}
    \end{subfigure}

    \caption{Similarity of rankings considering word and image embeddings, together with a baseline comparing the ranking of image embeddings and a randomly shuffled counterpart. A Word2Vec model for word embeddings is considered in the top row (a-c), while a BERT-based multilingual model is considered in the bottom row (d-f).}
    \label{fig:rank_emb}
\end{figure}

\begin{figure}[htbp]
    \centering
    \includegraphics[width=0.6\textwidth]{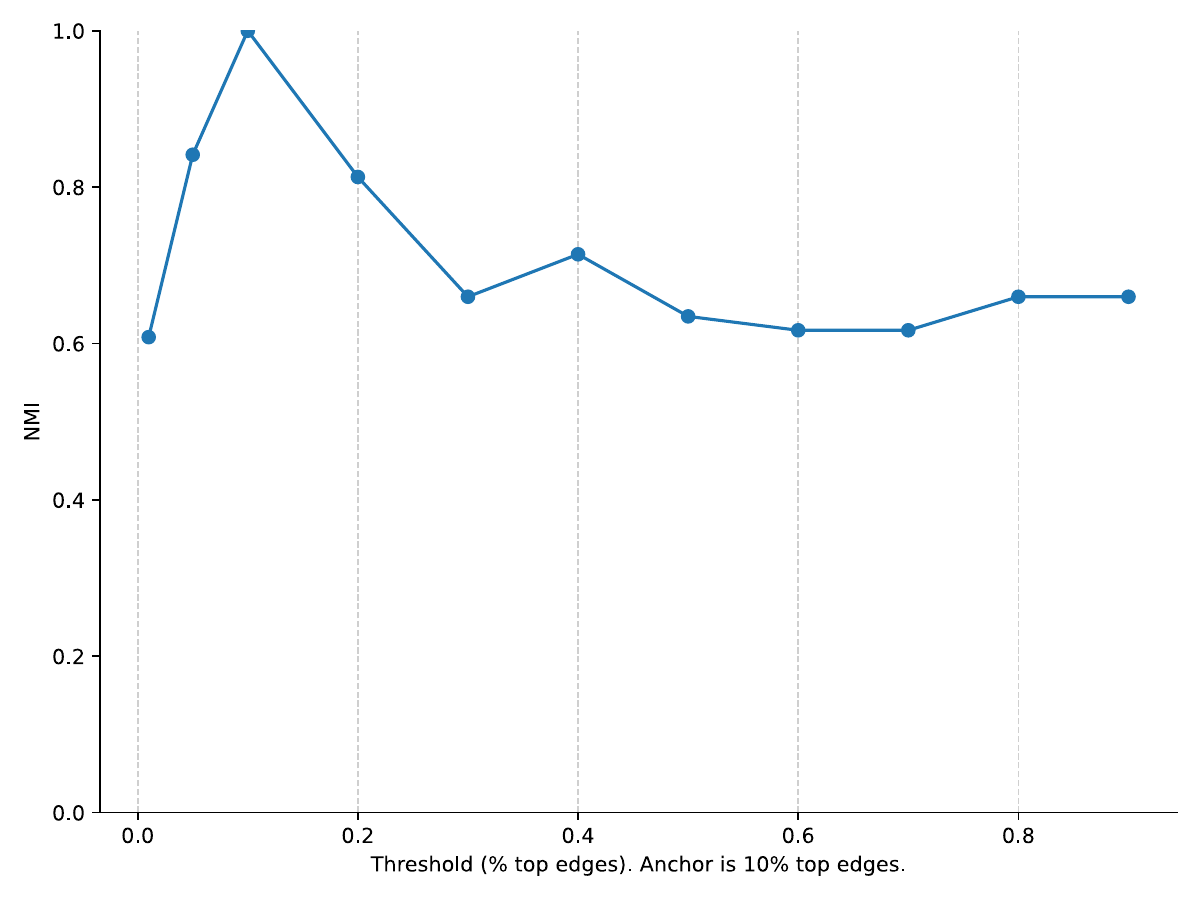} 
    \caption{NMI score for robustness of community structure across percentages of strongest edges in the network.}
    \label{fig:nmi_community-robust}
\end{figure}

When comparing networks of country similarities based on sketch production and language, we use a fixed number of nodes (top 100 countries by volume of sketches) in the main analysis. We then retain the top-10\% strongest edges from the complete similarity graph connecting these nodes. Supplementary Figure \ref{fig:nmi_community-robust} shows the robustness of Louvain community structures across different percentage of strongest edges retained (1\%, 5\%, 10\%, 20\%, 30\%, 40\%, 50\%, 60\%, 70\%, 80\%, 90\%), using the Normalized Mutual Information (NMI) score for the comparison. Specifically, we consider community labels for the graph with the top-10\% edges as an anchor to compare community labels at other percentage values for edge strength. Values range from 0.60 to 1 (in the case of self-comparison), oscillating around 0.80-0.85 for the percentage of top edges closest to the anchor (i.e., 5\% and 20\%).

\begin{figure}[htbp]
    \centering

    \begin{subfigure}{0.38\textwidth}
        \includegraphics[width=\linewidth]{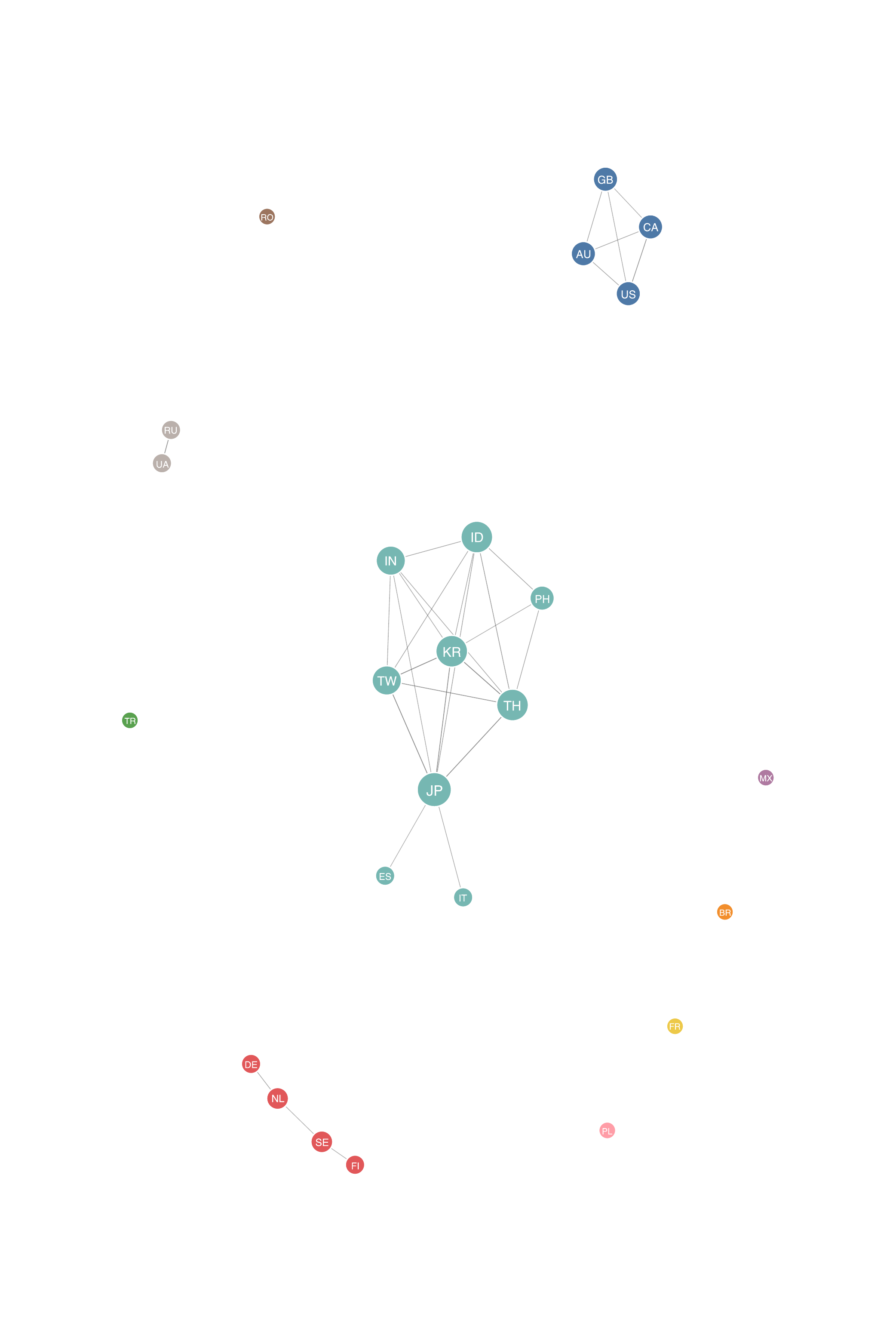} 
        \caption{Image-based, \textbf{25} nodes.}
    \end{subfigure}
    \hspace{0.5cm}
    \begin{subfigure}{0.38\textwidth}
        \includegraphics[width=\linewidth]{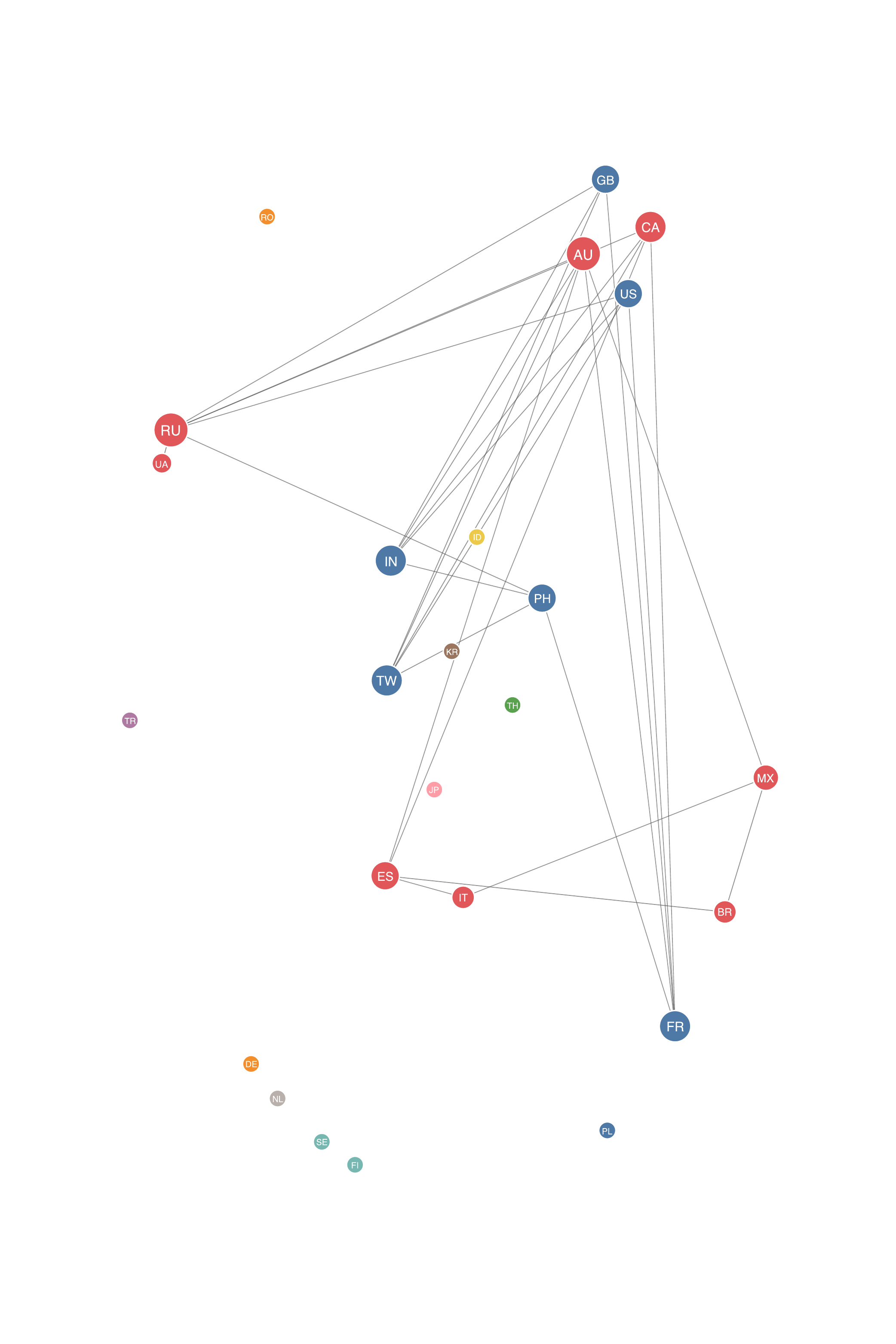}
        \caption{Language-based, \textbf{25} nodes.}
    \end{subfigure}
    \vspace{0.3cm}
    \begin{subfigure}{0.38\textwidth}
        \includegraphics[width=\linewidth]{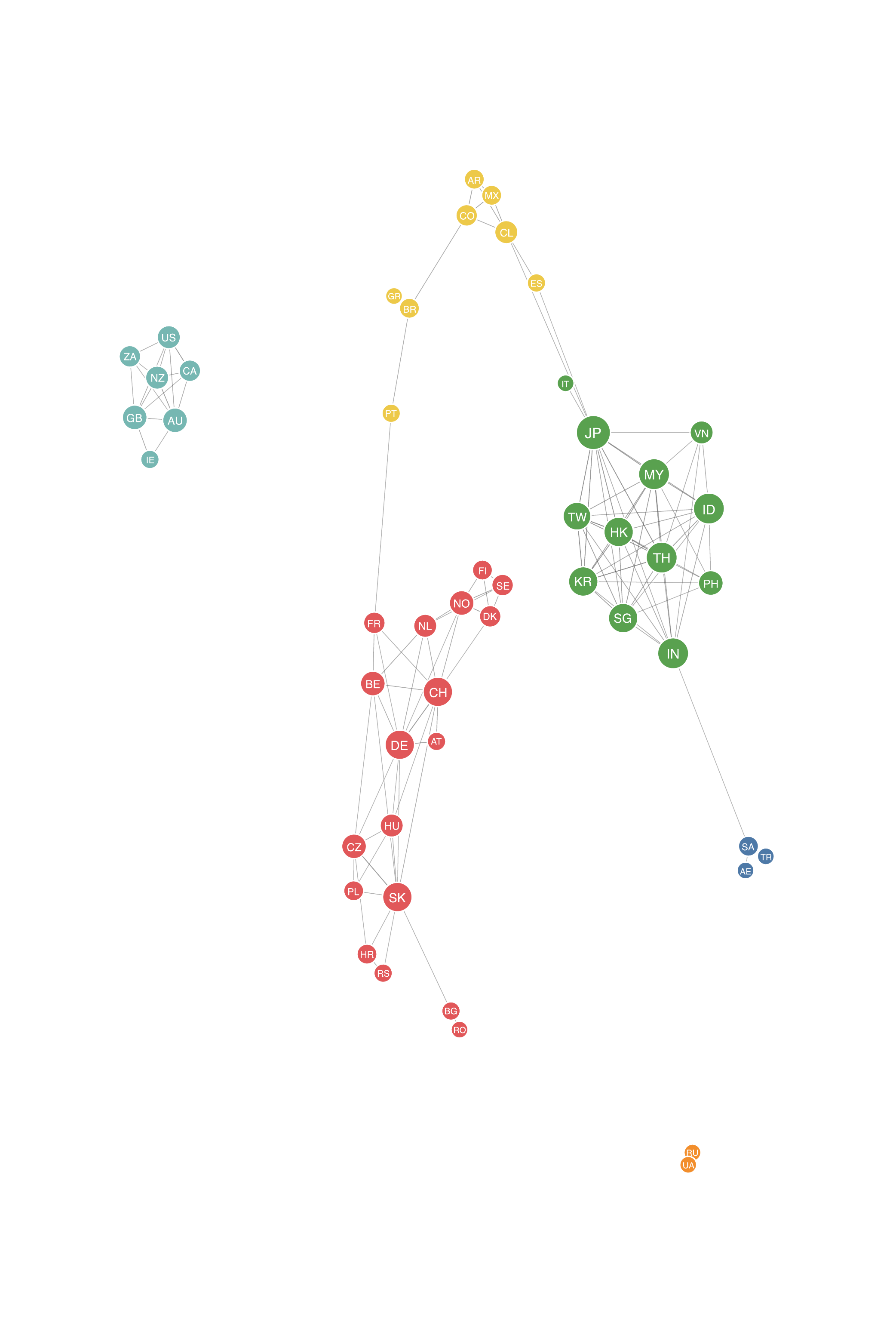}
        \caption{Image-based, \textbf{50} nodes.}
    \end{subfigure}
    \hspace{0.5cm}
    \begin{subfigure}{0.38\textwidth}
        \includegraphics[width=\linewidth]{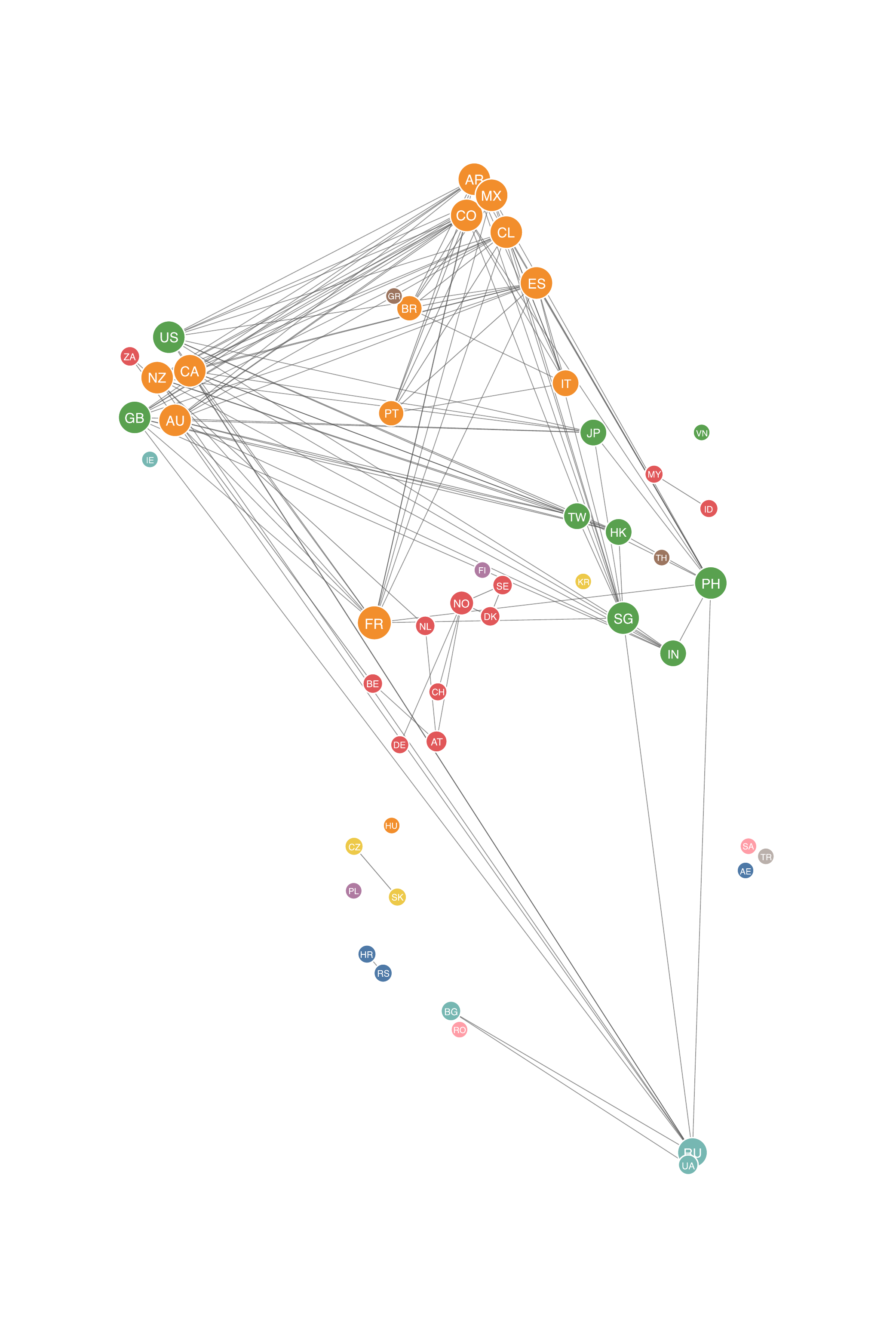}
        \caption{Language-based, \textbf{50} nodes.}
    \end{subfigure}

    \caption{Comparison of network countries' similarity based on image and words. Each row corresponds to a different number of top-$n$ nodes retained. The word-based network is mapped to the coordinates of the image-based one to facilitate comparison.}
    \label{fig:network-robust}
\end{figure}

To test the robustness of our findings to the choice of number of countries considered in the network, we also evaluate alternative configurations with 25 and 50 nodes (25 and 50 most prominent countries, respectively). In each case, we adjust the number of edges to preserve the same network density as in the reference image-based network with 100 nodes and top-10\% strongest edges. Supplementary Figure~\ref{fig:network-robust} shows that the main identified communities (e.g., Asia, countries with English as their primary language, Europe, Post-Soviet Eurasia) remain stable across these different configurations. 
The NMI computed on the intersection of nodes between the 100-node network and the 25-node subset is 0.94, and between the 100-node and 50-node networks is 0.87.

We also test the robustness of the network comparison metrics between the image-based, language-based, and culture-based networks by varying the number of nodes (25 and 50) and the percentage of strongest edges retained (1\%, 5\%, and 10–90\% in steps of 10).
The results, shown in Supplementary Figure \ref{fig:network_metrics-robust}, indicate that the similarity ratios are greater than 1 across different number of nodes (the median ratio across metrics is 1.034 for the networks with 25 nodes and 1.038 for the ones with 50 nodes), confirming that the image-based network is more similar to the culture-based graph than the language-based network is.

\begin{figure}[htbp]
    \centering
    \begin{subfigure}{0.45\textwidth}
        \includegraphics[width=\linewidth]{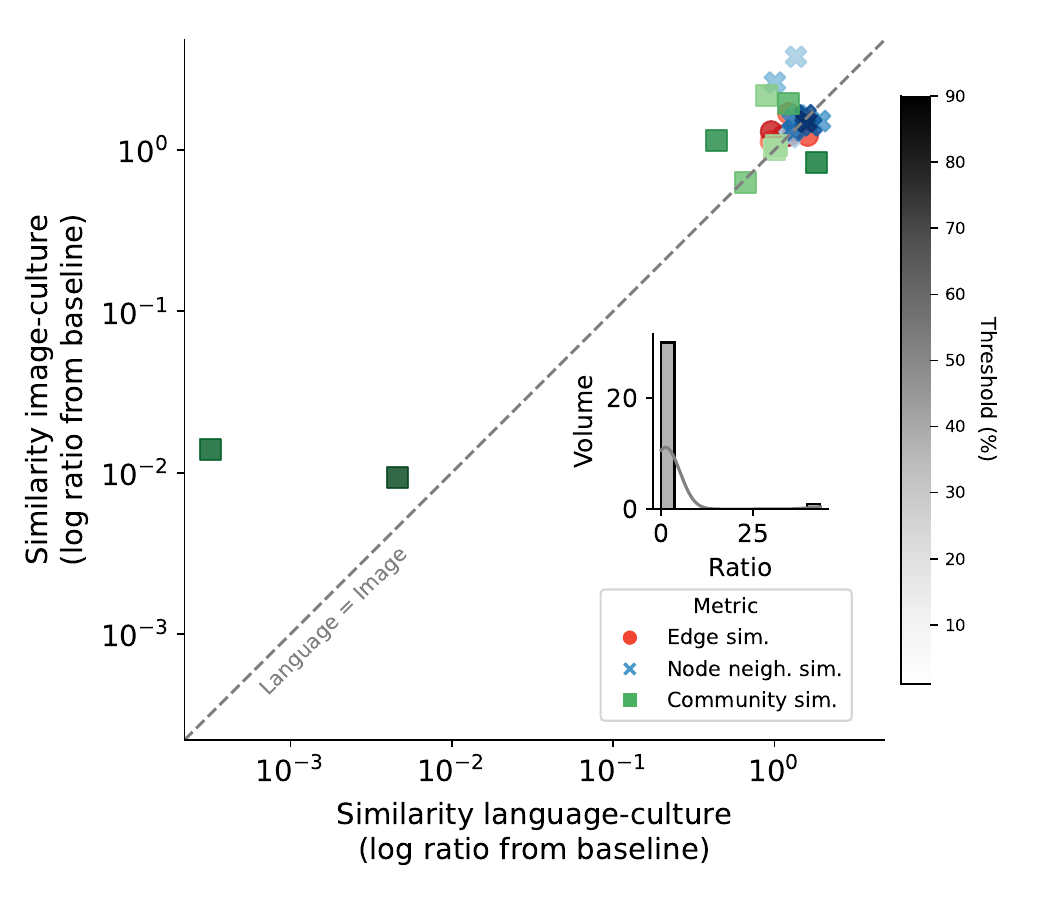} 
        \caption{\textbf{25} nodes.}
    \end{subfigure}
    \hspace{0.2cm}
    \begin{subfigure}{0.45\textwidth}
        \includegraphics[width=\linewidth]{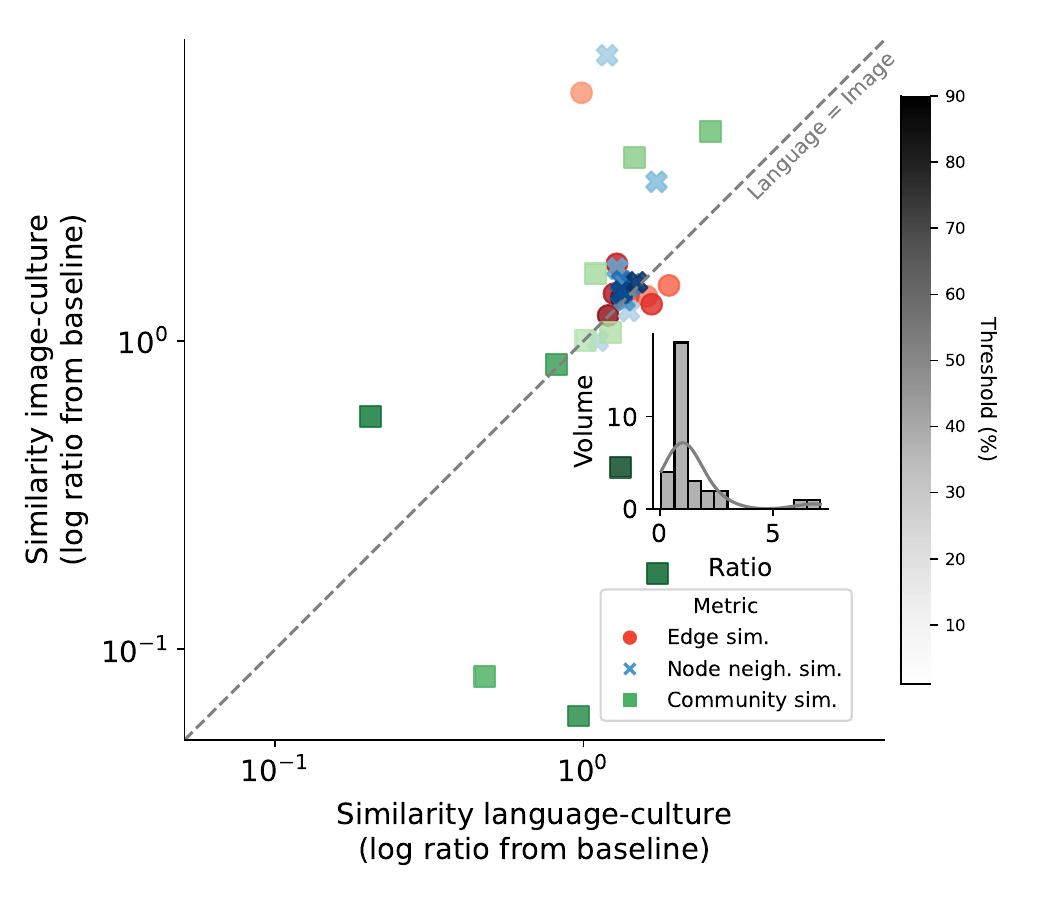}
        \caption{\textbf{50} nodes.}
    \end{subfigure}
    \caption{Comparison between image- and language-based network similarity with the cultural network, across metrics, considering the ratio with respect to a baseline.}
    The analysis focuses on the 25 (left) and 50 (right) most represented countries in the dataset; 
    The diagonal represents equality between image- and language-based comparisons, values above the diagonal show a higher similarity for the image network and values below show a higher similarity for the language network.
    The color gradient refers to the threshold considered (1\%, 5\% and from 10\% to 90\%, with a step of 10: lighter for lower thresholds, darker for higher ones. This is reflected in the colormap reported in the plot in grey scale. The embedded histograms show the distribution of the ratio between image-based and language-based similarity with the cultural network across all metrics and thresholds.
    \label{fig:network_metrics-robust}
\end{figure}

A partial Mantel test confirms this result by showing that image similarity remains significantly associated with cultural similarity even after controlling for language similarity (significant and positive $r$), whereas language similarity is no longer significantly associated with cultural similarity once image similarity is accounted for (cf. Supplementary Table \ref{tab:mantel_results}).

\begin{table}[htbp]
\centering
\begin{tabular}{lccc}
\hline
\textbf{N. nodes} & \textbf{Test of} & \textbf{r} & \textbf{p} \\
\hline
25  &  culture--image    & 0.164 & 0.024 \\
25  &  culture--language & 0.061 & 0.293 \\
50  &  culture--image    & 0.130 & 0.024 \\
50  &  culture--language & 0.159 & 0.030 \\
100 &  culture--image    & 0.229 & 0.001 \\
100 &  culture--language & 0.064 & 0.161 \\
\hline
\end{tabular}
\caption{Partial Mantel test results across network sizes.}
\label{tab:mantel_results}
\end{table}

\begin{figure}[htbp]
    \centering
    \begin{subfigure}{0.3\textwidth}
        \includegraphics[width=\linewidth]{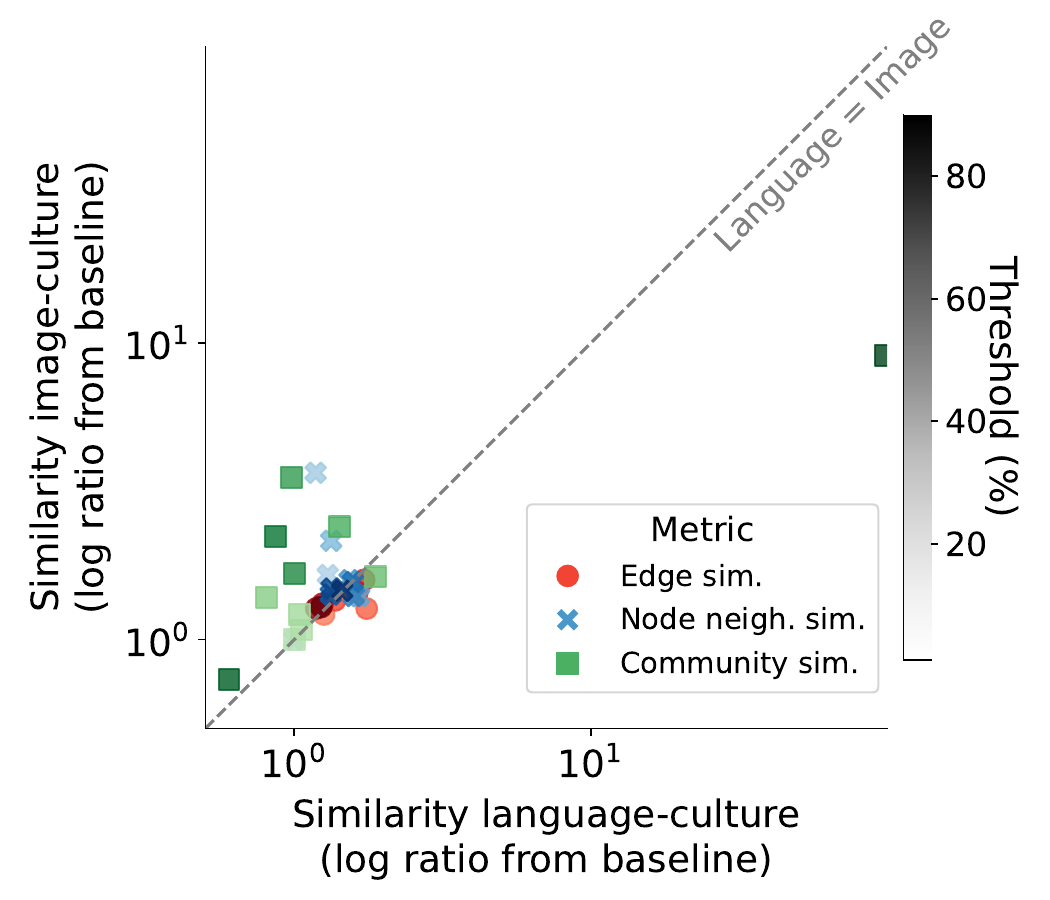}
        \caption{\textbf{25} nodes.}
    \end{subfigure}
    \hspace{0.1cm}
    \begin{subfigure}{0.3\textwidth}
        \includegraphics[width=\linewidth]{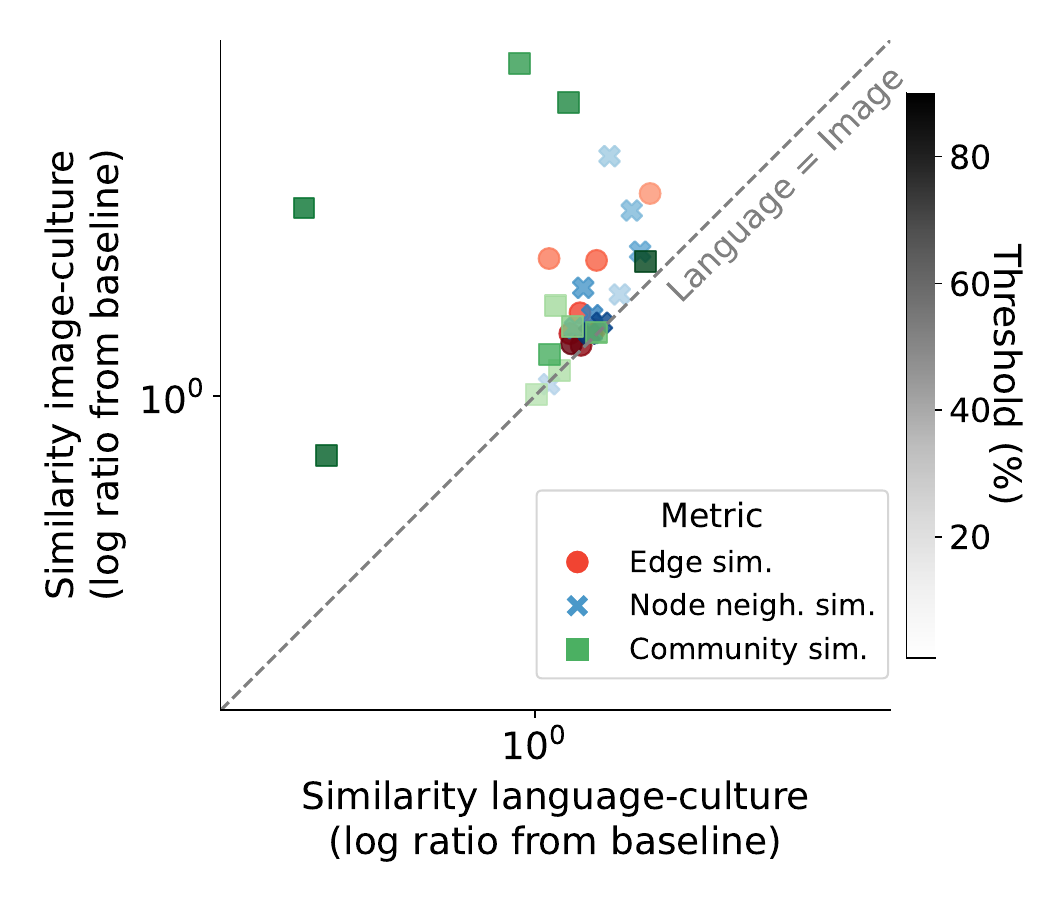}
        \caption{\textbf{50} nodes.}
    \end{subfigure}
    \hspace{0.1cm}
    \begin{subfigure}{0.3\textwidth}
        \includegraphics[width=\linewidth]{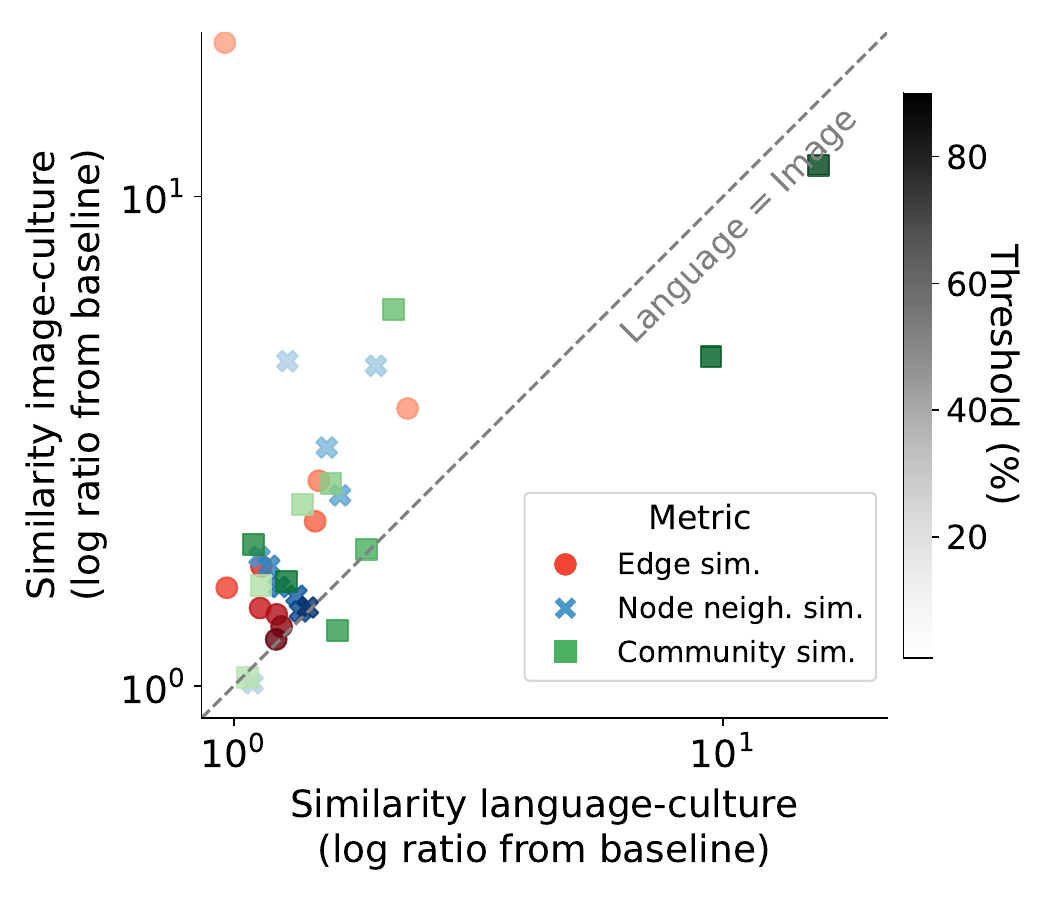}
        \caption{\textbf{100} nodes.}
    \end{subfigure}
    \caption{Comparison between image- and language-based network similarity with the cultural network, across metrics, considering the ratio with respect to a baseline.}
    The analysis focuses on the 25 (left), 50 (middle), and 100 (right) most represented countries in the dataset.
    \label{fig:network_metrics-all_waves}
\end{figure}

Supplementary Figure \ref{fig:network_metrics-all_waves} shows robustness to matching the most represented countries using all WVS waves, rather than the most recent wave alone: similarity ratios remain greater than 1 across all number of nodes (the median ratio across metrics is 1.03 for 25 nodes, 1.17 for 50 nodes, and 1.45 for 100 nodes).

\begin{figure}[htbp]
    \centering
    \begin{subfigure}{0.23\textwidth}
        \includegraphics[width=\linewidth]{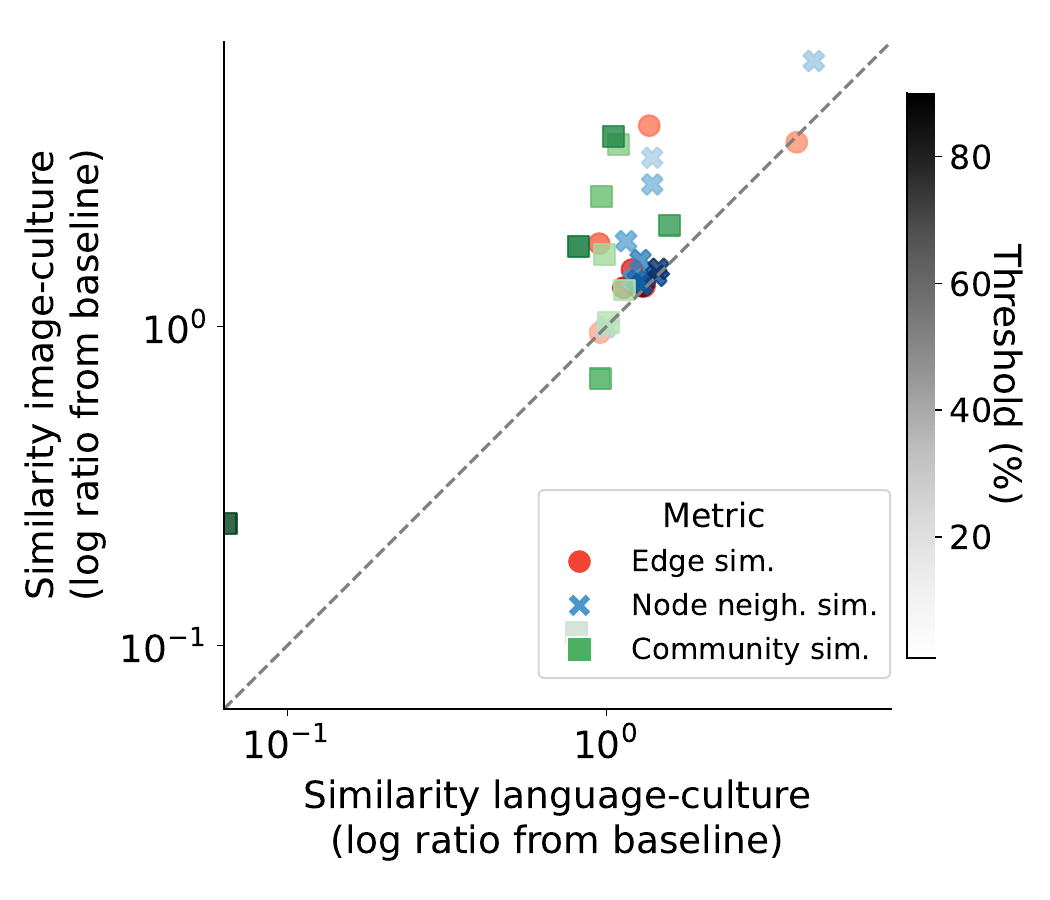}
        \caption{Altruism.}
    \end{subfigure}
    \hfill
    \begin{subfigure}{0.23\textwidth}
        \includegraphics[width=\linewidth]{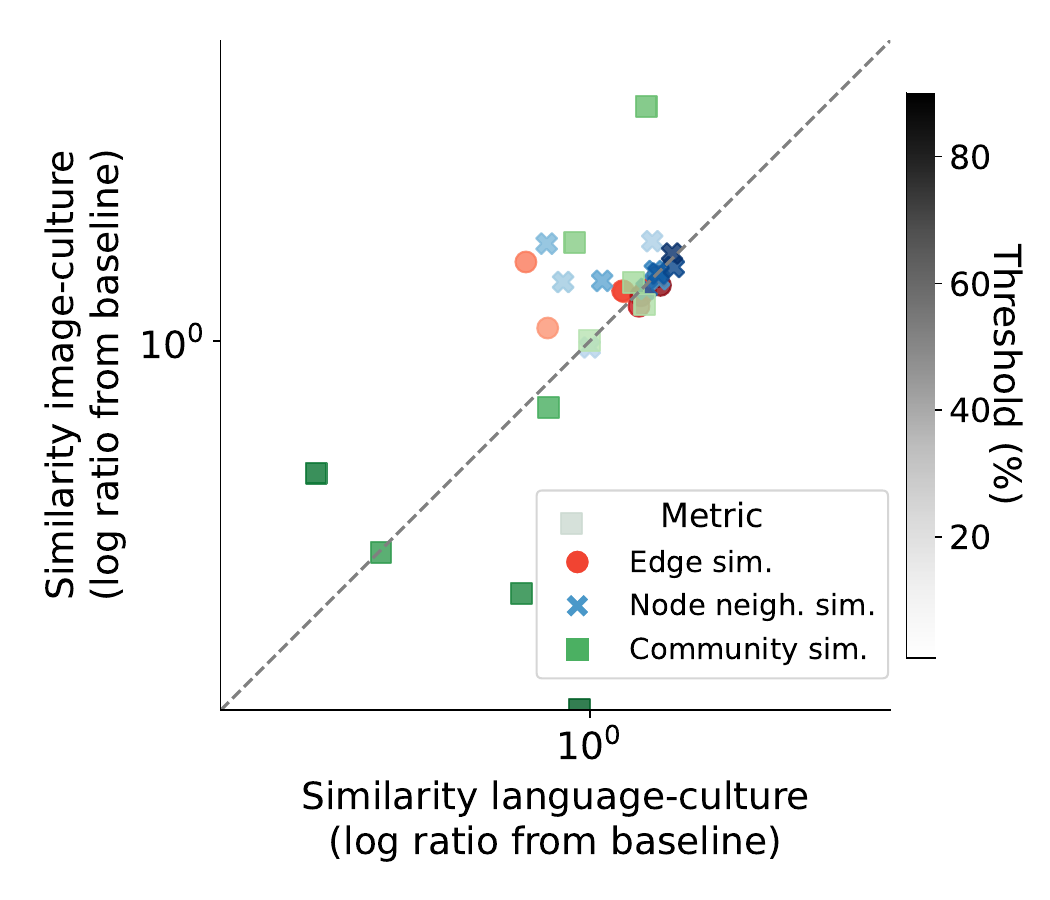}
        \caption{Arts-creativity}
    \end{subfigure}
    \hfill
    \begin{subfigure}{0.23\textwidth}
        \includegraphics[width=\linewidth]{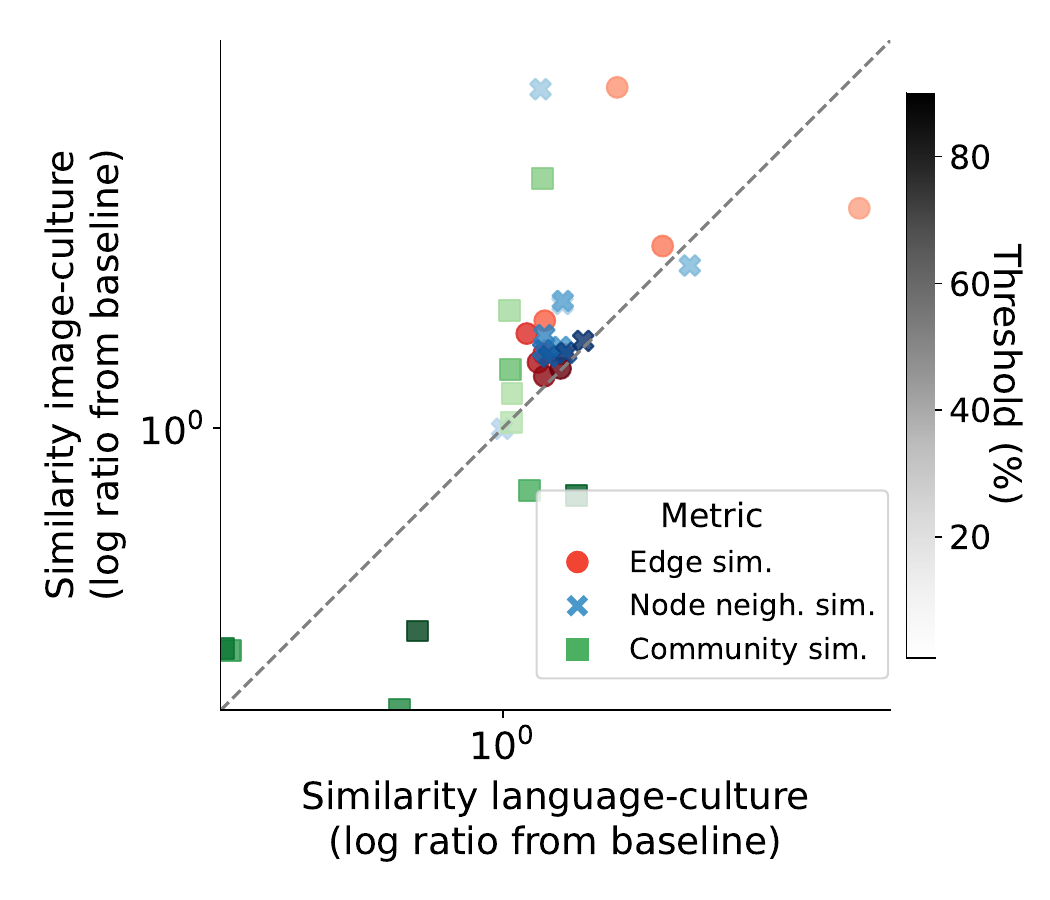}
        \caption{Beliefs}
    \end{subfigure}
    \hfill
    \begin{subfigure}{0.23\textwidth}
        \includegraphics[width=\linewidth]{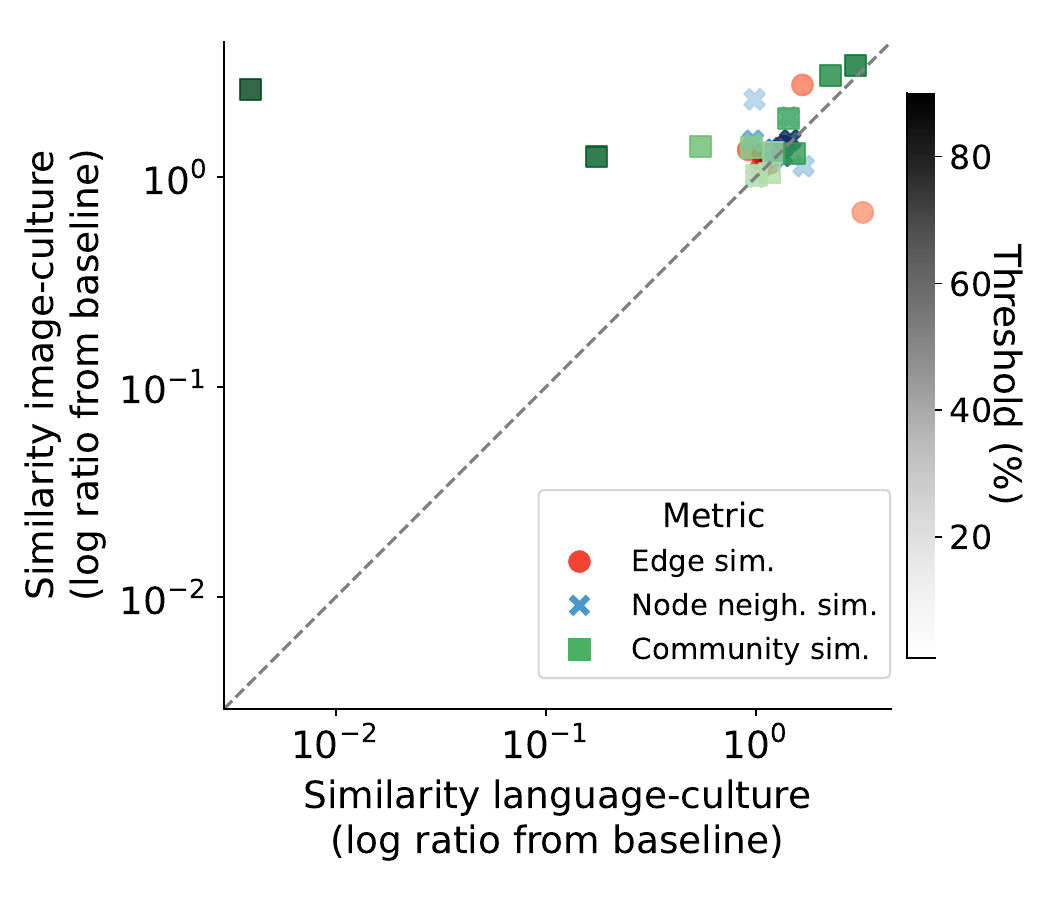}
        \caption{Env.-conservation.}
    \end{subfigure}

    \vspace{0.3cm}
    \begin{subfigure}{0.23\textwidth}
        \includegraphics[width=\linewidth]{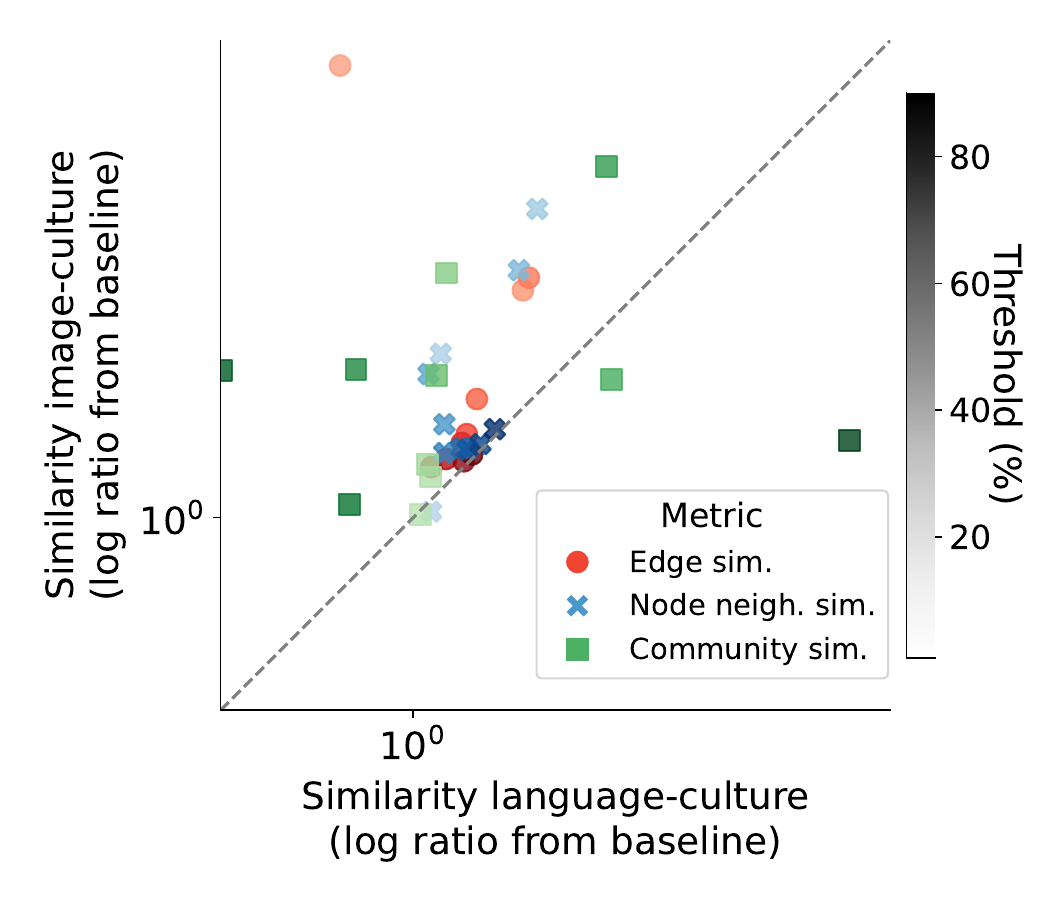}
        \caption{Financial.}
    \end{subfigure}
    \hfill
    \begin{subfigure}{0.23\textwidth}
        \includegraphics[width=\linewidth]{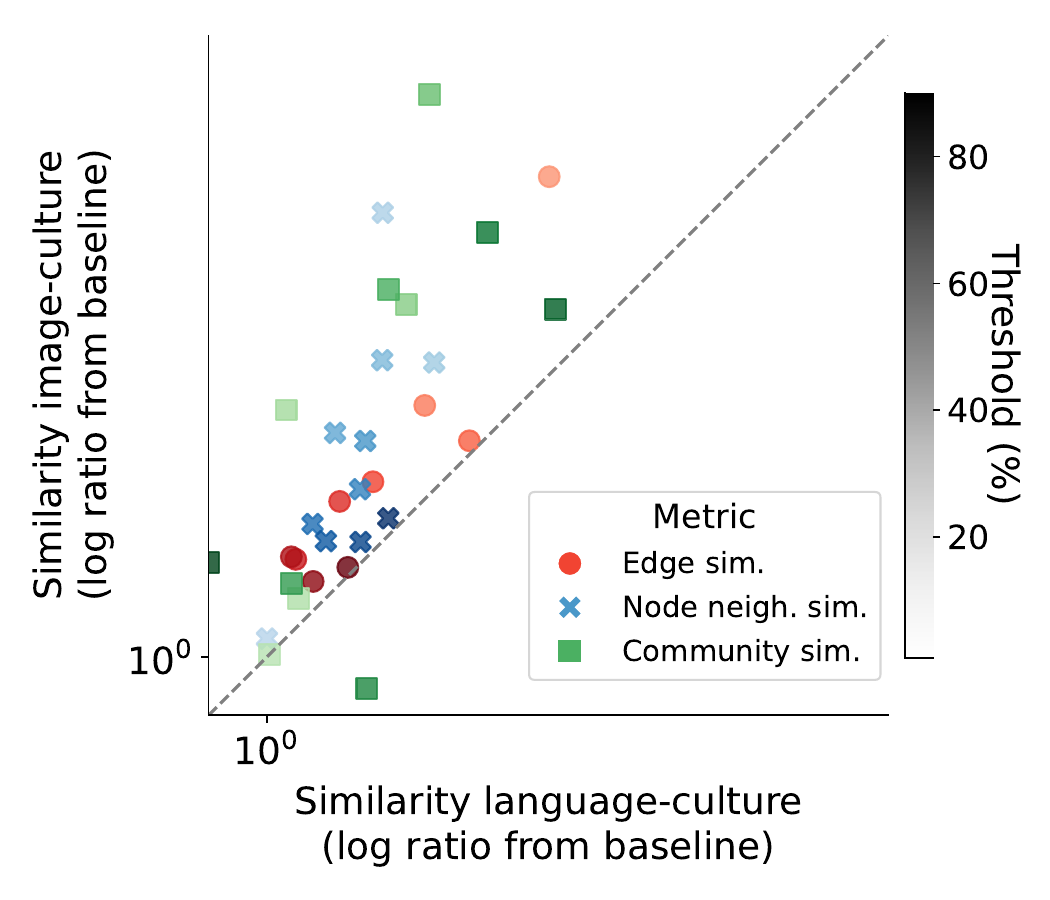}
        \caption{Group-memb.}
    \end{subfigure}
    \hfill
    \begin{subfigure}{0.23\textwidth}
        \includegraphics[width=\linewidth]{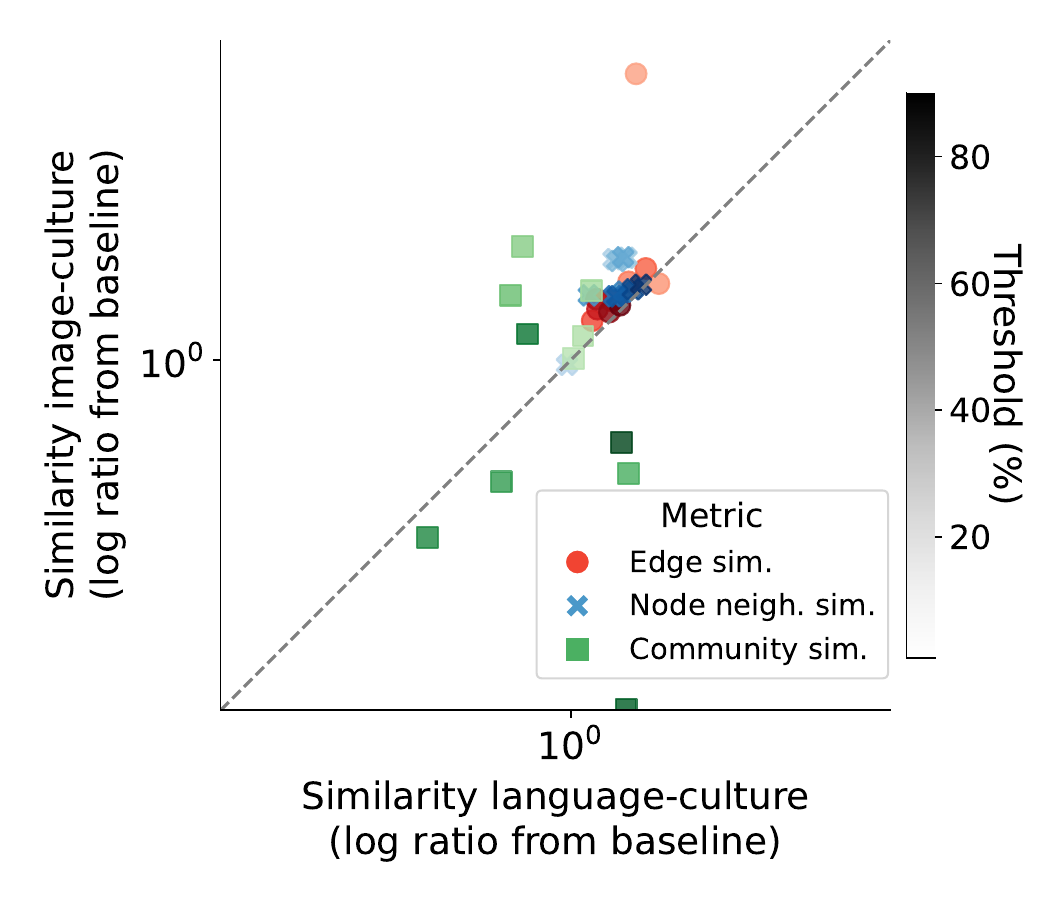}
        \caption{Law.}
    \end{subfigure}
    \hfill
    \begin{subfigure}{0.23\textwidth}
        \includegraphics[width=\linewidth]{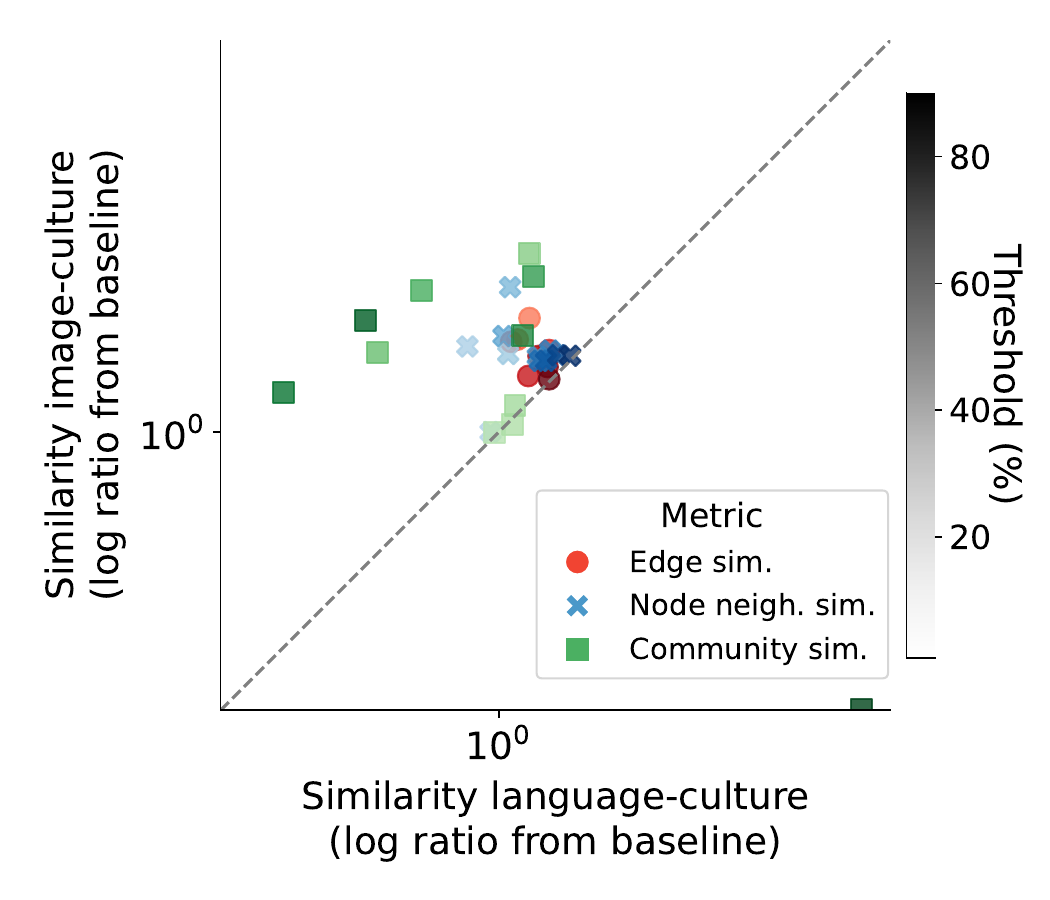}
        \caption{Leisure.}
    \end{subfigure}

    \vspace{0.3cm}
    \begin{subfigure}{0.23\textwidth}
        \includegraphics[width=\linewidth]{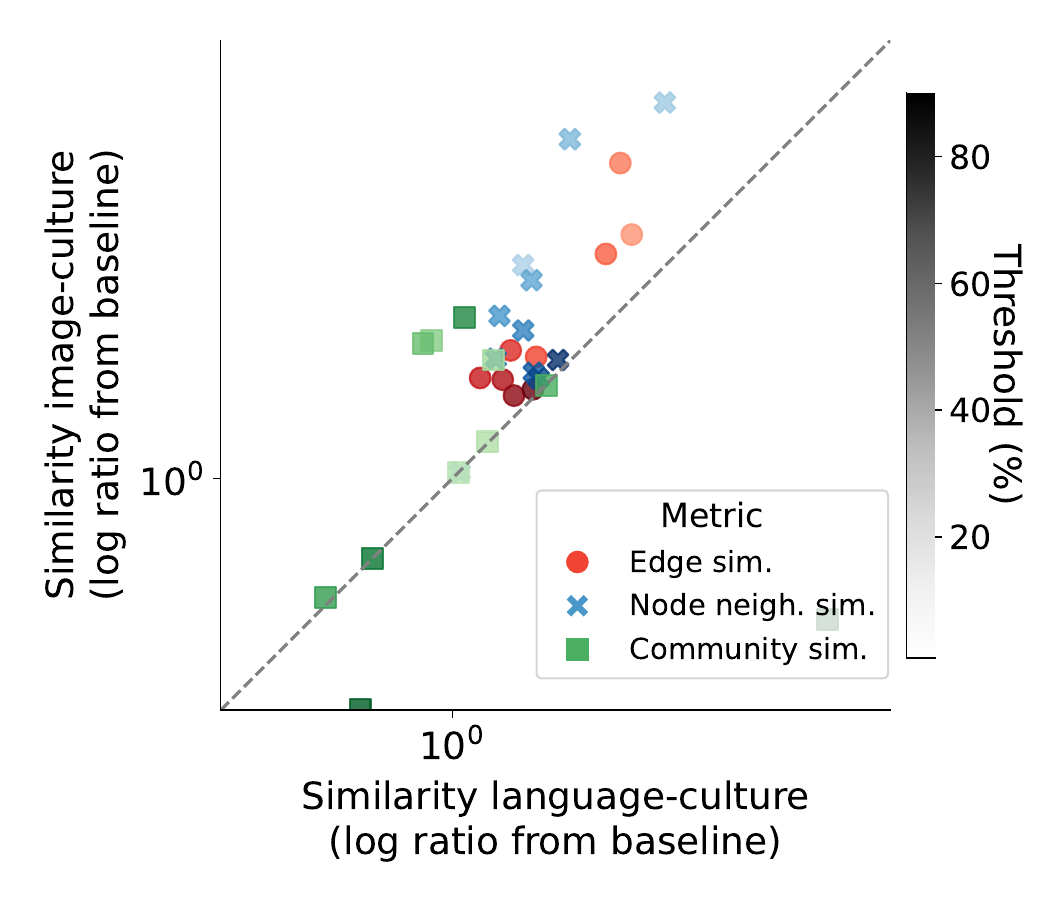}
        \caption{Political.}
    \end{subfigure}
    \hfill
    \begin{subfigure}{0.23\textwidth}
        \includegraphics[width=\linewidth]{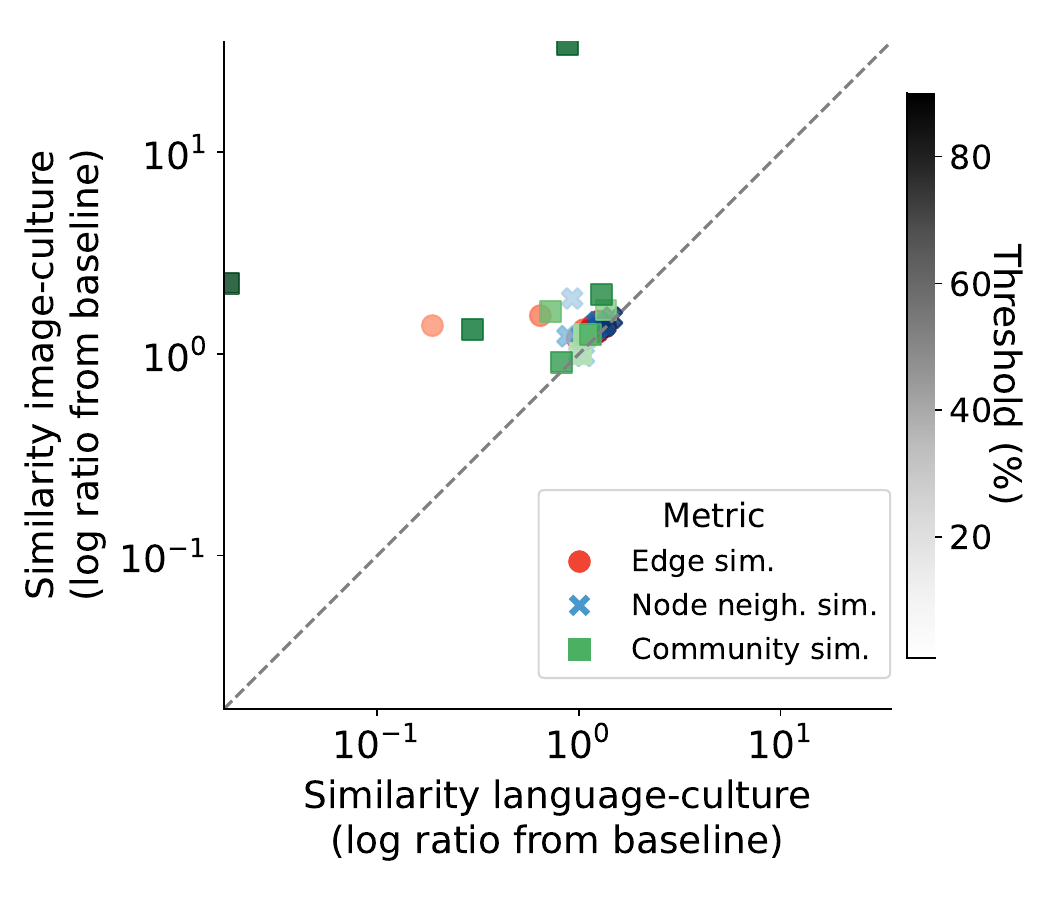}
        \caption{Science-innovation.}
    \end{subfigure}
    \hfill
    \begin{subfigure}{0.23\textwidth}
        \includegraphics[width=\linewidth]{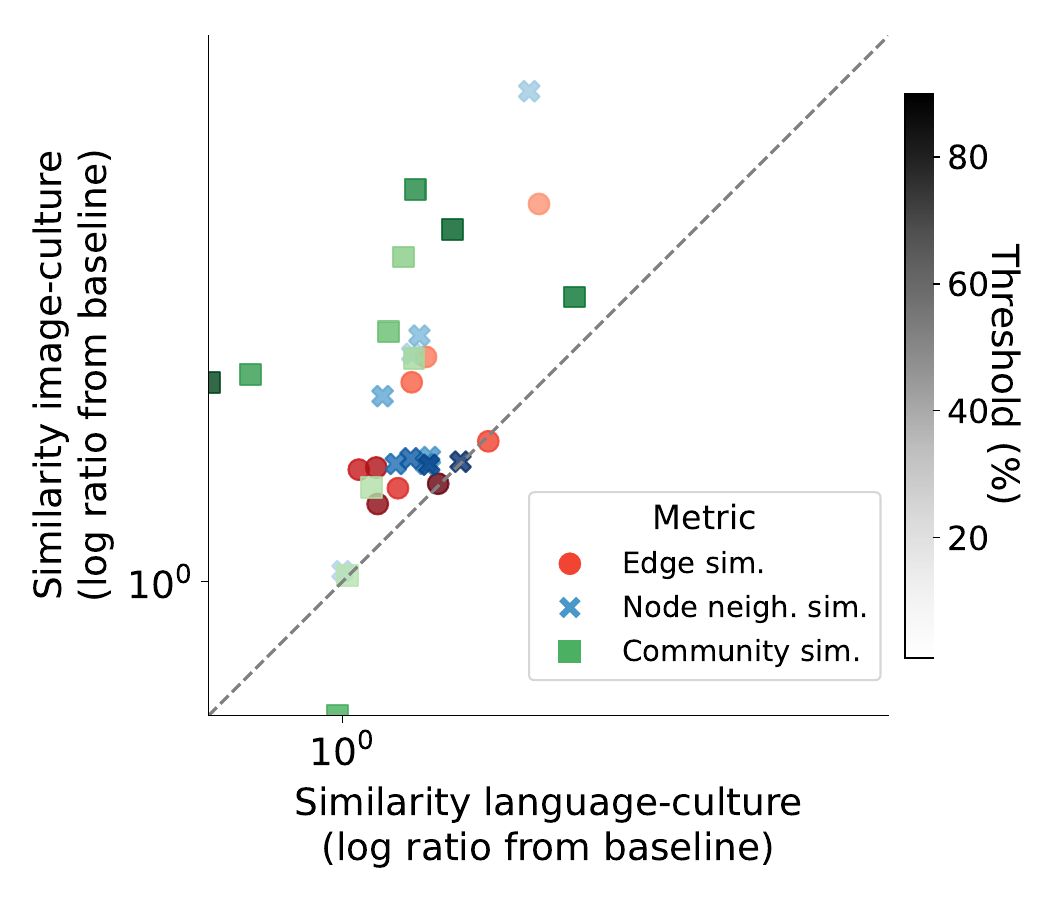}
        \caption{Sexuality.}
    \end{subfigure}
    \hfill
    \begin{subfigure}{0.23\textwidth}
        \includegraphics[width=\linewidth]{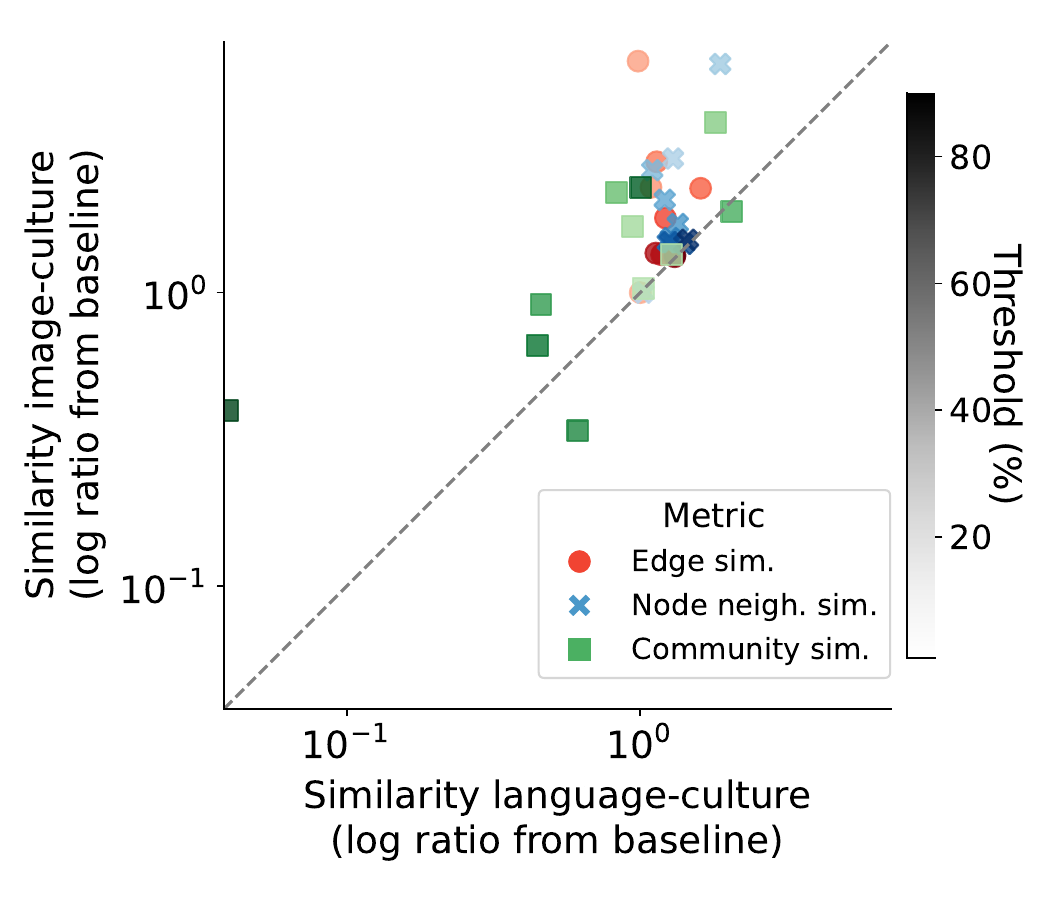}
        \caption{Social relations.}
    \end{subfigure}

    \vspace{0.3cm}
    \begin{subfigure}{0.23\textwidth}
        \includegraphics[width=\linewidth]{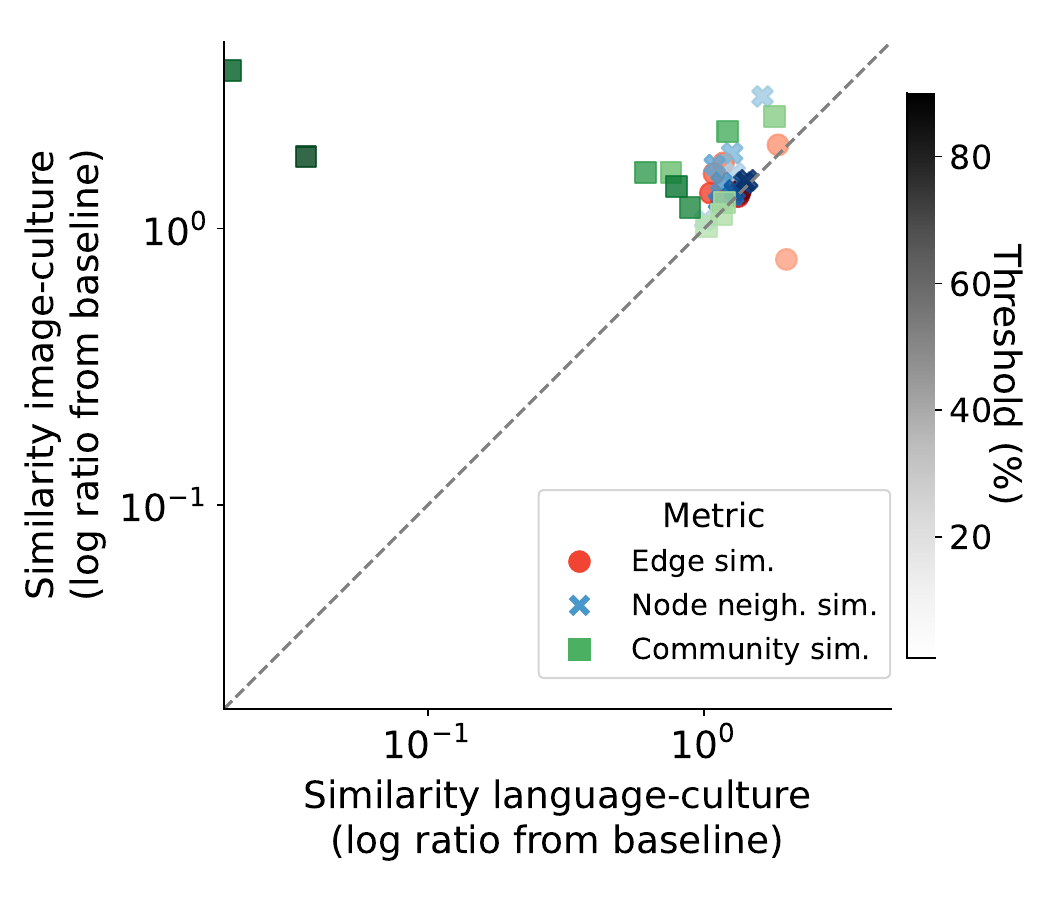}
        \caption{Survival}
    \end{subfigure}

    \caption{Comparison between image- and language-based network similarity with the cultural network, across metrics, considering the ratio with respect to a baseline.
    The analysis focuses on thematic dimensions of the World Values Survey.}
    \label{fig:network_metrics-robust_subdims}
\end{figure}

For a more fine-grained perspective, we also examine WVS sub-dimensions, namely altruism, arts-creativity, beliefs, environment-conservation, financial, group-membership, law, leisure-recreation-sports-hobbies, political, science-innovation, sexuality, social-relations, and survival. Supplementary Figure \ref{fig:network_metrics-robust_subdims} shows results for the 100-node network: median similarity ratios across metrics are greater than 1 for all sub-dimensions (ranging 1--1.30), confirming that the image-based network is more similar to the culture-based graph than the language-based network is across sub-dimensions of culture.

Moreover, we test the similarity between the image-based, language-based, and culture-based networks using an alternative edge-filtering procedure. Specifically, we apply disparity filtering to the image-based network.
We then construct the language-based and culture-based networks by retaining a number of edges matching the density of this reference network. The resulting ratios between image-culture and language-culture similarities (relative to a random baseline) are 1.68 for edge similarity, 1.96 for node-neighborhood similarity, and 1.80 for community similarity. These ratios greater than 1 confirm that the higher similarity of the image-based network to the culture-based graph, compared to the language-based one, is robust to the choice of network filtering method.

\end{document}